\PassOptionsToPackage{table}{xcolor} %
\documentclass[letterpaper,twocolumn,10pt]{article}
\usepackage{usenix-2020-09}

\usepackage[T1]{fontenc}
\usepackage[utf8]{inputenc} %
\usepackage[english]{babel}
\usepackage{adjustbox} %
\usepackage[table]{xcolor} %
\usepackage{tcolorbox} %
\usepackage{arrayjobx} %
\usepackage{xargs} %
\usepackage{amsmath} 
\usepackage{amssymb}

\definecolor{lightgray}{gray}{0.95}

\expandafter\let\csname c@tblerows\endcsname\rownum

\usepackage[
    backend=biber,
    isbn=false,
    doi=false,
    sorting=none,
    url=false,
    eprint=false,
    defernumbers=true,
    mincitenames=1,
    maxcitenames=1,
    maxbibnames=99, %
    mincrossrefs=99, %
    uniquelist=false,
    style=numeric-comp]{biblatex}

\usepackage[style=american, threshold=2]{csquotes}

\AtBeginEnvironment{quote}{\vspace{-0.3\baselineskip}}
\AtEndEnvironment{quote}{\vspace{-0.3\baselineskip}}

\usepackage[inline]{enumitem}
\usepackage{svg}
\usepackage{graphicx}
\usepackage{subcaption}
\usepackage{float}
\usepackage{tabularx}
\usepackage{booktabs}
\usepackage[para, flushleft]{threeparttable}
\usepackage{multirow}
\usepackage{makecell}
\usepackage{tikz}
\usetikzlibrary{shapes,arrows,fit,matrix,positioning}
\usepackage{hyperref}

\usepackage[gen]{eurosym}

\newlength\bubblesize
\usepackage[usetwoside=true]{mdframed}
\usepackage{tcolorbox}
\tcbuselibrary{breakable}
\tcbuselibrary{skins}
\usepackage{enumitem}
\definecolor{darkgray}{gray}{0.3}
\tcbset{}
\newtcolorbox{summaryBox}[2][]
{
    enhanced,
    breakable,
    frame hidden,
    borderline west = {3pt}{0pt}{lightgray},
    colback         = white,
    size            = fbox,
    left            = 0.3em,
    coltitle        = black,
    title           = {\color{darkgray} #2. },
    attach title to upper,
    #1,
}

\makeatletter
\newcommand\boldparagraph{%
    \@startsection{boldparagraph}{4}{0\parindent}{2pt plus 1pt minus 1pt}{0pt}
    {\noindent\normalfont\normalsize\bfseries\maybe@addperiod}*%
}
\newcommand{\maybe@addperiod}[1]{%
    \let\@period\@empty%
    \def\@IEEEsectpunct{}%
    #1\@addpunct{.}\enspace%
}
\makeatother

\newcommand{\eg}{e.\,g.,}

\newcommand{\cf}{cf.}

\newcommand{\availability}{\emph{Availability} section}

\newcommand\COMMENTX[2]{}

\newcommand\nicolas[1]{}
\newcommand\alex[1]{}
\newcommand\dominik[1]{}
\newcommand\sascha[1]{}
\newcommand\jan[1]{}
\newcommand\todo[1]{}
\newcommand\yas[1]{}

\newcommand\revision[1]{#1}
\newcommand\revisionII[1]{#1}

\newcommand\definevar[2]{%
  \expandafter\newcommand\csname var#1var\endcsname{#2}%
}
\newcommand{\var}[1]{\ifcsname var#1var\endcsname%
        \csname var#1var\endcsname%
    \else\PackageWarning{Var}{`#1' does not exist.}%
        TODO%
    \fi%
}


\usepackage[skip=5pt]{caption}
\usepackage[]{hyperref}
\hypersetup{
    hidelinks,%
}

\usepackage{orcidlink}

\usepackage[acronym, %
    nohypertypes={acronym}, %
]{glossaries}
\newacronym{aic}{AIC}{Akaike Information Criterion}
\newacronym{vcs}{VCS}{version control system}
\newacronym{irb}{IRB}{institutional review board}
\newacronym{erb}{ERB}{ethical review board}
\newacronym{gdpr}{GDPR}{General Data Protection Regulation}
\newacronym{api}{API}{application programming interface}
\newacronym{cvs}{CVS}{Concurrent Versions System}
\newacronym{tfvc}{TFVC}{Microsoft Team Foundation Version Control}
\newacronym{aws}{AWS}{Amazon Web Services}
\newacronym{awscc}{AWSCC}{Amazon Web Services Code Commit}
\newacronym{gc}{GC}{Google Cloud}
\newacronym[firstplural=Common Vulnerabilities and Exposures (CVEs)]{cve}{CVE}{Common Vulnerabilities and Exposures}
\newacronym{svn}{SVN}{Apache Subversion}
\newacronym{dos}{DoS}{Denial of Service}
\newacronym{pii}{PII}{personally identifiable information}
\newacronym{mturk}{MTurk}{Amazon Mechanical Turk}
\newacronym{saas}{SaaS}{Secret-as-a-Service}
\newacronym{irr}{IRR}{Inter-rater Reliability}

\definevar{survey.participants.github.started}{101}
\definevar{survey.participants.upwork.started}{52}
\definevar{survey.participants.total.started}{153}
\definevar{survey.participants.github.finished}{59}
\definevar{survey.participants.upwork.finished}{51}
\definevar{survey.participants.total.finished}{110}
\definevar{interviews.participants.total}{14}
\definevar{survey.participants.github}{59}
\definevar{survey.participants.upwork}{50}
\definevar{survey.participants.total}{109}
\definevar{survey.participants.github.started.text}{101}
\definevar{survey.participants.github.started.Text}{101}
\definevar{survey.participants.upwork.started.text}{52}
\definevar{survey.participants.upwork.started.Text}{52}
\definevar{survey.participants.total.started.text}{153}
\definevar{survey.participants.total.started.Text}{153}
\definevar{survey.participants.github.finished.text}{59}
\definevar{survey.participants.github.finished.Text}{59}
\definevar{survey.participants.upwork.finished.text}{51}
\definevar{survey.participants.upwork.finished.Text}{51}
\definevar{survey.participants.total.finished.text}{110}
\definevar{survey.participants.total.finished.Text}{110}
\definevar{interviews.participants.total.text}{14}
\definevar{interviews.participants.total.Text}{14}
\definevar{survey.participants.github.text}{59}
\definevar{survey.participants.github.Text}{59}
\definevar{survey.participants.upwork.text}{50}
\definevar{survey.participants.upwork.Text}{50}
\definevar{survey.participants.total.text}{109}
\definevar{survey.participants.total.Text}{109}

\definevar{demo.all.gender.n/a}{0}
\definevar{demo.all.gender.n/a.perc}{0.0\%}
\definevar{demo.all.gender.Man}{95}
\definevar{demo.all.gender.Woman}{6}
\definevar{demo.all.gender.Prefernottodisclose}{4}
\definevar{demo.all.gender.Non-binary}{4}
\definevar{demo.all.gender.Man.perc}{87.2\%}
\definevar{demo.all.gender.Woman.perc}{5.5\%}
\definevar{demo.all.gender.Prefernottodisclose.perc}{3.7\%}
\definevar{demo.all.gender.Non-binary.perc}{3.7\%}
\definevar{demo.upwork.gender.n/a}{0}
\definevar{demo.upwork.gender.n/a.perc}{0.0\%}
\definevar{demo.upwork.gender.Man}{43}
\definevar{demo.upwork.gender.Woman}{5}
\definevar{demo.upwork.gender.Prefernottodisclose}{2}
\definevar{demo.upwork.gender.Man.perc}{86.0\%}
\definevar{demo.upwork.gender.Woman.perc}{10.0\%}
\definevar{demo.upwork.gender.Prefernottodisclose.perc}{4.0\%}
\definevar{demo.github.gender.n/a}{0}
\definevar{demo.github.gender.n/a.perc}{0.0\%}
\definevar{demo.github.gender.Man}{52}
\definevar{demo.github.gender.Non-binary}{4}
\definevar{demo.github.gender.Prefernottodisclose}{2}
\definevar{demo.github.gender.Woman}{1}
\definevar{demo.github.gender.Man.perc}{88.1\%}
\definevar{demo.github.gender.Non-binary.perc}{6.8\%}
\definevar{demo.github.gender.Prefernottodisclose.perc}{3.4\%}
\definevar{demo.github.gender.Woman.perc}{1.7\%}
\definevar{demo.interviews.gender.n/a}{0}
\definevar{demo.interviews.gender.n/a.perc}{0.0\%}
\definevar{demo.interviews.gender.Man}{13}
\definevar{demo.interviews.gender.Non-binary}{1}
\definevar{demo.interviews.gender.Man.perc}{92.9\%}
\definevar{demo.interviews.gender.Non-binary.perc}{7.1\%}
\definevar{demo.all.age.n/a}{1}
\definevar{demo.all.age.n/a.perc}{0.9\%}
\definevar{demo.all.age.count}{108.0}
\definevar{demo.all.age.mean}{33.2}
\definevar{demo.all.age.std}{11.1}
\definevar{demo.all.age.min}{18.0}
\definevar{demo.all.age.25percentile}{25.0}
\definevar{demo.all.age.50percentile}{30.0}
\definevar{demo.all.age.75percentile}{38.2}
\definevar{demo.all.age.max}{70.0}
\definevar{demo.upwork.age.n/a}{0}
\definevar{demo.upwork.age.n/a.perc}{0.0\%}
\definevar{demo.upwork.age.count}{50.0}
\definevar{demo.upwork.age.mean}{31.3}
\definevar{demo.upwork.age.std}{8.6}
\definevar{demo.upwork.age.min}{21.0}
\definevar{demo.upwork.age.25percentile}{26.0}
\definevar{demo.upwork.age.50percentile}{29.0}
\definevar{demo.upwork.age.75percentile}{34.8}
\definevar{demo.upwork.age.max}{70.0}
\definevar{demo.github.age.n/a}{1}
\definevar{demo.github.age.n/a.perc}{1.7\%}
\definevar{demo.github.age.count}{58.0}
\definevar{demo.github.age.mean}{34.9}
\definevar{demo.github.age.std}{12.7}
\definevar{demo.github.age.min}{18.0}
\definevar{demo.github.age.25percentile}{25.0}
\definevar{demo.github.age.50percentile}{33.0}
\definevar{demo.github.age.75percentile}{42.8}
\definevar{demo.github.age.max}{65.0}
\definevar{demo.interviews.age.n/a}{0}
\definevar{demo.interviews.age.n/a.perc}{0.0\%}
\definevar{demo.interviews.age.count}{14.0}
\definevar{demo.interviews.age.mean}{32.1}
\definevar{demo.interviews.age.std}{11.8}
\definevar{demo.interviews.age.min}{19.0}
\definevar{demo.interviews.age.25percentile}{24.0}
\definevar{demo.interviews.age.50percentile}{28.0}
\definevar{demo.interviews.age.75percentile}{39.8}
\definevar{demo.interviews.age.max}{62.0}
\definevar{demo.all.countries.Prefernottodisclose}{3}
\definevar{demo.all.countries.Prefernottodisclose.perc}{2.8\%}
\definevar{demo.all.countries.UnitedStatesofAmerica}{20}
\definevar{demo.all.countries.India}{12}
\definevar{demo.all.countries.Germany}{11}
\definevar{demo.all.countries.Pakistan}{9}
\definevar{demo.all.countries.RussianFederation}{4}
\definevar{demo.all.countries.Nigeria}{4}
\definevar{demo.all.countries.Canada}{4}
\definevar{demo.all.countries.UnitedKingdomofGreatBritainandNorthernIreland}{4}
\definevar{demo.all.countries.Kenya}{4}
\definevar{demo.all.countries.nan}{3}
\definevar{demo.all.countries.France}{3}
\definevar{demo.all.countries.Denmark}{3}
\definevar{demo.all.countries.Sweden}{2}
\definevar{demo.all.countries.SriLanka}{2}
\definevar{demo.all.countries.Indonesia}{2}
\definevar{demo.all.countries.Egypt}{2}
\definevar{demo.all.countries.Brazil}{2}
\definevar{demo.all.countries.Poland}{2}
\definevar{demo.all.countries.Singapore}{1}
\definevar{demo.all.countries.SouthKorea}{1}
\definevar{demo.all.countries.HongKong(S.A.R.)}{1}
\definevar{demo.all.countries.TheformerYugoslavRepublicofMacedonia}{1}
\definevar{demo.all.countries.Thailand}{1}
\definevar{demo.all.countries.Hungary}{1}
\definevar{demo.all.countries.China}{1}
\definevar{demo.all.countries.Albania}{1}
\definevar{demo.all.countries.Israel}{1}
\definevar{demo.all.countries.Portugal}{1}
\definevar{demo.all.countries.DominicanRepublic}{1}
\definevar{demo.all.countries.Netherlands}{1}
\definevar{demo.all.countries.Georgia}{1}
\definevar{demo.all.countries.Latvia}{1}
\definevar{demo.all.countries.Mexico}{1}
\definevar{demo.all.countries.Australia}{1}
\definevar{demo.all.countries.UnitedStatesofAmerica.perc}{18.3\%}
\definevar{demo.all.countries.India.perc}{11.0\%}
\definevar{demo.all.countries.Germany.perc}{10.1\%}
\definevar{demo.all.countries.Pakistan.perc}{8.3\%}
\definevar{demo.all.countries.RussianFederation.perc}{3.7\%}
\definevar{demo.all.countries.Nigeria.perc}{3.7\%}
\definevar{demo.all.countries.Canada.perc}{3.7\%}
\definevar{demo.all.countries.UnitedKingdomofGreatBritainandNorthernIreland.perc}{3.7\%}
\definevar{demo.all.countries.Kenya.perc}{3.7\%}
\definevar{demo.all.countries.nan.perc}{2.8\%}
\definevar{demo.all.countries.France.perc}{2.8\%}
\definevar{demo.all.countries.Denmark.perc}{2.8\%}
\definevar{demo.all.countries.Sweden.perc}{1.8\%}
\definevar{demo.all.countries.SriLanka.perc}{1.8\%}
\definevar{demo.all.countries.Indonesia.perc}{1.8\%}
\definevar{demo.all.countries.Egypt.perc}{1.8\%}
\definevar{demo.all.countries.Brazil.perc}{1.8\%}
\definevar{demo.all.countries.Poland.perc}{1.8\%}
\definevar{demo.all.countries.Singapore.perc}{0.9\%}
\definevar{demo.all.countries.SouthKorea.perc}{0.9\%}
\definevar{demo.all.countries.HongKong(S.A.R.).perc}{0.9\%}
\definevar{demo.all.countries.TheformerYugoslavRepublicofMacedonia.perc}{0.9\%}
\definevar{demo.all.countries.Thailand.perc}{0.9\%}
\definevar{demo.all.countries.Hungary.perc}{0.9\%}
\definevar{demo.all.countries.China.perc}{0.9\%}
\definevar{demo.all.countries.Albania.perc}{0.9\%}
\definevar{demo.all.countries.Israel.perc}{0.9\%}
\definevar{demo.all.countries.Portugal.perc}{0.9\%}
\definevar{demo.all.countries.DominicanRepublic.perc}{0.9\%}
\definevar{demo.all.countries.Netherlands.perc}{0.9\%}
\definevar{demo.all.countries.Georgia.perc}{0.9\%}
\definevar{demo.all.countries.Latvia.perc}{0.9\%}
\definevar{demo.all.countries.Mexico.perc}{0.9\%}
\definevar{demo.all.countries.Australia.perc}{0.9\%}
\definevar{demo.upwork.countries.Prefernottodisclose}{2}
\definevar{demo.upwork.countries.Prefernottodisclose.perc}{4.0\%}
\definevar{demo.upwork.countries.India}{10}
\definevar{demo.upwork.countries.Pakistan}{7}
\definevar{demo.upwork.countries.Kenya}{4}
\definevar{demo.upwork.countries.Canada}{2}
\definevar{demo.upwork.countries.nan}{2}
\definevar{demo.upwork.countries.Nigeria}{2}
\definevar{demo.upwork.countries.SriLanka}{2}
\definevar{demo.upwork.countries.Egypt}{2}
\definevar{demo.upwork.countries.Indonesia}{2}
\definevar{demo.upwork.countries.UnitedStatesofAmerica}{1}
\definevar{demo.upwork.countries.Netherlands}{1}
\definevar{demo.upwork.countries.Denmark}{1}
\definevar{demo.upwork.countries.RussianFederation}{1}
\definevar{demo.upwork.countries.Portugal}{1}
\definevar{demo.upwork.countries.Sweden}{1}
\definevar{demo.upwork.countries.Israel}{1}
\definevar{demo.upwork.countries.DominicanRepublic}{1}
\definevar{demo.upwork.countries.Poland}{1}
\definevar{demo.upwork.countries.Mexico}{1}
\definevar{demo.upwork.countries.UnitedKingdomofGreatBritainandNorthernIreland}{1}
\definevar{demo.upwork.countries.Albania}{1}
\definevar{demo.upwork.countries.Latvia}{1}
\definevar{demo.upwork.countries.Georgia}{1}
\definevar{demo.upwork.countries.Singapore}{1}
\definevar{demo.upwork.countries.China}{1}
\definevar{demo.upwork.countries.France}{1}
\definevar{demo.upwork.countries.India.perc}{20.0\%}
\definevar{demo.upwork.countries.Pakistan.perc}{14.0\%}
\definevar{demo.upwork.countries.Kenya.perc}{8.0\%}
\definevar{demo.upwork.countries.Canada.perc}{4.0\%}
\definevar{demo.upwork.countries.nan.perc}{4.0\%}
\definevar{demo.upwork.countries.Nigeria.perc}{4.0\%}
\definevar{demo.upwork.countries.SriLanka.perc}{4.0\%}
\definevar{demo.upwork.countries.Egypt.perc}{4.0\%}
\definevar{demo.upwork.countries.Indonesia.perc}{4.0\%}
\definevar{demo.upwork.countries.UnitedStatesofAmerica.perc}{2.0\%}
\definevar{demo.upwork.countries.Netherlands.perc}{2.0\%}
\definevar{demo.upwork.countries.Denmark.perc}{2.0\%}
\definevar{demo.upwork.countries.RussianFederation.perc}{2.0\%}
\definevar{demo.upwork.countries.Portugal.perc}{2.0\%}
\definevar{demo.upwork.countries.Sweden.perc}{2.0\%}
\definevar{demo.upwork.countries.Israel.perc}{2.0\%}
\definevar{demo.upwork.countries.DominicanRepublic.perc}{2.0\%}
\definevar{demo.upwork.countries.Poland.perc}{2.0\%}
\definevar{demo.upwork.countries.Mexico.perc}{2.0\%}
\definevar{demo.upwork.countries.UnitedKingdomofGreatBritainandNorthernIreland.perc}{2.0\%}
\definevar{demo.upwork.countries.Albania.perc}{2.0\%}
\definevar{demo.upwork.countries.Latvia.perc}{2.0\%}
\definevar{demo.upwork.countries.Georgia.perc}{2.0\%}
\definevar{demo.upwork.countries.Singapore.perc}{2.0\%}
\definevar{demo.upwork.countries.China.perc}{2.0\%}
\definevar{demo.upwork.countries.France.perc}{2.0\%}
\definevar{demo.github.countries.Prefernottodisclose}{1}
\definevar{demo.github.countries.Prefernottodisclose.perc}{1.7\%}
\definevar{demo.github.countries.UnitedStatesofAmerica}{19}
\definevar{demo.github.countries.Germany}{11}
\definevar{demo.github.countries.UnitedKingdomofGreatBritainandNorthernIreland}{3}
\definevar{demo.github.countries.RussianFederation}{3}
\definevar{demo.github.countries.India}{2}
\definevar{demo.github.countries.Brazil}{2}
\definevar{demo.github.countries.France}{2}
\definevar{demo.github.countries.Canada}{2}
\definevar{demo.github.countries.Denmark}{2}
\definevar{demo.github.countries.Nigeria}{2}
\definevar{demo.github.countries.Pakistan}{2}
\definevar{demo.github.countries.TheformerYugoslavRepublicofMacedonia}{1}
\definevar{demo.github.countries.Australia}{1}
\definevar{demo.github.countries.SouthKorea}{1}
\definevar{demo.github.countries.HongKong(S.A.R.)}{1}
\definevar{demo.github.countries.Poland}{1}
\definevar{demo.github.countries.Sweden}{1}
\definevar{demo.github.countries.Hungary}{1}
\definevar{demo.github.countries.nan}{1}
\definevar{demo.github.countries.Thailand}{1}
\definevar{demo.github.countries.UnitedStatesofAmerica.perc}{32.2\%}
\definevar{demo.github.countries.Germany.perc}{18.6\%}
\definevar{demo.github.countries.UnitedKingdomofGreatBritainandNorthernIreland.perc}{5.1\%}
\definevar{demo.github.countries.RussianFederation.perc}{5.1\%}
\definevar{demo.github.countries.India.perc}{3.4\%}
\definevar{demo.github.countries.Brazil.perc}{3.4\%}
\definevar{demo.github.countries.France.perc}{3.4\%}
\definevar{demo.github.countries.Canada.perc}{3.4\%}
\definevar{demo.github.countries.Denmark.perc}{3.4\%}
\definevar{demo.github.countries.Nigeria.perc}{3.4\%}
\definevar{demo.github.countries.Pakistan.perc}{3.4\%}
\definevar{demo.github.countries.TheformerYugoslavRepublicofMacedonia.perc}{1.7\%}
\definevar{demo.github.countries.Australia.perc}{1.7\%}
\definevar{demo.github.countries.SouthKorea.perc}{1.7\%}
\definevar{demo.github.countries.HongKong(S.A.R.).perc}{1.7\%}
\definevar{demo.github.countries.Poland.perc}{1.7\%}
\definevar{demo.github.countries.Sweden.perc}{1.7\%}
\definevar{demo.github.countries.Hungary.perc}{1.7\%}
\definevar{demo.github.countries.nan.perc}{1.7\%}
\definevar{demo.github.countries.Thailand.perc}{1.7\%}
\definevar{demo.interviews.countries.Prefernottodisclose}{0}
\definevar{demo.interviews.countries.Prefernottodisclose.perc}{0.0\%}
\definevar{demo.interviews.countries.India}{3}
\definevar{demo.interviews.countries.UnitedStatesofAmerica}{3}
\definevar{demo.interviews.countries.Pakistan}{2}
\definevar{demo.interviews.countries.NewZealand}{1}
\definevar{demo.interviews.countries.Canada}{1}
\definevar{demo.interviews.countries.Italy}{1}
\definevar{demo.interviews.countries.Belarus}{1}
\definevar{demo.interviews.countries.Brazil}{1}
\definevar{demo.interviews.countries.Kenya}{1}
\definevar{demo.interviews.countries.India.perc}{21.4\%}
\definevar{demo.interviews.countries.UnitedStatesofAmerica.perc}{21.4\%}
\definevar{demo.interviews.countries.Pakistan.perc}{14.3\%}
\definevar{demo.interviews.countries.NewZealand.perc}{7.1\%}
\definevar{demo.interviews.countries.Canada.perc}{7.1\%}
\definevar{demo.interviews.countries.Italy.perc}{7.1\%}
\definevar{demo.interviews.countries.Belarus.perc}{7.1\%}
\definevar{demo.interviews.countries.Brazil.perc}{7.1\%}
\definevar{demo.interviews.countries.Kenya.perc}{7.1\%}
\definevar{demo.all.countries.distinct}{33}
\definevar{demo.all.countries.other}{54}
\definevar{demo.all.countries.other.perc}{49.5\%}
\definevar{demo.upwork.countries.other}{30}
\definevar{demo.upwork.countries.other.perc}{60.0\%}
\definevar{demo.github.countries.other}{24}
\definevar{demo.github.countries.other.perc}{40.7\%}
\definevar{demo.all.edu.Prefernottodisclose}{0}
\definevar{demo.all.edu.Prefernottodisclose.perc}{0.0\%}
\definevar{demo.all.edu.Bachelor}{51}
\definevar{demo.all.edu.Master}{26}
\definevar{demo.all.edu.Somecollege,currentlyenrolledincollegeortwo-yearassociate’sdegree}{15}
\definevar{demo.all.edu.Doctoral}{7}
\definevar{demo.all.edu.Highschoolorequivalent}{6}
\definevar{demo.all.edu.Somegraduateschoolorcurrentlyenrolledingraduateschool}{4}
\definevar{demo.all.edu.Bachelor.perc}{46.8\%}
\definevar{demo.all.edu.Master.perc}{23.9\%}
\definevar{demo.all.edu.Somecollege,currentlyenrolledincollegeortwo-yearassociate’sdegree.perc}{13.8\%}
\definevar{demo.all.edu.Doctoral.perc}{6.4\%}
\definevar{demo.all.edu.Highschoolorequivalent.perc}{5.5\%}
\definevar{demo.all.edu.Somegraduateschoolorcurrentlyenrolledingraduateschool.perc}{3.7\%}
\definevar{demo.upwork.edu.Prefernottodisclose}{0}
\definevar{demo.upwork.edu.Prefernottodisclose.perc}{0.0\%}
\definevar{demo.upwork.edu.Bachelor}{30}
\definevar{demo.upwork.edu.Master}{14}
\definevar{demo.upwork.edu.Somecollege,currentlyenrolledincollegeortwo-yearassociate’sdegree}{3}
\definevar{demo.upwork.edu.Somegraduateschoolorcurrentlyenrolledingraduateschool}{2}
\definevar{demo.upwork.edu.Doctoral}{1}
\definevar{demo.upwork.edu.Bachelor.perc}{60.0\%}
\definevar{demo.upwork.edu.Master.perc}{28.0\%}
\definevar{demo.upwork.edu.Somecollege,currentlyenrolledincollegeortwo-yearassociate’sdegree.perc}{6.0\%}
\definevar{demo.upwork.edu.Somegraduateschoolorcurrentlyenrolledingraduateschool.perc}{4.0\%}
\definevar{demo.upwork.edu.Doctoral.perc}{2.0\%}
\definevar{demo.github.edu.Prefernottodisclose}{0}
\definevar{demo.github.edu.Prefernottodisclose.perc}{0.0\%}
\definevar{demo.github.edu.Bachelor}{21}
\definevar{demo.github.edu.Somecollege,currentlyenrolledincollegeortwo-yearassociate’sdegree}{12}
\definevar{demo.github.edu.Master}{12}
\definevar{demo.github.edu.Doctoral}{6}
\definevar{demo.github.edu.Highschoolorequivalent}{6}
\definevar{demo.github.edu.Somegraduateschoolorcurrentlyenrolledingraduateschool}{2}
\definevar{demo.github.edu.Bachelor.perc}{35.6\%}
\definevar{demo.github.edu.Somecollege,currentlyenrolledincollegeortwo-yearassociate’sdegree.perc}{20.3\%}
\definevar{demo.github.edu.Master.perc}{20.3\%}
\definevar{demo.github.edu.Doctoral.perc}{10.2\%}
\definevar{demo.github.edu.Highschoolorequivalent.perc}{10.2\%}
\definevar{demo.github.edu.Somegraduateschoolorcurrentlyenrolledingraduateschool.perc}{3.4\%}
\definevar{demo.all.employment.Prefernottodisclose}{2}
\definevar{demo.all.employment.Prefernottodisclose.perc}{1.8\%}
\definevar{demo.all.employment.Employedfull-time}{71}
\definevar{demo.all.employment.Self-employed/Freelancer}{33}
\definevar{demo.all.employment.Student}{12}
\definevar{demo.all.employment.Employedpart-time}{11}
\definevar{demo.all.employment.Lookingforwork}{4}
\definevar{demo.all.employment.Retired}{3}
\definevar{demo.all.employment.Employedfull-time.perc}{65.1\%}
\definevar{demo.all.employment.Self-employed/Freelancer.perc}{30.3\%}
\definevar{demo.all.employment.Student.perc}{11.0\%}
\definevar{demo.all.employment.Employedpart-time.perc}{10.1\%}
\definevar{demo.all.employment.Lookingforwork.perc}{3.7\%}
\definevar{demo.all.employment.Retired.perc}{2.8\%}
\definevar{demo.upwork.employment.Prefernottodisclose}{2}
\definevar{demo.upwork.employment.Prefernottodisclose.perc}{4.0\%}
\definevar{demo.upwork.employment.Employedfull-time}{37}
\definevar{demo.upwork.employment.Self-employed/Freelancer}{18}
\definevar{demo.upwork.employment.Employedpart-time}{3}
\definevar{demo.upwork.employment.Student}{2}
\definevar{demo.upwork.employment.Prefertoself-describe}{2}
\definevar{demo.upwork.employment.Lookingforwork}{1}
\definevar{demo.upwork.employment.Employedfull-time.perc}{74.0\%}
\definevar{demo.upwork.employment.Self-employed/Freelancer.perc}{36.0\%}
\definevar{demo.upwork.employment.Employedpart-time.perc}{6.0\%}
\definevar{demo.upwork.employment.Student.perc}{4.0\%}
\definevar{demo.upwork.employment.Prefertoself-describe.perc}{4.0\%}
\definevar{demo.upwork.employment.Lookingforwork.perc}{2.0\%}
\definevar{demo.github.employment.Prefernottodisclose}{0}
\definevar{demo.github.employment.Prefernottodisclose.perc}{0.0\%}
\definevar{demo.github.employment.Employedfull-time}{34}
\definevar{demo.github.employment.Self-employed/Freelancer}{15}
\definevar{demo.github.employment.Student}{10}
\definevar{demo.github.employment.Employedpart-time}{8}
\definevar{demo.github.employment.Retired}{3}
\definevar{demo.github.employment.Lookingforwork}{2}
\definevar{demo.github.employment.Employedfull-time.perc}{57.6\%}
\definevar{demo.github.employment.Self-employed/Freelancer.perc}{25.4\%}
\definevar{demo.github.employment.Student.perc}{16.9\%}
\definevar{demo.github.employment.Employedpart-time.perc}{13.6\%}
\definevar{demo.github.employment.Retired.perc}{5.1\%}
\definevar{demo.github.employment.Lookingforwork.perc}{3.4\%}
\definevar{demo.all.worked.Prefernottodisclose}{0}
\definevar{demo.all.worked.Prefernottodisclose.perc}{0.0\%}
\definevar{demo.all.worked.Company}{90}
\definevar{demo.all.worked.Solo}{76}
\definevar{demo.all.worked.OpenSourceProject}{60}
\definevar{demo.all.worked.Prefertoself-describe}{4}
\definevar{demo.all.worked.Company.perc}{82.6\%}
\definevar{demo.all.worked.Solo.perc}{69.7\%}
\definevar{demo.all.worked.OpenSourceProject.perc}{55.0\%}
\definevar{demo.all.worked.Prefertoself-describe.perc}{3.7\%}
\definevar{demo.upwork.worked.Prefernottodisclose}{0}
\definevar{demo.upwork.worked.Prefernottodisclose.perc}{0.0\%}
\definevar{demo.upwork.worked.Company}{48}
\definevar{demo.upwork.worked.Solo}{32}
\definevar{demo.upwork.worked.OpenSourceProject}{14}
\definevar{demo.upwork.worked.Prefertoself-describe}{1}
\definevar{demo.upwork.worked.Company.perc}{96.0\%}
\definevar{demo.upwork.worked.Solo.perc}{64.0\%}
\definevar{demo.upwork.worked.OpenSourceProject.perc}{28.0\%}
\definevar{demo.upwork.worked.Prefertoself-describe.perc}{2.0\%}
\definevar{demo.github.worked.Prefernottodisclose}{0}
\definevar{demo.github.worked.Prefernottodisclose.perc}{0.0\%}
\definevar{demo.github.worked.OpenSourceProject}{46}
\definevar{demo.github.worked.Solo}{44}
\definevar{demo.github.worked.Company}{42}
\definevar{demo.github.worked.Prefertoself-describe}{3}
\definevar{demo.github.worked.OpenSourceProject.perc}{78.0\%}
\definevar{demo.github.worked.Solo.perc}{74.6\%}
\definevar{demo.github.worked.Company.perc}{71.2\%}
\definevar{demo.github.worked.Prefertoself-describe.perc}{5.1\%}
\definevar{demo.all.progexp.years.Prefernottodisclose}{0}
\definevar{demo.all.progexp.years.Prefernottodisclose.perc}{0.0\%}
\definevar{demo.all.progexp.years.>5Years}{59}
\definevar{demo.all.progexp.years.2-5Years}{36}
\definevar{demo.all.progexp.years.1-2Years}{10}
\definevar{demo.all.progexp.years.<1Year}{4}
\definevar{demo.all.progexp.years.>5Years.perc}{54.1\%}
\definevar{demo.all.progexp.years.2-5Years.perc}{33.0\%}
\definevar{demo.all.progexp.years.1-2Years.perc}{9.2\%}
\definevar{demo.all.progexp.years.<1Year.perc}{3.7\%}
\definevar{demo.upwork.progexp.years.Prefernottodisclose}{0}
\definevar{demo.upwork.progexp.years.Prefernottodisclose.perc}{0.0\%}
\definevar{demo.upwork.progexp.years.2-5Years}{21}
\definevar{demo.upwork.progexp.years.>5Years}{18}
\definevar{demo.upwork.progexp.years.1-2Years}{8}
\definevar{demo.upwork.progexp.years.<1Year}{3}
\definevar{demo.upwork.progexp.years.2-5Years.perc}{42.0\%}
\definevar{demo.upwork.progexp.years.>5Years.perc}{36.0\%}
\definevar{demo.upwork.progexp.years.1-2Years.perc}{16.0\%}
\definevar{demo.upwork.progexp.years.<1Year.perc}{6.0\%}
\definevar{demo.github.progexp.years.Prefernottodisclose}{0}
\definevar{demo.github.progexp.years.Prefernottodisclose.perc}{0.0\%}
\definevar{demo.github.progexp.years.>5Years}{41}
\definevar{demo.github.progexp.years.2-5Years}{15}
\definevar{demo.github.progexp.years.1-2Years}{2}
\definevar{demo.github.progexp.years.<1Year}{1}
\definevar{demo.github.progexp.years.>5Years.perc}{69.5\%}
\definevar{demo.github.progexp.years.2-5Years.perc}{25.4\%}
\definevar{demo.github.progexp.years.1-2Years.perc}{3.4\%}
\definevar{demo.github.progexp.years.<1Year.perc}{1.7\%}
\definevar{demo.all.progexp.cloudengineer.Prefernottodisclose}{31}
\definevar{demo.all.progexp.cloudengineer.Prefernottodisclose.perc}{28.4\%}
\definevar{demo.all.progexp.cloudengineer.2-5Years}{38}
\definevar{demo.all.progexp.cloudengineer.1-2Years}{15}
\definevar{demo.all.progexp.cloudengineer.>5Years}{14}
\definevar{demo.all.progexp.cloudengineer.<1Year}{11}
\definevar{demo.all.progexp.cloudengineer.2-5Years.perc}{48.7\%}
\definevar{demo.all.progexp.cloudengineer.1-2Years.perc}{19.2\%}
\definevar{demo.all.progexp.cloudengineer.>5Years.perc}{17.9\%}
\definevar{demo.all.progexp.cloudengineer.<1Year.perc}{14.1\%}
\definevar{demo.upwork.progexp.cloudengineer.Prefernottodisclose}{0}
\definevar{demo.upwork.progexp.cloudengineer.Prefernottodisclose.perc}{0.0\%}
\definevar{demo.upwork.progexp.cloudengineer.2-5Years}{29}
\definevar{demo.upwork.progexp.cloudengineer.1-2Years}{12}
\definevar{demo.upwork.progexp.cloudengineer.<1Year}{5}
\definevar{demo.upwork.progexp.cloudengineer.>5Years}{4}
\definevar{demo.upwork.progexp.cloudengineer.2-5Years.perc}{58.0\%}
\definevar{demo.upwork.progexp.cloudengineer.1-2Years.perc}{24.0\%}
\definevar{demo.upwork.progexp.cloudengineer.<1Year.perc}{10.0\%}
\definevar{demo.upwork.progexp.cloudengineer.>5Years.perc}{8.0\%}
\definevar{demo.github.progexp.cloudengineer.Prefernottodisclose}{31}
\definevar{demo.github.progexp.cloudengineer.Prefernottodisclose.perc}{52.5\%}
\definevar{demo.github.progexp.cloudengineer.>5Years}{10}
\definevar{demo.github.progexp.cloudengineer.2-5Years}{9}
\definevar{demo.github.progexp.cloudengineer.<1Year}{6}
\definevar{demo.github.progexp.cloudengineer.1-2Years}{3}
\definevar{demo.github.progexp.cloudengineer.>5Years.perc}{35.7\%}
\definevar{demo.github.progexp.cloudengineer.2-5Years.perc}{32.1\%}
\definevar{demo.github.progexp.cloudengineer.<1Year.perc}{21.4\%}
\definevar{demo.github.progexp.cloudengineer.1-2Years.perc}{10.7\%}
\definevar{demo.all.progexp.projects.Prefernottodisclose}{0}
\definevar{demo.all.progexp.projects.Prefernottodisclose.perc}{0.0\%}
\definevar{demo.all.progexp.projects.1-5Projects}{33}
\definevar{demo.all.progexp.projects.5-10Projects}{31}
\definevar{demo.all.progexp.projects.10-25Projects}{23}
\definevar{demo.all.progexp.projects.≥25Projects}{22}
\definevar{demo.all.progexp.projects.1-5Projects.perc}{30.3\%}
\definevar{demo.all.progexp.projects.5-10Projects.perc}{28.4\%}
\definevar{demo.all.progexp.projects.10-25Projects.perc}{21.1\%}
\definevar{demo.all.progexp.projects.≥25Projects.perc}{20.2\%}
\definevar{demo.upwork.progexp.projects.Prefernottodisclose}{0}
\definevar{demo.upwork.progexp.projects.Prefernottodisclose.perc}{0.0\%}
\definevar{demo.upwork.progexp.projects.5-10Projects}{19}
\definevar{demo.upwork.progexp.projects.1-5Projects}{17}
\definevar{demo.upwork.progexp.projects.10-25Projects}{10}
\definevar{demo.upwork.progexp.projects.≥25Projects}{4}
\definevar{demo.upwork.progexp.projects.5-10Projects.perc}{38.0\%}
\definevar{demo.upwork.progexp.projects.1-5Projects.perc}{34.0\%}
\definevar{demo.upwork.progexp.projects.10-25Projects.perc}{20.0\%}
\definevar{demo.upwork.progexp.projects.≥25Projects.perc}{8.0\%}
\definevar{demo.github.progexp.projects.Prefernottodisclose}{0}
\definevar{demo.github.progexp.projects.Prefernottodisclose.perc}{0.0\%}
\definevar{demo.github.progexp.projects.≥25Projects}{18}
\definevar{demo.github.progexp.projects.1-5Projects}{16}
\definevar{demo.github.progexp.projects.10-25Projects}{13}
\definevar{demo.github.progexp.projects.5-10Projects}{12}
\definevar{demo.github.progexp.projects.≥25Projects.perc}{30.5\%}
\definevar{demo.github.progexp.projects.1-5Projects.perc}{27.1\%}
\definevar{demo.github.progexp.projects.10-25Projects.perc}{22.0\%}
\definevar{demo.github.progexp.projects.5-10Projects.perc}{20.3\%}
\definevar{demo.all.progexp.learning.Prefernottodisclose}{0}
\definevar{demo.all.progexp.learning.Prefernottodisclose.perc}{0.0\%}
\definevar{demo.all.progexp.learning.Self-taught}{102}
\definevar{demo.all.progexp.learning.College/University}{64}
\definevar{demo.all.progexp.learning.On-the-jobtraining}{61}
\definevar{demo.all.progexp.learning.Onlineclass}{47}
\definevar{demo.all.progexp.learning.Codingcamp}{14}
\definevar{demo.all.progexp.learning.Other}{5}
\definevar{demo.all.progexp.learning.Self-taught.perc}{93.6\%}
\definevar{demo.all.progexp.learning.College/University.perc}{58.7\%}
\definevar{demo.all.progexp.learning.On-the-jobtraining.perc}{56.0\%}
\definevar{demo.all.progexp.learning.Onlineclass.perc}{43.1\%}
\definevar{demo.all.progexp.learning.Codingcamp.perc}{12.8\%}
\definevar{demo.all.progexp.learning.Other.perc}{4.6\%}
\definevar{demo.upwork.progexp.learning.Prefernottodisclose}{0}
\definevar{demo.upwork.progexp.learning.Prefernottodisclose.perc}{0.0\%}
\definevar{demo.upwork.progexp.learning.Self-taught}{46}
\definevar{demo.upwork.progexp.learning.On-the-jobtraining}{36}
\definevar{demo.upwork.progexp.learning.Onlineclass}{30}
\definevar{demo.upwork.progexp.learning.College/University}{27}
\definevar{demo.upwork.progexp.learning.Codingcamp}{9}
\definevar{demo.upwork.progexp.learning.Other}{3}
\definevar{demo.upwork.progexp.learning.Self-taught.perc}{92.0\%}
\definevar{demo.upwork.progexp.learning.On-the-jobtraining.perc}{72.0\%}
\definevar{demo.upwork.progexp.learning.Onlineclass.perc}{60.0\%}
\definevar{demo.upwork.progexp.learning.College/University.perc}{54.0\%}
\definevar{demo.upwork.progexp.learning.Codingcamp.perc}{18.0\%}
\definevar{demo.upwork.progexp.learning.Other.perc}{6.0\%}
\definevar{demo.github.progexp.learning.Prefernottodisclose}{0}
\definevar{demo.github.progexp.learning.Prefernottodisclose.perc}{0.0\%}
\definevar{demo.github.progexp.learning.Self-taught}{56}
\definevar{demo.github.progexp.learning.College/University}{37}
\definevar{demo.github.progexp.learning.On-the-jobtraining}{25}
\definevar{demo.github.progexp.learning.Onlineclass}{17}
\definevar{demo.github.progexp.learning.Codingcamp}{5}
\definevar{demo.github.progexp.learning.Other}{2}
\definevar{demo.github.progexp.learning.Self-taught.perc}{94.9\%}
\definevar{demo.github.progexp.learning.College/University.perc}{62.7\%}
\definevar{demo.github.progexp.learning.On-the-jobtraining.perc}{42.4\%}
\definevar{demo.github.progexp.learning.Onlineclass.perc}{28.8\%}
\definevar{demo.github.progexp.learning.Codingcamp.perc}{8.5\%}
\definevar{demo.github.progexp.learning.Other.perc}{3.4\%}
\definevar{demo.interviews.progexp.learning.Prefernottodisclose}{0}
\definevar{demo.interviews.progexp.learning.Prefernottodisclose.perc}{0.0\%}
\definevar{demo.interviews.progexp.learning.Self-taught}{12}
\definevar{demo.interviews.progexp.learning.College/University}{10}
\definevar{demo.interviews.progexp.learning.On-the-jobtraining}{8}
\definevar{demo.interviews.progexp.learning.Onlineclass}{5}
\definevar{demo.interviews.progexp.learning.Other}{1}
\definevar{demo.interviews.progexp.learning.Self-taught.perc}{85.7\%}
\definevar{demo.interviews.progexp.learning.College/University.perc}{71.4\%}
\definevar{demo.interviews.progexp.learning.On-the-jobtraining.perc}{57.1\%}
\definevar{demo.interviews.progexp.learning.Onlineclass.perc}{35.7\%}
\definevar{demo.interviews.progexp.learning.Other.perc}{7.1\%}
\definevar{demo.all.secexp.responsible.Prefernottodisclose}{0}
\definevar{demo.all.secexp.responsible.Prefernottodisclose.perc}{0.0\%}
\definevar{demo.all.secexp.responsible.Agree}{41}
\definevar{demo.all.secexp.responsible.StronglyAgree}{38}
\definevar{demo.all.secexp.responsible.Neutral}{18}
\definevar{demo.all.secexp.responsible.Stronglydisagree}{6}
\definevar{demo.all.secexp.responsible.Disagree}{6}
\definevar{demo.all.secexp.responsible.Agree.perc}{37.6\%}
\definevar{demo.all.secexp.responsible.StronglyAgree.perc}{34.9\%}
\definevar{demo.all.secexp.responsible.Neutral.perc}{16.5\%}
\definevar{demo.all.secexp.responsible.Stronglydisagree.perc}{5.5\%}
\definevar{demo.all.secexp.responsible.Disagree.perc}{5.5\%}
\definevar{demo.all.secexp.responsible.AgreeOrStronglyAgree}{79}
\definevar{demo.all.secexp.responsible.AgreeOrStronglyAgree.perc}{72.5\%}
\definevar{demo.all.secexp.takecare.Prefernottodisclose}{0}
\definevar{demo.all.secexp.takecare.Prefernottodisclose.perc}{0.0\%}
\definevar{demo.all.secexp.takecare.Agree}{51}
\definevar{demo.all.secexp.takecare.StronglyAgree}{46}
\definevar{demo.all.secexp.takecare.Neutral}{6}
\definevar{demo.all.secexp.takecare.Disagree}{3}
\definevar{demo.all.secexp.takecare.Stronglydisagree}{3}
\definevar{demo.all.secexp.takecare.Agree.perc}{46.8\%}
\definevar{demo.all.secexp.takecare.StronglyAgree.perc}{42.2\%}
\definevar{demo.all.secexp.takecare.Neutral.perc}{5.5\%}
\definevar{demo.all.secexp.takecare.Disagree.perc}{2.8\%}
\definevar{demo.all.secexp.takecare.Stronglydisagree.perc}{2.8\%}
\definevar{demo.all.secexp.takecare.AgreeOrStronglyAgree}{97}
\definevar{demo.all.secexp.takecare.AgreeOrStronglyAgree.perc}{89.0\%}
\definevar{demo.all.secexp.confident.Prefernottodisclose}{0}
\definevar{demo.all.secexp.confident.Prefernottodisclose.perc}{0.0\%}
\definevar{demo.all.secexp.confident.Agree}{40}
\definevar{demo.all.secexp.confident.Neutral}{30}
\definevar{demo.all.secexp.confident.StronglyAgree}{19}
\definevar{demo.all.secexp.confident.Disagree}{16}
\definevar{demo.all.secexp.confident.Stronglydisagree}{4}
\definevar{demo.all.secexp.confident.Agree.perc}{36.7\%}
\definevar{demo.all.secexp.confident.Neutral.perc}{27.5\%}
\definevar{demo.all.secexp.confident.StronglyAgree.perc}{17.4\%}
\definevar{demo.all.secexp.confident.Disagree.perc}{14.7\%}
\definevar{demo.all.secexp.confident.Stronglydisagree.perc}{3.7\%}
\definevar{demo.all.secexp.confident.AgreeOrStronglyAgree}{59}
\definevar{demo.all.secexp.confident.AgreeOrStronglyAgree.perc}{54.1\%}
\definevar{demo.upwork.secexp.responsible.Prefernottodisclose}{0}
\definevar{demo.upwork.secexp.responsible.Prefernottodisclose.perc}{0.0\%}
\definevar{demo.upwork.secexp.responsible.Agree}{20}
\definevar{demo.upwork.secexp.responsible.StronglyAgree}{16}
\definevar{demo.upwork.secexp.responsible.Neutral}{11}
\definevar{demo.upwork.secexp.responsible.Stronglydisagree}{2}
\definevar{demo.upwork.secexp.responsible.Disagree}{1}
\definevar{demo.upwork.secexp.responsible.Agree.perc}{40.0\%}
\definevar{demo.upwork.secexp.responsible.StronglyAgree.perc}{32.0\%}
\definevar{demo.upwork.secexp.responsible.Neutral.perc}{22.0\%}
\definevar{demo.upwork.secexp.responsible.Stronglydisagree.perc}{4.0\%}
\definevar{demo.upwork.secexp.responsible.Disagree.perc}{2.0\%}
\definevar{demo.upwork.secexp.responsible.AgreeOrStronglyAgree}{36}
\definevar{demo.upwork.secexp.responsible.AgreeOrStronglyAgree.perc}{72.0\%}
\definevar{demo.upwork.secexp.takecare.Prefernottodisclose}{0}
\definevar{demo.upwork.secexp.takecare.Prefernottodisclose.perc}{0.0\%}
\definevar{demo.upwork.secexp.takecare.Agree}{26}
\definevar{demo.upwork.secexp.takecare.StronglyAgree}{19}
\definevar{demo.upwork.secexp.takecare.Neutral}{2}
\definevar{demo.upwork.secexp.takecare.Stronglydisagree}{2}
\definevar{demo.upwork.secexp.takecare.Disagree}{1}
\definevar{demo.upwork.secexp.takecare.Agree.perc}{52.0\%}
\definevar{demo.upwork.secexp.takecare.StronglyAgree.perc}{38.0\%}
\definevar{demo.upwork.secexp.takecare.Neutral.perc}{4.0\%}
\definevar{demo.upwork.secexp.takecare.Stronglydisagree.perc}{4.0\%}
\definevar{demo.upwork.secexp.takecare.Disagree.perc}{2.0\%}
\definevar{demo.upwork.secexp.takecare.AgreeOrStronglyAgree}{45}
\definevar{demo.upwork.secexp.takecare.AgreeOrStronglyAgree.perc}{90.0\%}
\definevar{demo.upwork.secexp.confident.Prefernottodisclose}{0}
\definevar{demo.upwork.secexp.confident.Prefernottodisclose.perc}{0.0\%}
\definevar{demo.upwork.secexp.confident.Agree}{24}
\definevar{demo.upwork.secexp.confident.Neutral}{13}
\definevar{demo.upwork.secexp.confident.StronglyAgree}{8}
\definevar{demo.upwork.secexp.confident.Disagree}{4}
\definevar{demo.upwork.secexp.confident.Stronglydisagree}{1}
\definevar{demo.upwork.secexp.confident.Agree.perc}{48.0\%}
\definevar{demo.upwork.secexp.confident.Neutral.perc}{26.0\%}
\definevar{demo.upwork.secexp.confident.StronglyAgree.perc}{16.0\%}
\definevar{demo.upwork.secexp.confident.Disagree.perc}{8.0\%}
\definevar{demo.upwork.secexp.confident.Stronglydisagree.perc}{2.0\%}
\definevar{demo.upwork.secexp.confident.AgreeOrStronglyAgree}{32}
\definevar{demo.upwork.secexp.confident.AgreeOrStronglyAgree.perc}{64.0\%}
\definevar{demo.github.secexp.responsible.Prefernottodisclose}{0}
\definevar{demo.github.secexp.responsible.Prefernottodisclose.perc}{0.0\%}
\definevar{demo.github.secexp.responsible.StronglyAgree}{22}
\definevar{demo.github.secexp.responsible.Agree}{21}
\definevar{demo.github.secexp.responsible.Neutral}{7}
\definevar{demo.github.secexp.responsible.Disagree}{5}
\definevar{demo.github.secexp.responsible.Stronglydisagree}{4}
\definevar{demo.github.secexp.responsible.StronglyAgree.perc}{37.3\%}
\definevar{demo.github.secexp.responsible.Agree.perc}{35.6\%}
\definevar{demo.github.secexp.responsible.Neutral.perc}{11.9\%}
\definevar{demo.github.secexp.responsible.Disagree.perc}{8.5\%}
\definevar{demo.github.secexp.responsible.Stronglydisagree.perc}{6.8\%}
\definevar{demo.github.secexp.responsible.AgreeOrStronglyAgree}{43}
\definevar{demo.github.secexp.responsible.AgreeOrStronglyAgree.perc}{72.9\%}
\definevar{demo.github.secexp.takecare.Prefernottodisclose}{0}
\definevar{demo.github.secexp.takecare.Prefernottodisclose.perc}{0.0\%}
\definevar{demo.github.secexp.takecare.StronglyAgree}{27}
\definevar{demo.github.secexp.takecare.Agree}{25}
\definevar{demo.github.secexp.takecare.Neutral}{4}
\definevar{demo.github.secexp.takecare.Disagree}{2}
\definevar{demo.github.secexp.takecare.Stronglydisagree}{1}
\definevar{demo.github.secexp.takecare.StronglyAgree.perc}{45.8\%}
\definevar{demo.github.secexp.takecare.Agree.perc}{42.4\%}
\definevar{demo.github.secexp.takecare.Neutral.perc}{6.8\%}
\definevar{demo.github.secexp.takecare.Disagree.perc}{3.4\%}
\definevar{demo.github.secexp.takecare.Stronglydisagree.perc}{1.7\%}
\definevar{demo.github.secexp.takecare.AgreeOrStronglyAgree}{52}
\definevar{demo.github.secexp.takecare.AgreeOrStronglyAgree.perc}{88.1\%}
\definevar{demo.github.secexp.confident.Prefernottodisclose}{0}
\definevar{demo.github.secexp.confident.Prefernottodisclose.perc}{0.0\%}
\definevar{demo.github.secexp.confident.Neutral}{17}
\definevar{demo.github.secexp.confident.Agree}{16}
\definevar{demo.github.secexp.confident.Disagree}{12}
\definevar{demo.github.secexp.confident.StronglyAgree}{11}
\definevar{demo.github.secexp.confident.Stronglydisagree}{3}
\definevar{demo.github.secexp.confident.Neutral.perc}{28.8\%}
\definevar{demo.github.secexp.confident.Agree.perc}{27.1\%}
\definevar{demo.github.secexp.confident.Disagree.perc}{20.3\%}
\definevar{demo.github.secexp.confident.StronglyAgree.perc}{18.6\%}
\definevar{demo.github.secexp.confident.Stronglydisagree.perc}{5.1\%}
\definevar{demo.github.secexp.confident.AgreeOrStronglyAgree}{27}
\definevar{demo.github.secexp.confident.AgreeOrStronglyAgree.perc}{45.8\%}
\definevar{demo.all.secexp.training.Prefernottodisclose}{0}
\definevar{demo.all.secexp.training.Prefernottodisclose.perc}{0.0\%}
\definevar{demo.all.secexp.training.Self-taught}{86}
\definevar{demo.all.secexp.training.On-the-jobtraining}{55}
\definevar{demo.all.secexp.training.Onlineclass}{39}
\definevar{demo.all.secexp.training.College/University}{24}
\definevar{demo.all.secexp.training.None}{17}
\definevar{demo.all.secexp.training.Other}{7}
\definevar{demo.all.secexp.training.Codingcamp}{6}
\definevar{demo.all.secexp.training.Self-taught.perc}{78.9\%}
\definevar{demo.all.secexp.training.On-the-jobtraining.perc}{50.5\%}
\definevar{demo.all.secexp.training.Onlineclass.perc}{35.8\%}
\definevar{demo.all.secexp.training.College/University.perc}{22.0\%}
\definevar{demo.all.secexp.training.None.perc}{15.6\%}
\definevar{demo.all.secexp.training.Other.perc}{6.4\%}
\definevar{demo.all.secexp.training.Codingcamp.perc}{5.5\%}
\definevar{demo.upwork.secexp.training.Prefernottodisclose}{0}
\definevar{demo.upwork.secexp.training.Prefernottodisclose.perc}{0.0\%}
\definevar{demo.upwork.secexp.training.On-the-jobtraining}{38}
\definevar{demo.upwork.secexp.training.Self-taught}{36}
\definevar{demo.upwork.secexp.training.Onlineclass}{27}
\definevar{demo.upwork.secexp.training.College/University}{16}
\definevar{demo.upwork.secexp.training.Codingcamp}{5}
\definevar{demo.upwork.secexp.training.None}{4}
\definevar{demo.upwork.secexp.training.Other}{4}
\definevar{demo.upwork.secexp.training.On-the-jobtraining.perc}{76.0\%}
\definevar{demo.upwork.secexp.training.Self-taught.perc}{72.0\%}
\definevar{demo.upwork.secexp.training.Onlineclass.perc}{54.0\%}
\definevar{demo.upwork.secexp.training.College/University.perc}{32.0\%}
\definevar{demo.upwork.secexp.training.Codingcamp.perc}{10.0\%}
\definevar{demo.upwork.secexp.training.None.perc}{8.0\%}
\definevar{demo.upwork.secexp.training.Other.perc}{8.0\%}
\definevar{demo.github.secexp.training.Prefernottodisclose}{0}
\definevar{demo.github.secexp.training.Prefernottodisclose.perc}{0.0\%}
\definevar{demo.github.secexp.training.Self-taught}{50}
\definevar{demo.github.secexp.training.On-the-jobtraining}{17}
\definevar{demo.github.secexp.training.None}{13}
\definevar{demo.github.secexp.training.Onlineclass}{12}
\definevar{demo.github.secexp.training.College/University}{8}
\definevar{demo.github.secexp.training.Other}{3}
\definevar{demo.github.secexp.training.Codingcamp}{1}
\definevar{demo.github.secexp.training.Self-taught.perc}{84.7\%}
\definevar{demo.github.secexp.training.On-the-jobtraining.perc}{28.8\%}
\definevar{demo.github.secexp.training.None.perc}{22.0\%}
\definevar{demo.github.secexp.training.Onlineclass.perc}{20.3\%}
\definevar{demo.github.secexp.training.College/University.perc}{13.6\%}
\definevar{demo.github.secexp.training.Other.perc}{5.1\%}
\definevar{demo.github.secexp.training.Codingcamp.perc}{1.7\%}
\definevar{demo.all.gender.n/a.text}{zero}
\definevar{demo.all.gender.n/a.Text}{Zero}
\definevar{demo.all.gender.n/a.perc.text}{zero}
\definevar{demo.all.gender.n/a.perc.Text}{Zero}
\definevar{demo.all.gender.Man.text}{95}
\definevar{demo.all.gender.Man.Text}{95}
\definevar{demo.all.gender.Woman.text}{six}
\definevar{demo.all.gender.Woman.Text}{Six}
\definevar{demo.all.gender.Prefernottodisclose.text}{four}
\definevar{demo.all.gender.Prefernottodisclose.Text}{Four}
\definevar{demo.all.gender.Non-binary.text}{four}
\definevar{demo.all.gender.Non-binary.Text}{Four}
\definevar{demo.all.gender.Man.perc.text}{zero}
\definevar{demo.all.gender.Man.perc.Text}{Zero}
\definevar{demo.all.gender.Woman.perc.text}{zero}
\definevar{demo.all.gender.Woman.perc.Text}{Zero}
\definevar{demo.all.gender.Prefernottodisclose.perc.text}{zero}
\definevar{demo.all.gender.Prefernottodisclose.perc.Text}{Zero}
\definevar{demo.all.gender.Non-binary.perc.text}{zero}
\definevar{demo.all.gender.Non-binary.perc.Text}{Zero}
\definevar{demo.upwork.gender.n/a.text}{zero}
\definevar{demo.upwork.gender.n/a.Text}{Zero}
\definevar{demo.upwork.gender.n/a.perc.text}{zero}
\definevar{demo.upwork.gender.n/a.perc.Text}{Zero}
\definevar{demo.upwork.gender.Man.text}{43}
\definevar{demo.upwork.gender.Man.Text}{43}
\definevar{demo.upwork.gender.Woman.text}{five}
\definevar{demo.upwork.gender.Woman.Text}{Five}
\definevar{demo.upwork.gender.Prefernottodisclose.text}{two}
\definevar{demo.upwork.gender.Prefernottodisclose.Text}{Two}
\definevar{demo.upwork.gender.Man.perc.text}{zero}
\definevar{demo.upwork.gender.Man.perc.Text}{Zero}
\definevar{demo.upwork.gender.Woman.perc.text}{zero}
\definevar{demo.upwork.gender.Woman.perc.Text}{Zero}
\definevar{demo.upwork.gender.Prefernottodisclose.perc.text}{zero}
\definevar{demo.upwork.gender.Prefernottodisclose.perc.Text}{Zero}
\definevar{demo.github.gender.n/a.text}{zero}
\definevar{demo.github.gender.n/a.Text}{Zero}
\definevar{demo.github.gender.n/a.perc.text}{zero}
\definevar{demo.github.gender.n/a.perc.Text}{Zero}
\definevar{demo.github.gender.Man.text}{52}
\definevar{demo.github.gender.Man.Text}{52}
\definevar{demo.github.gender.Non-binary.text}{four}
\definevar{demo.github.gender.Non-binary.Text}{Four}
\definevar{demo.github.gender.Prefernottodisclose.text}{two}
\definevar{demo.github.gender.Prefernottodisclose.Text}{Two}
\definevar{demo.github.gender.Woman.text}{one}
\definevar{demo.github.gender.Woman.Text}{One}
\definevar{demo.github.gender.Man.perc.text}{zero}
\definevar{demo.github.gender.Man.perc.Text}{Zero}
\definevar{demo.github.gender.Non-binary.perc.text}{zero}
\definevar{demo.github.gender.Non-binary.perc.Text}{Zero}
\definevar{demo.github.gender.Prefernottodisclose.perc.text}{zero}
\definevar{demo.github.gender.Prefernottodisclose.perc.Text}{Zero}
\definevar{demo.github.gender.Woman.perc.text}{zero}
\definevar{demo.github.gender.Woman.perc.Text}{Zero}
\definevar{demo.interviews.gender.n/a.text}{zero}
\definevar{demo.interviews.gender.n/a.Text}{Zero}
\definevar{demo.interviews.gender.n/a.perc.text}{zero}
\definevar{demo.interviews.gender.n/a.perc.Text}{Zero}
\definevar{demo.interviews.gender.Man.text}{13}
\definevar{demo.interviews.gender.Man.Text}{13}
\definevar{demo.interviews.gender.Non-binary.text}{one}
\definevar{demo.interviews.gender.Non-binary.Text}{One}
\definevar{demo.interviews.gender.Man.perc.text}{zero}
\definevar{demo.interviews.gender.Man.perc.Text}{Zero}
\definevar{demo.interviews.gender.Non-binary.perc.text}{zero}
\definevar{demo.interviews.gender.Non-binary.perc.Text}{Zero}
\definevar{demo.all.age.n/a.text}{one}
\definevar{demo.all.age.n/a.Text}{One}
\definevar{demo.all.age.n/a.perc.text}{zero}
\definevar{demo.all.age.n/a.perc.Text}{Zero}
\definevar{demo.all.age.count.text}{108.0}
\definevar{demo.all.age.count.Text}{108.0}
\definevar{demo.all.age.mean.text}{33.2}
\definevar{demo.all.age.mean.Text}{33.2}
\definevar{demo.all.age.std.text}{eleven}
\definevar{demo.all.age.std.Text}{Eleven}
\definevar{demo.all.age.min.text}{18.0}
\definevar{demo.all.age.min.Text}{18.0}
\definevar{demo.all.age.25percentile.text}{25.0}
\definevar{demo.all.age.25percentile.Text}{25.0}
\definevar{demo.all.age.50percentile.text}{30.0}
\definevar{demo.all.age.50percentile.Text}{30.0}
\definevar{demo.all.age.75percentile.text}{38.2}
\definevar{demo.all.age.75percentile.Text}{38.2}
\definevar{demo.all.age.max.text}{70.0}
\definevar{demo.all.age.max.Text}{70.0}
\definevar{demo.upwork.age.n/a.text}{zero}
\definevar{demo.upwork.age.n/a.Text}{Zero}
\definevar{demo.upwork.age.n/a.perc.text}{zero}
\definevar{demo.upwork.age.n/a.perc.Text}{Zero}
\definevar{demo.upwork.age.count.text}{50.0}
\definevar{demo.upwork.age.count.Text}{50.0}
\definevar{demo.upwork.age.mean.text}{31.3}
\definevar{demo.upwork.age.mean.Text}{31.3}
\definevar{demo.upwork.age.std.text}{eight}
\definevar{demo.upwork.age.std.Text}{Eight}
\definevar{demo.upwork.age.min.text}{21.0}
\definevar{demo.upwork.age.min.Text}{21.0}
\definevar{demo.upwork.age.25percentile.text}{26.0}
\definevar{demo.upwork.age.25percentile.Text}{26.0}
\definevar{demo.upwork.age.50percentile.text}{29.0}
\definevar{demo.upwork.age.50percentile.Text}{29.0}
\definevar{demo.upwork.age.75percentile.text}{34.8}
\definevar{demo.upwork.age.75percentile.Text}{34.8}
\definevar{demo.upwork.age.max.text}{70.0}
\definevar{demo.upwork.age.max.Text}{70.0}
\definevar{demo.github.age.n/a.text}{one}
\definevar{demo.github.age.n/a.Text}{One}
\definevar{demo.github.age.n/a.perc.text}{zero}
\definevar{demo.github.age.n/a.perc.Text}{Zero}
\definevar{demo.github.age.count.text}{58.0}
\definevar{demo.github.age.count.Text}{58.0}
\definevar{demo.github.age.mean.text}{34.9}
\definevar{demo.github.age.mean.Text}{34.9}
\definevar{demo.github.age.std.text}{12.7}
\definevar{demo.github.age.std.Text}{12.7}
\definevar{demo.github.age.min.text}{18.0}
\definevar{demo.github.age.min.Text}{18.0}
\definevar{demo.github.age.25percentile.text}{25.0}
\definevar{demo.github.age.25percentile.Text}{25.0}
\definevar{demo.github.age.50percentile.text}{33.0}
\definevar{demo.github.age.50percentile.Text}{33.0}
\definevar{demo.github.age.75percentile.text}{42.8}
\definevar{demo.github.age.75percentile.Text}{42.8}
\definevar{demo.github.age.max.text}{65.0}
\definevar{demo.github.age.max.Text}{65.0}
\definevar{demo.interviews.age.n/a.text}{zero}
\definevar{demo.interviews.age.n/a.Text}{Zero}
\definevar{demo.interviews.age.n/a.perc.text}{zero}
\definevar{demo.interviews.age.n/a.perc.Text}{Zero}
\definevar{demo.interviews.age.count.text}{14.0}
\definevar{demo.interviews.age.count.Text}{14.0}
\definevar{demo.interviews.age.mean.text}{32.1}
\definevar{demo.interviews.age.mean.Text}{32.1}
\definevar{demo.interviews.age.std.text}{eleven}
\definevar{demo.interviews.age.std.Text}{Eleven}
\definevar{demo.interviews.age.min.text}{19.0}
\definevar{demo.interviews.age.min.Text}{19.0}
\definevar{demo.interviews.age.25percentile.text}{24.0}
\definevar{demo.interviews.age.25percentile.Text}{24.0}
\definevar{demo.interviews.age.50percentile.text}{28.0}
\definevar{demo.interviews.age.50percentile.Text}{28.0}
\definevar{demo.interviews.age.75percentile.text}{39.8}
\definevar{demo.interviews.age.75percentile.Text}{39.8}
\definevar{demo.interviews.age.max.text}{62.0}
\definevar{demo.interviews.age.max.Text}{62.0}
\definevar{demo.all.countries.Prefernottodisclose.text}{three}
\definevar{demo.all.countries.Prefernottodisclose.Text}{Three}
\definevar{demo.all.countries.Prefernottodisclose.perc.text}{zero}
\definevar{demo.all.countries.Prefernottodisclose.perc.Text}{Zero}
\definevar{demo.all.countries.UnitedStatesofAmerica.text}{20}
\definevar{demo.all.countries.UnitedStatesofAmerica.Text}{20}
\definevar{demo.all.countries.India.text}{12}
\definevar{demo.all.countries.India.Text}{12}
\definevar{demo.all.countries.Germany.text}{eleven}
\definevar{demo.all.countries.Germany.Text}{Eleven}
\definevar{demo.all.countries.Pakistan.text}{nine}
\definevar{demo.all.countries.Pakistan.Text}{Nine}
\definevar{demo.all.countries.RussianFederation.text}{four}
\definevar{demo.all.countries.RussianFederation.Text}{Four}
\definevar{demo.all.countries.Nigeria.text}{four}
\definevar{demo.all.countries.Nigeria.Text}{Four}
\definevar{demo.all.countries.Canada.text}{four}
\definevar{demo.all.countries.Canada.Text}{Four}
\definevar{demo.all.countries.UnitedKingdomofGreatBritainandNorthernIreland.text}{four}
\definevar{demo.all.countries.UnitedKingdomofGreatBritainandNorthernIreland.Text}{Four}
\definevar{demo.all.countries.Kenya.text}{four}
\definevar{demo.all.countries.Kenya.Text}{Four}
\definevar{demo.all.countries.nan.text}{three}
\definevar{demo.all.countries.nan.Text}{Three}
\definevar{demo.all.countries.France.text}{three}
\definevar{demo.all.countries.France.Text}{Three}
\definevar{demo.all.countries.Denmark.text}{three}
\definevar{demo.all.countries.Denmark.Text}{Three}
\definevar{demo.all.countries.Sweden.text}{two}
\definevar{demo.all.countries.Sweden.Text}{Two}
\definevar{demo.all.countries.SriLanka.text}{two}
\definevar{demo.all.countries.SriLanka.Text}{Two}
\definevar{demo.all.countries.Indonesia.text}{two}
\definevar{demo.all.countries.Indonesia.Text}{Two}
\definevar{demo.all.countries.Egypt.text}{two}
\definevar{demo.all.countries.Egypt.Text}{Two}
\definevar{demo.all.countries.Brazil.text}{two}
\definevar{demo.all.countries.Brazil.Text}{Two}
\definevar{demo.all.countries.Poland.text}{two}
\definevar{demo.all.countries.Poland.Text}{Two}
\definevar{demo.all.countries.Singapore.text}{one}
\definevar{demo.all.countries.Singapore.Text}{One}
\definevar{demo.all.countries.SouthKorea.text}{one}
\definevar{demo.all.countries.SouthKorea.Text}{One}
\definevar{demo.all.countries.HongKong(S.A.R.).text}{one}
\definevar{demo.all.countries.HongKong(S.A.R.).Text}{One}
\definevar{demo.all.countries.TheformerYugoslavRepublicofMacedonia.text}{one}
\definevar{demo.all.countries.TheformerYugoslavRepublicofMacedonia.Text}{One}
\definevar{demo.all.countries.Thailand.text}{one}
\definevar{demo.all.countries.Thailand.Text}{One}
\definevar{demo.all.countries.Hungary.text}{one}
\definevar{demo.all.countries.Hungary.Text}{One}
\definevar{demo.all.countries.China.text}{one}
\definevar{demo.all.countries.China.Text}{One}
\definevar{demo.all.countries.Albania.text}{one}
\definevar{demo.all.countries.Albania.Text}{One}
\definevar{demo.all.countries.Israel.text}{one}
\definevar{demo.all.countries.Israel.Text}{One}
\definevar{demo.all.countries.Portugal.text}{one}
\definevar{demo.all.countries.Portugal.Text}{One}
\definevar{demo.all.countries.DominicanRepublic.text}{one}
\definevar{demo.all.countries.DominicanRepublic.Text}{One}
\definevar{demo.all.countries.Netherlands.text}{one}
\definevar{demo.all.countries.Netherlands.Text}{One}
\definevar{demo.all.countries.Georgia.text}{one}
\definevar{demo.all.countries.Georgia.Text}{One}
\definevar{demo.all.countries.Latvia.text}{one}
\definevar{demo.all.countries.Latvia.Text}{One}
\definevar{demo.all.countries.Mexico.text}{one}
\definevar{demo.all.countries.Mexico.Text}{One}
\definevar{demo.all.countries.Australia.text}{one}
\definevar{demo.all.countries.Australia.Text}{One}
\definevar{demo.all.countries.UnitedStatesofAmerica.perc.text}{zero}
\definevar{demo.all.countries.UnitedStatesofAmerica.perc.Text}{Zero}
\definevar{demo.all.countries.India.perc.text}{zero}
\definevar{demo.all.countries.India.perc.Text}{Zero}
\definevar{demo.all.countries.Germany.perc.text}{zero}
\definevar{demo.all.countries.Germany.perc.Text}{Zero}
\definevar{demo.all.countries.Pakistan.perc.text}{zero}
\definevar{demo.all.countries.Pakistan.perc.Text}{Zero}
\definevar{demo.all.countries.RussianFederation.perc.text}{zero}
\definevar{demo.all.countries.RussianFederation.perc.Text}{Zero}
\definevar{demo.all.countries.Nigeria.perc.text}{zero}
\definevar{demo.all.countries.Nigeria.perc.Text}{Zero}
\definevar{demo.all.countries.Canada.perc.text}{zero}
\definevar{demo.all.countries.Canada.perc.Text}{Zero}
\definevar{demo.all.countries.UnitedKingdomofGreatBritainandNorthernIreland.perc.text}{zero}
\definevar{demo.all.countries.UnitedKingdomofGreatBritainandNorthernIreland.perc.Text}{Zero}
\definevar{demo.all.countries.Kenya.perc.text}{zero}
\definevar{demo.all.countries.Kenya.perc.Text}{Zero}
\definevar{demo.all.countries.nan.perc.text}{zero}
\definevar{demo.all.countries.nan.perc.Text}{Zero}
\definevar{demo.all.countries.France.perc.text}{zero}
\definevar{demo.all.countries.France.perc.Text}{Zero}
\definevar{demo.all.countries.Denmark.perc.text}{zero}
\definevar{demo.all.countries.Denmark.perc.Text}{Zero}
\definevar{demo.all.countries.Sweden.perc.text}{zero}
\definevar{demo.all.countries.Sweden.perc.Text}{Zero}
\definevar{demo.all.countries.SriLanka.perc.text}{zero}
\definevar{demo.all.countries.SriLanka.perc.Text}{Zero}
\definevar{demo.all.countries.Indonesia.perc.text}{zero}
\definevar{demo.all.countries.Indonesia.perc.Text}{Zero}
\definevar{demo.all.countries.Egypt.perc.text}{zero}
\definevar{demo.all.countries.Egypt.perc.Text}{Zero}
\definevar{demo.all.countries.Brazil.perc.text}{zero}
\definevar{demo.all.countries.Brazil.perc.Text}{Zero}
\definevar{demo.all.countries.Poland.perc.text}{zero}
\definevar{demo.all.countries.Poland.perc.Text}{Zero}
\definevar{demo.all.countries.Singapore.perc.text}{zero}
\definevar{demo.all.countries.Singapore.perc.Text}{Zero}
\definevar{demo.all.countries.SouthKorea.perc.text}{zero}
\definevar{demo.all.countries.SouthKorea.perc.Text}{Zero}
\definevar{demo.all.countries.HongKong(S.A.R.).perc.text}{zero}
\definevar{demo.all.countries.HongKong(S.A.R.).perc.Text}{Zero}
\definevar{demo.all.countries.TheformerYugoslavRepublicofMacedonia.perc.text}{zero}
\definevar{demo.all.countries.TheformerYugoslavRepublicofMacedonia.perc.Text}{Zero}
\definevar{demo.all.countries.Thailand.perc.text}{zero}
\definevar{demo.all.countries.Thailand.perc.Text}{Zero}
\definevar{demo.all.countries.Hungary.perc.text}{zero}
\definevar{demo.all.countries.Hungary.perc.Text}{Zero}
\definevar{demo.all.countries.China.perc.text}{zero}
\definevar{demo.all.countries.China.perc.Text}{Zero}
\definevar{demo.all.countries.Albania.perc.text}{zero}
\definevar{demo.all.countries.Albania.perc.Text}{Zero}
\definevar{demo.all.countries.Israel.perc.text}{zero}
\definevar{demo.all.countries.Israel.perc.Text}{Zero}
\definevar{demo.all.countries.Portugal.perc.text}{zero}
\definevar{demo.all.countries.Portugal.perc.Text}{Zero}
\definevar{demo.all.countries.DominicanRepublic.perc.text}{zero}
\definevar{demo.all.countries.DominicanRepublic.perc.Text}{Zero}
\definevar{demo.all.countries.Netherlands.perc.text}{zero}
\definevar{demo.all.countries.Netherlands.perc.Text}{Zero}
\definevar{demo.all.countries.Georgia.perc.text}{zero}
\definevar{demo.all.countries.Georgia.perc.Text}{Zero}
\definevar{demo.all.countries.Latvia.perc.text}{zero}
\definevar{demo.all.countries.Latvia.perc.Text}{Zero}
\definevar{demo.all.countries.Mexico.perc.text}{zero}
\definevar{demo.all.countries.Mexico.perc.Text}{Zero}
\definevar{demo.all.countries.Australia.perc.text}{zero}
\definevar{demo.all.countries.Australia.perc.Text}{Zero}
\definevar{demo.upwork.countries.Prefernottodisclose.text}{two}
\definevar{demo.upwork.countries.Prefernottodisclose.Text}{Two}
\definevar{demo.upwork.countries.Prefernottodisclose.perc.text}{zero}
\definevar{demo.upwork.countries.Prefernottodisclose.perc.Text}{Zero}
\definevar{demo.upwork.countries.India.text}{ten}
\definevar{demo.upwork.countries.India.Text}{Ten}
\definevar{demo.upwork.countries.Pakistan.text}{seven}
\definevar{demo.upwork.countries.Pakistan.Text}{Seven}
\definevar{demo.upwork.countries.Kenya.text}{four}
\definevar{demo.upwork.countries.Kenya.Text}{Four}
\definevar{demo.upwork.countries.Canada.text}{two}
\definevar{demo.upwork.countries.Canada.Text}{Two}
\definevar{demo.upwork.countries.nan.text}{two}
\definevar{demo.upwork.countries.nan.Text}{Two}
\definevar{demo.upwork.countries.Nigeria.text}{two}
\definevar{demo.upwork.countries.Nigeria.Text}{Two}
\definevar{demo.upwork.countries.SriLanka.text}{two}
\definevar{demo.upwork.countries.SriLanka.Text}{Two}
\definevar{demo.upwork.countries.Egypt.text}{two}
\definevar{demo.upwork.countries.Egypt.Text}{Two}
\definevar{demo.upwork.countries.Indonesia.text}{two}
\definevar{demo.upwork.countries.Indonesia.Text}{Two}
\definevar{demo.upwork.countries.UnitedStatesofAmerica.text}{one}
\definevar{demo.upwork.countries.UnitedStatesofAmerica.Text}{One}
\definevar{demo.upwork.countries.Netherlands.text}{one}
\definevar{demo.upwork.countries.Netherlands.Text}{One}
\definevar{demo.upwork.countries.Denmark.text}{one}
\definevar{demo.upwork.countries.Denmark.Text}{One}
\definevar{demo.upwork.countries.RussianFederation.text}{one}
\definevar{demo.upwork.countries.RussianFederation.Text}{One}
\definevar{demo.upwork.countries.Portugal.text}{one}
\definevar{demo.upwork.countries.Portugal.Text}{One}
\definevar{demo.upwork.countries.Sweden.text}{one}
\definevar{demo.upwork.countries.Sweden.Text}{One}
\definevar{demo.upwork.countries.Israel.text}{one}
\definevar{demo.upwork.countries.Israel.Text}{One}
\definevar{demo.upwork.countries.DominicanRepublic.text}{one}
\definevar{demo.upwork.countries.DominicanRepublic.Text}{One}
\definevar{demo.upwork.countries.Poland.text}{one}
\definevar{demo.upwork.countries.Poland.Text}{One}
\definevar{demo.upwork.countries.Mexico.text}{one}
\definevar{demo.upwork.countries.Mexico.Text}{One}
\definevar{demo.upwork.countries.UnitedKingdomofGreatBritainandNorthernIreland.text}{one}
\definevar{demo.upwork.countries.UnitedKingdomofGreatBritainandNorthernIreland.Text}{One}
\definevar{demo.upwork.countries.Albania.text}{one}
\definevar{demo.upwork.countries.Albania.Text}{One}
\definevar{demo.upwork.countries.Latvia.text}{one}
\definevar{demo.upwork.countries.Latvia.Text}{One}
\definevar{demo.upwork.countries.Georgia.text}{one}
\definevar{demo.upwork.countries.Georgia.Text}{One}
\definevar{demo.upwork.countries.Singapore.text}{one}
\definevar{demo.upwork.countries.Singapore.Text}{One}
\definevar{demo.upwork.countries.China.text}{one}
\definevar{demo.upwork.countries.China.Text}{One}
\definevar{demo.upwork.countries.France.text}{one}
\definevar{demo.upwork.countries.France.Text}{One}
\definevar{demo.upwork.countries.India.perc.text}{zero}
\definevar{demo.upwork.countries.India.perc.Text}{Zero}
\definevar{demo.upwork.countries.Pakistan.perc.text}{zero}
\definevar{demo.upwork.countries.Pakistan.perc.Text}{Zero}
\definevar{demo.upwork.countries.Kenya.perc.text}{zero}
\definevar{demo.upwork.countries.Kenya.perc.Text}{Zero}
\definevar{demo.upwork.countries.Canada.perc.text}{zero}
\definevar{demo.upwork.countries.Canada.perc.Text}{Zero}
\definevar{demo.upwork.countries.nan.perc.text}{zero}
\definevar{demo.upwork.countries.nan.perc.Text}{Zero}
\definevar{demo.upwork.countries.Nigeria.perc.text}{zero}
\definevar{demo.upwork.countries.Nigeria.perc.Text}{Zero}
\definevar{demo.upwork.countries.SriLanka.perc.text}{zero}
\definevar{demo.upwork.countries.SriLanka.perc.Text}{Zero}
\definevar{demo.upwork.countries.Egypt.perc.text}{zero}
\definevar{demo.upwork.countries.Egypt.perc.Text}{Zero}
\definevar{demo.upwork.countries.Indonesia.perc.text}{zero}
\definevar{demo.upwork.countries.Indonesia.perc.Text}{Zero}
\definevar{demo.upwork.countries.UnitedStatesofAmerica.perc.text}{zero}
\definevar{demo.upwork.countries.UnitedStatesofAmerica.perc.Text}{Zero}
\definevar{demo.upwork.countries.Netherlands.perc.text}{zero}
\definevar{demo.upwork.countries.Netherlands.perc.Text}{Zero}
\definevar{demo.upwork.countries.Denmark.perc.text}{zero}
\definevar{demo.upwork.countries.Denmark.perc.Text}{Zero}
\definevar{demo.upwork.countries.RussianFederation.perc.text}{zero}
\definevar{demo.upwork.countries.RussianFederation.perc.Text}{Zero}
\definevar{demo.upwork.countries.Portugal.perc.text}{zero}
\definevar{demo.upwork.countries.Portugal.perc.Text}{Zero}
\definevar{demo.upwork.countries.Sweden.perc.text}{zero}
\definevar{demo.upwork.countries.Sweden.perc.Text}{Zero}
\definevar{demo.upwork.countries.Israel.perc.text}{zero}
\definevar{demo.upwork.countries.Israel.perc.Text}{Zero}
\definevar{demo.upwork.countries.DominicanRepublic.perc.text}{zero}
\definevar{demo.upwork.countries.DominicanRepublic.perc.Text}{Zero}
\definevar{demo.upwork.countries.Poland.perc.text}{zero}
\definevar{demo.upwork.countries.Poland.perc.Text}{Zero}
\definevar{demo.upwork.countries.Mexico.perc.text}{zero}
\definevar{demo.upwork.countries.Mexico.perc.Text}{Zero}
\definevar{demo.upwork.countries.UnitedKingdomofGreatBritainandNorthernIreland.perc.text}{zero}
\definevar{demo.upwork.countries.UnitedKingdomofGreatBritainandNorthernIreland.perc.Text}{Zero}
\definevar{demo.upwork.countries.Albania.perc.text}{zero}
\definevar{demo.upwork.countries.Albania.perc.Text}{Zero}
\definevar{demo.upwork.countries.Latvia.perc.text}{zero}
\definevar{demo.upwork.countries.Latvia.perc.Text}{Zero}
\definevar{demo.upwork.countries.Georgia.perc.text}{zero}
\definevar{demo.upwork.countries.Georgia.perc.Text}{Zero}
\definevar{demo.upwork.countries.Singapore.perc.text}{zero}
\definevar{demo.upwork.countries.Singapore.perc.Text}{Zero}
\definevar{demo.upwork.countries.China.perc.text}{zero}
\definevar{demo.upwork.countries.China.perc.Text}{Zero}
\definevar{demo.upwork.countries.France.perc.text}{zero}
\definevar{demo.upwork.countries.France.perc.Text}{Zero}
\definevar{demo.github.countries.Prefernottodisclose.text}{one}
\definevar{demo.github.countries.Prefernottodisclose.Text}{One}
\definevar{demo.github.countries.Prefernottodisclose.perc.text}{zero}
\definevar{demo.github.countries.Prefernottodisclose.perc.Text}{Zero}
\definevar{demo.github.countries.UnitedStatesofAmerica.text}{19}
\definevar{demo.github.countries.UnitedStatesofAmerica.Text}{19}
\definevar{demo.github.countries.Germany.text}{eleven}
\definevar{demo.github.countries.Germany.Text}{Eleven}
\definevar{demo.github.countries.UnitedKingdomofGreatBritainandNorthernIreland.text}{three}
\definevar{demo.github.countries.UnitedKingdomofGreatBritainandNorthernIreland.Text}{Three}
\definevar{demo.github.countries.RussianFederation.text}{three}
\definevar{demo.github.countries.RussianFederation.Text}{Three}
\definevar{demo.github.countries.India.text}{two}
\definevar{demo.github.countries.India.Text}{Two}
\definevar{demo.github.countries.Brazil.text}{two}
\definevar{demo.github.countries.Brazil.Text}{Two}
\definevar{demo.github.countries.France.text}{two}
\definevar{demo.github.countries.France.Text}{Two}
\definevar{demo.github.countries.Canada.text}{two}
\definevar{demo.github.countries.Canada.Text}{Two}
\definevar{demo.github.countries.Denmark.text}{two}
\definevar{demo.github.countries.Denmark.Text}{Two}
\definevar{demo.github.countries.Nigeria.text}{two}
\definevar{demo.github.countries.Nigeria.Text}{Two}
\definevar{demo.github.countries.Pakistan.text}{two}
\definevar{demo.github.countries.Pakistan.Text}{Two}
\definevar{demo.github.countries.TheformerYugoslavRepublicofMacedonia.text}{one}
\definevar{demo.github.countries.TheformerYugoslavRepublicofMacedonia.Text}{One}
\definevar{demo.github.countries.Australia.text}{one}
\definevar{demo.github.countries.Australia.Text}{One}
\definevar{demo.github.countries.SouthKorea.text}{one}
\definevar{demo.github.countries.SouthKorea.Text}{One}
\definevar{demo.github.countries.HongKong(S.A.R.).text}{one}
\definevar{demo.github.countries.HongKong(S.A.R.).Text}{One}
\definevar{demo.github.countries.Poland.text}{one}
\definevar{demo.github.countries.Poland.Text}{One}
\definevar{demo.github.countries.Sweden.text}{one}
\definevar{demo.github.countries.Sweden.Text}{One}
\definevar{demo.github.countries.Hungary.text}{one}
\definevar{demo.github.countries.Hungary.Text}{One}
\definevar{demo.github.countries.nan.text}{one}
\definevar{demo.github.countries.nan.Text}{One}
\definevar{demo.github.countries.Thailand.text}{one}
\definevar{demo.github.countries.Thailand.Text}{One}
\definevar{demo.github.countries.UnitedStatesofAmerica.perc.text}{zero}
\definevar{demo.github.countries.UnitedStatesofAmerica.perc.Text}{Zero}
\definevar{demo.github.countries.Germany.perc.text}{zero}
\definevar{demo.github.countries.Germany.perc.Text}{Zero}
\definevar{demo.github.countries.UnitedKingdomofGreatBritainandNorthernIreland.perc.text}{zero}
\definevar{demo.github.countries.UnitedKingdomofGreatBritainandNorthernIreland.perc.Text}{Zero}
\definevar{demo.github.countries.RussianFederation.perc.text}{zero}
\definevar{demo.github.countries.RussianFederation.perc.Text}{Zero}
\definevar{demo.github.countries.India.perc.text}{zero}
\definevar{demo.github.countries.India.perc.Text}{Zero}
\definevar{demo.github.countries.Brazil.perc.text}{zero}
\definevar{demo.github.countries.Brazil.perc.Text}{Zero}
\definevar{demo.github.countries.France.perc.text}{zero}
\definevar{demo.github.countries.France.perc.Text}{Zero}
\definevar{demo.github.countries.Canada.perc.text}{zero}
\definevar{demo.github.countries.Canada.perc.Text}{Zero}
\definevar{demo.github.countries.Denmark.perc.text}{zero}
\definevar{demo.github.countries.Denmark.perc.Text}{Zero}
\definevar{demo.github.countries.Nigeria.perc.text}{zero}
\definevar{demo.github.countries.Nigeria.perc.Text}{Zero}
\definevar{demo.github.countries.Pakistan.perc.text}{zero}
\definevar{demo.github.countries.Pakistan.perc.Text}{Zero}
\definevar{demo.github.countries.TheformerYugoslavRepublicofMacedonia.perc.text}{zero}
\definevar{demo.github.countries.TheformerYugoslavRepublicofMacedonia.perc.Text}{Zero}
\definevar{demo.github.countries.Australia.perc.text}{zero}
\definevar{demo.github.countries.Australia.perc.Text}{Zero}
\definevar{demo.github.countries.SouthKorea.perc.text}{zero}
\definevar{demo.github.countries.SouthKorea.perc.Text}{Zero}
\definevar{demo.github.countries.HongKong(S.A.R.).perc.text}{zero}
\definevar{demo.github.countries.HongKong(S.A.R.).perc.Text}{Zero}
\definevar{demo.github.countries.Poland.perc.text}{zero}
\definevar{demo.github.countries.Poland.perc.Text}{Zero}
\definevar{demo.github.countries.Sweden.perc.text}{zero}
\definevar{demo.github.countries.Sweden.perc.Text}{Zero}
\definevar{demo.github.countries.Hungary.perc.text}{zero}
\definevar{demo.github.countries.Hungary.perc.Text}{Zero}
\definevar{demo.github.countries.nan.perc.text}{zero}
\definevar{demo.github.countries.nan.perc.Text}{Zero}
\definevar{demo.github.countries.Thailand.perc.text}{zero}
\definevar{demo.github.countries.Thailand.perc.Text}{Zero}
\definevar{demo.interviews.countries.Prefernottodisclose.text}{zero}
\definevar{demo.interviews.countries.Prefernottodisclose.Text}{Zero}
\definevar{demo.interviews.countries.Prefernottodisclose.perc.text}{zero}
\definevar{demo.interviews.countries.Prefernottodisclose.perc.Text}{Zero}
\definevar{demo.interviews.countries.India.text}{three}
\definevar{demo.interviews.countries.India.Text}{Three}
\definevar{demo.interviews.countries.UnitedStatesofAmerica.text}{three}
\definevar{demo.interviews.countries.UnitedStatesofAmerica.Text}{Three}
\definevar{demo.interviews.countries.Pakistan.text}{two}
\definevar{demo.interviews.countries.Pakistan.Text}{Two}
\definevar{demo.interviews.countries.NewZealand.text}{one}
\definevar{demo.interviews.countries.NewZealand.Text}{One}
\definevar{demo.interviews.countries.Canada.text}{one}
\definevar{demo.interviews.countries.Canada.Text}{One}
\definevar{demo.interviews.countries.Italy.text}{one}
\definevar{demo.interviews.countries.Italy.Text}{One}
\definevar{demo.interviews.countries.Belarus.text}{one}
\definevar{demo.interviews.countries.Belarus.Text}{One}
\definevar{demo.interviews.countries.Brazil.text}{one}
\definevar{demo.interviews.countries.Brazil.Text}{One}
\definevar{demo.interviews.countries.Kenya.text}{one}
\definevar{demo.interviews.countries.Kenya.Text}{One}
\definevar{demo.interviews.countries.India.perc.text}{zero}
\definevar{demo.interviews.countries.India.perc.Text}{Zero}
\definevar{demo.interviews.countries.UnitedStatesofAmerica.perc.text}{zero}
\definevar{demo.interviews.countries.UnitedStatesofAmerica.perc.Text}{Zero}
\definevar{demo.interviews.countries.Pakistan.perc.text}{zero}
\definevar{demo.interviews.countries.Pakistan.perc.Text}{Zero}
\definevar{demo.interviews.countries.NewZealand.perc.text}{zero}
\definevar{demo.interviews.countries.NewZealand.perc.Text}{Zero}
\definevar{demo.interviews.countries.Canada.perc.text}{zero}
\definevar{demo.interviews.countries.Canada.perc.Text}{Zero}
\definevar{demo.interviews.countries.Italy.perc.text}{zero}
\definevar{demo.interviews.countries.Italy.perc.Text}{Zero}
\definevar{demo.interviews.countries.Belarus.perc.text}{zero}
\definevar{demo.interviews.countries.Belarus.perc.Text}{Zero}
\definevar{demo.interviews.countries.Brazil.perc.text}{zero}
\definevar{demo.interviews.countries.Brazil.perc.Text}{Zero}
\definevar{demo.interviews.countries.Kenya.perc.text}{zero}
\definevar{demo.interviews.countries.Kenya.perc.Text}{Zero}
\definevar{demo.all.countries.distinct.text}{33}
\definevar{demo.all.countries.distinct.Text}{33}
\definevar{demo.all.countries.other.text}{54}
\definevar{demo.all.countries.other.Text}{54}
\definevar{demo.all.countries.other.perc.text}{zero}
\definevar{demo.all.countries.other.perc.Text}{Zero}
\definevar{demo.upwork.countries.other.text}{30}
\definevar{demo.upwork.countries.other.Text}{30}
\definevar{demo.upwork.countries.other.perc.text}{zero}
\definevar{demo.upwork.countries.other.perc.Text}{Zero}
\definevar{demo.github.countries.other.text}{24}
\definevar{demo.github.countries.other.Text}{24}
\definevar{demo.github.countries.other.perc.text}{zero}
\definevar{demo.github.countries.other.perc.Text}{Zero}
\definevar{demo.all.edu.Prefernottodisclose.text}{zero}
\definevar{demo.all.edu.Prefernottodisclose.Text}{Zero}
\definevar{demo.all.edu.Prefernottodisclose.perc.text}{zero}
\definevar{demo.all.edu.Prefernottodisclose.perc.Text}{Zero}
\definevar{demo.all.edu.Bachelor.text}{51}
\definevar{demo.all.edu.Bachelor.Text}{51}
\definevar{demo.all.edu.Master.text}{26}
\definevar{demo.all.edu.Master.Text}{26}
\definevar{demo.all.edu.Somecollege,currentlyenrolledincollegeortwo-yearassociate’sdegree.text}{15}
\definevar{demo.all.edu.Somecollege,currentlyenrolledincollegeortwo-yearassociate’sdegree.Text}{15}
\definevar{demo.all.edu.Doctoral.text}{seven}
\definevar{demo.all.edu.Doctoral.Text}{Seven}
\definevar{demo.all.edu.Highschoolorequivalent.text}{six}
\definevar{demo.all.edu.Highschoolorequivalent.Text}{Six}
\definevar{demo.all.edu.Somegraduateschoolorcurrentlyenrolledingraduateschool.text}{four}
\definevar{demo.all.edu.Somegraduateschoolorcurrentlyenrolledingraduateschool.Text}{Four}
\definevar{demo.all.edu.Bachelor.perc.text}{zero}
\definevar{demo.all.edu.Bachelor.perc.Text}{Zero}
\definevar{demo.all.edu.Master.perc.text}{zero}
\definevar{demo.all.edu.Master.perc.Text}{Zero}
\definevar{demo.all.edu.Somecollege,currentlyenrolledincollegeortwo-yearassociate’sdegree.perc.text}{zero}
\definevar{demo.all.edu.Somecollege,currentlyenrolledincollegeortwo-yearassociate’sdegree.perc.Text}{Zero}
\definevar{demo.all.edu.Doctoral.perc.text}{zero}
\definevar{demo.all.edu.Doctoral.perc.Text}{Zero}
\definevar{demo.all.edu.Highschoolorequivalent.perc.text}{zero}
\definevar{demo.all.edu.Highschoolorequivalent.perc.Text}{Zero}
\definevar{demo.all.edu.Somegraduateschoolorcurrentlyenrolledingraduateschool.perc.text}{zero}
\definevar{demo.all.edu.Somegraduateschoolorcurrentlyenrolledingraduateschool.perc.Text}{Zero}
\definevar{demo.upwork.edu.Prefernottodisclose.text}{zero}
\definevar{demo.upwork.edu.Prefernottodisclose.Text}{Zero}
\definevar{demo.upwork.edu.Prefernottodisclose.perc.text}{zero}
\definevar{demo.upwork.edu.Prefernottodisclose.perc.Text}{Zero}
\definevar{demo.upwork.edu.Bachelor.text}{30}
\definevar{demo.upwork.edu.Bachelor.Text}{30}
\definevar{demo.upwork.edu.Master.text}{14}
\definevar{demo.upwork.edu.Master.Text}{14}
\definevar{demo.upwork.edu.Somecollege,currentlyenrolledincollegeortwo-yearassociate’sdegree.text}{three}
\definevar{demo.upwork.edu.Somecollege,currentlyenrolledincollegeortwo-yearassociate’sdegree.Text}{Three}
\definevar{demo.upwork.edu.Somegraduateschoolorcurrentlyenrolledingraduateschool.text}{two}
\definevar{demo.upwork.edu.Somegraduateschoolorcurrentlyenrolledingraduateschool.Text}{Two}
\definevar{demo.upwork.edu.Doctoral.text}{one}
\definevar{demo.upwork.edu.Doctoral.Text}{One}
\definevar{demo.upwork.edu.Bachelor.perc.text}{zero}
\definevar{demo.upwork.edu.Bachelor.perc.Text}{Zero}
\definevar{demo.upwork.edu.Master.perc.text}{zero}
\definevar{demo.upwork.edu.Master.perc.Text}{Zero}
\definevar{demo.upwork.edu.Somecollege,currentlyenrolledincollegeortwo-yearassociate’sdegree.perc.text}{zero}
\definevar{demo.upwork.edu.Somecollege,currentlyenrolledincollegeortwo-yearassociate’sdegree.perc.Text}{Zero}
\definevar{demo.upwork.edu.Somegraduateschoolorcurrentlyenrolledingraduateschool.perc.text}{zero}
\definevar{demo.upwork.edu.Somegraduateschoolorcurrentlyenrolledingraduateschool.perc.Text}{Zero}
\definevar{demo.upwork.edu.Doctoral.perc.text}{zero}
\definevar{demo.upwork.edu.Doctoral.perc.Text}{Zero}
\definevar{demo.github.edu.Prefernottodisclose.text}{zero}
\definevar{demo.github.edu.Prefernottodisclose.Text}{Zero}
\definevar{demo.github.edu.Prefernottodisclose.perc.text}{zero}
\definevar{demo.github.edu.Prefernottodisclose.perc.Text}{Zero}
\definevar{demo.github.edu.Bachelor.text}{21}
\definevar{demo.github.edu.Bachelor.Text}{21}
\definevar{demo.github.edu.Somecollege,currentlyenrolledincollegeortwo-yearassociate’sdegree.text}{12}
\definevar{demo.github.edu.Somecollege,currentlyenrolledincollegeortwo-yearassociate’sdegree.Text}{12}
\definevar{demo.github.edu.Master.text}{12}
\definevar{demo.github.edu.Master.Text}{12}
\definevar{demo.github.edu.Doctoral.text}{six}
\definevar{demo.github.edu.Doctoral.Text}{Six}
\definevar{demo.github.edu.Highschoolorequivalent.text}{six}
\definevar{demo.github.edu.Highschoolorequivalent.Text}{Six}
\definevar{demo.github.edu.Somegraduateschoolorcurrentlyenrolledingraduateschool.text}{two}
\definevar{demo.github.edu.Somegraduateschoolorcurrentlyenrolledingraduateschool.Text}{Two}
\definevar{demo.github.edu.Bachelor.perc.text}{zero}
\definevar{demo.github.edu.Bachelor.perc.Text}{Zero}
\definevar{demo.github.edu.Somecollege,currentlyenrolledincollegeortwo-yearassociate’sdegree.perc.text}{zero}
\definevar{demo.github.edu.Somecollege,currentlyenrolledincollegeortwo-yearassociate’sdegree.perc.Text}{Zero}
\definevar{demo.github.edu.Master.perc.text}{zero}
\definevar{demo.github.edu.Master.perc.Text}{Zero}
\definevar{demo.github.edu.Doctoral.perc.text}{zero}
\definevar{demo.github.edu.Doctoral.perc.Text}{Zero}
\definevar{demo.github.edu.Highschoolorequivalent.perc.text}{zero}
\definevar{demo.github.edu.Highschoolorequivalent.perc.Text}{Zero}
\definevar{demo.github.edu.Somegraduateschoolorcurrentlyenrolledingraduateschool.perc.text}{zero}
\definevar{demo.github.edu.Somegraduateschoolorcurrentlyenrolledingraduateschool.perc.Text}{Zero}
\definevar{demo.all.employment.Prefernottodisclose.text}{two}
\definevar{demo.all.employment.Prefernottodisclose.Text}{Two}
\definevar{demo.all.employment.Prefernottodisclose.perc.text}{zero}
\definevar{demo.all.employment.Prefernottodisclose.perc.Text}{Zero}
\definevar{demo.all.employment.Employedfull-time.text}{71}
\definevar{demo.all.employment.Employedfull-time.Text}{71}
\definevar{demo.all.employment.Self-employed/Freelancer.text}{33}
\definevar{demo.all.employment.Self-employed/Freelancer.Text}{33}
\definevar{demo.all.employment.Student.text}{12}
\definevar{demo.all.employment.Student.Text}{12}
\definevar{demo.all.employment.Employedpart-time.text}{eleven}
\definevar{demo.all.employment.Employedpart-time.Text}{Eleven}
\definevar{demo.all.employment.Lookingforwork.text}{four}
\definevar{demo.all.employment.Lookingforwork.Text}{Four}
\definevar{demo.all.employment.Retired.text}{three}
\definevar{demo.all.employment.Retired.Text}{Three}
\definevar{demo.all.employment.Employedfull-time.perc.text}{zero}
\definevar{demo.all.employment.Employedfull-time.perc.Text}{Zero}
\definevar{demo.all.employment.Self-employed/Freelancer.perc.text}{zero}
\definevar{demo.all.employment.Self-employed/Freelancer.perc.Text}{Zero}
\definevar{demo.all.employment.Student.perc.text}{zero}
\definevar{demo.all.employment.Student.perc.Text}{Zero}
\definevar{demo.all.employment.Employedpart-time.perc.text}{zero}
\definevar{demo.all.employment.Employedpart-time.perc.Text}{Zero}
\definevar{demo.all.employment.Lookingforwork.perc.text}{zero}
\definevar{demo.all.employment.Lookingforwork.perc.Text}{Zero}
\definevar{demo.all.employment.Retired.perc.text}{zero}
\definevar{demo.all.employment.Retired.perc.Text}{Zero}
\definevar{demo.upwork.employment.Prefernottodisclose.text}{two}
\definevar{demo.upwork.employment.Prefernottodisclose.Text}{Two}
\definevar{demo.upwork.employment.Prefernottodisclose.perc.text}{zero}
\definevar{demo.upwork.employment.Prefernottodisclose.perc.Text}{Zero}
\definevar{demo.upwork.employment.Employedfull-time.text}{37}
\definevar{demo.upwork.employment.Employedfull-time.Text}{37}
\definevar{demo.upwork.employment.Self-employed/Freelancer.text}{18}
\definevar{demo.upwork.employment.Self-employed/Freelancer.Text}{18}
\definevar{demo.upwork.employment.Employedpart-time.text}{three}
\definevar{demo.upwork.employment.Employedpart-time.Text}{Three}
\definevar{demo.upwork.employment.Student.text}{two}
\definevar{demo.upwork.employment.Student.Text}{Two}
\definevar{demo.upwork.employment.Prefertoself-describe.text}{two}
\definevar{demo.upwork.employment.Prefertoself-describe.Text}{Two}
\definevar{demo.upwork.employment.Lookingforwork.text}{one}
\definevar{demo.upwork.employment.Lookingforwork.Text}{One}
\definevar{demo.upwork.employment.Employedfull-time.perc.text}{zero}
\definevar{demo.upwork.employment.Employedfull-time.perc.Text}{Zero}
\definevar{demo.upwork.employment.Self-employed/Freelancer.perc.text}{zero}
\definevar{demo.upwork.employment.Self-employed/Freelancer.perc.Text}{Zero}
\definevar{demo.upwork.employment.Employedpart-time.perc.text}{zero}
\definevar{demo.upwork.employment.Employedpart-time.perc.Text}{Zero}
\definevar{demo.upwork.employment.Student.perc.text}{zero}
\definevar{demo.upwork.employment.Student.perc.Text}{Zero}
\definevar{demo.upwork.employment.Prefertoself-describe.perc.text}{zero}
\definevar{demo.upwork.employment.Prefertoself-describe.perc.Text}{Zero}
\definevar{demo.upwork.employment.Lookingforwork.perc.text}{zero}
\definevar{demo.upwork.employment.Lookingforwork.perc.Text}{Zero}
\definevar{demo.github.employment.Prefernottodisclose.text}{zero}
\definevar{demo.github.employment.Prefernottodisclose.Text}{Zero}
\definevar{demo.github.employment.Prefernottodisclose.perc.text}{zero}
\definevar{demo.github.employment.Prefernottodisclose.perc.Text}{Zero}
\definevar{demo.github.employment.Employedfull-time.text}{34}
\definevar{demo.github.employment.Employedfull-time.Text}{34}
\definevar{demo.github.employment.Self-employed/Freelancer.text}{15}
\definevar{demo.github.employment.Self-employed/Freelancer.Text}{15}
\definevar{demo.github.employment.Student.text}{ten}
\definevar{demo.github.employment.Student.Text}{Ten}
\definevar{demo.github.employment.Employedpart-time.text}{eight}
\definevar{demo.github.employment.Employedpart-time.Text}{Eight}
\definevar{demo.github.employment.Retired.text}{three}
\definevar{demo.github.employment.Retired.Text}{Three}
\definevar{demo.github.employment.Lookingforwork.text}{two}
\definevar{demo.github.employment.Lookingforwork.Text}{Two}
\definevar{demo.github.employment.Employedfull-time.perc.text}{zero}
\definevar{demo.github.employment.Employedfull-time.perc.Text}{Zero}
\definevar{demo.github.employment.Self-employed/Freelancer.perc.text}{zero}
\definevar{demo.github.employment.Self-employed/Freelancer.perc.Text}{Zero}
\definevar{demo.github.employment.Student.perc.text}{zero}
\definevar{demo.github.employment.Student.perc.Text}{Zero}
\definevar{demo.github.employment.Employedpart-time.perc.text}{zero}
\definevar{demo.github.employment.Employedpart-time.perc.Text}{Zero}
\definevar{demo.github.employment.Retired.perc.text}{zero}
\definevar{demo.github.employment.Retired.perc.Text}{Zero}
\definevar{demo.github.employment.Lookingforwork.perc.text}{zero}
\definevar{demo.github.employment.Lookingforwork.perc.Text}{Zero}
\definevar{demo.all.worked.Prefernottodisclose.text}{zero}
\definevar{demo.all.worked.Prefernottodisclose.Text}{Zero}
\definevar{demo.all.worked.Prefernottodisclose.perc.text}{zero}
\definevar{demo.all.worked.Prefernottodisclose.perc.Text}{Zero}
\definevar{demo.all.worked.Company.text}{90}
\definevar{demo.all.worked.Company.Text}{90}
\definevar{demo.all.worked.Solo.text}{76}
\definevar{demo.all.worked.Solo.Text}{76}
\definevar{demo.all.worked.OpenSourceProject.text}{60}
\definevar{demo.all.worked.OpenSourceProject.Text}{60}
\definevar{demo.all.worked.Prefertoself-describe.text}{four}
\definevar{demo.all.worked.Prefertoself-describe.Text}{Four}
\definevar{demo.all.worked.Company.perc.text}{zero}
\definevar{demo.all.worked.Company.perc.Text}{Zero}
\definevar{demo.all.worked.Solo.perc.text}{zero}
\definevar{demo.all.worked.Solo.perc.Text}{Zero}
\definevar{demo.all.worked.OpenSourceProject.perc.text}{zero}
\definevar{demo.all.worked.OpenSourceProject.perc.Text}{Zero}
\definevar{demo.all.worked.Prefertoself-describe.perc.text}{zero}
\definevar{demo.all.worked.Prefertoself-describe.perc.Text}{Zero}
\definevar{demo.upwork.worked.Prefernottodisclose.text}{zero}
\definevar{demo.upwork.worked.Prefernottodisclose.Text}{Zero}
\definevar{demo.upwork.worked.Prefernottodisclose.perc.text}{zero}
\definevar{demo.upwork.worked.Prefernottodisclose.perc.Text}{Zero}
\definevar{demo.upwork.worked.Company.text}{48}
\definevar{demo.upwork.worked.Company.Text}{48}
\definevar{demo.upwork.worked.Solo.text}{32}
\definevar{demo.upwork.worked.Solo.Text}{32}
\definevar{demo.upwork.worked.OpenSourceProject.text}{14}
\definevar{demo.upwork.worked.OpenSourceProject.Text}{14}
\definevar{demo.upwork.worked.Prefertoself-describe.text}{one}
\definevar{demo.upwork.worked.Prefertoself-describe.Text}{One}
\definevar{demo.upwork.worked.Company.perc.text}{zero}
\definevar{demo.upwork.worked.Company.perc.Text}{Zero}
\definevar{demo.upwork.worked.Solo.perc.text}{zero}
\definevar{demo.upwork.worked.Solo.perc.Text}{Zero}
\definevar{demo.upwork.worked.OpenSourceProject.perc.text}{zero}
\definevar{demo.upwork.worked.OpenSourceProject.perc.Text}{Zero}
\definevar{demo.upwork.worked.Prefertoself-describe.perc.text}{zero}
\definevar{demo.upwork.worked.Prefertoself-describe.perc.Text}{Zero}
\definevar{demo.github.worked.Prefernottodisclose.text}{zero}
\definevar{demo.github.worked.Prefernottodisclose.Text}{Zero}
\definevar{demo.github.worked.Prefernottodisclose.perc.text}{zero}
\definevar{demo.github.worked.Prefernottodisclose.perc.Text}{Zero}
\definevar{demo.github.worked.OpenSourceProject.text}{46}
\definevar{demo.github.worked.OpenSourceProject.Text}{46}
\definevar{demo.github.worked.Solo.text}{44}
\definevar{demo.github.worked.Solo.Text}{44}
\definevar{demo.github.worked.Company.text}{42}
\definevar{demo.github.worked.Company.Text}{42}
\definevar{demo.github.worked.Prefertoself-describe.text}{three}
\definevar{demo.github.worked.Prefertoself-describe.Text}{Three}
\definevar{demo.github.worked.OpenSourceProject.perc.text}{zero}
\definevar{demo.github.worked.OpenSourceProject.perc.Text}{Zero}
\definevar{demo.github.worked.Solo.perc.text}{zero}
\definevar{demo.github.worked.Solo.perc.Text}{Zero}
\definevar{demo.github.worked.Company.perc.text}{zero}
\definevar{demo.github.worked.Company.perc.Text}{Zero}
\definevar{demo.github.worked.Prefertoself-describe.perc.text}{zero}
\definevar{demo.github.worked.Prefertoself-describe.perc.Text}{Zero}
\definevar{demo.all.progexp.years.Prefernottodisclose.text}{zero}
\definevar{demo.all.progexp.years.Prefernottodisclose.Text}{Zero}
\definevar{demo.all.progexp.years.Prefernottodisclose.perc.text}{zero}
\definevar{demo.all.progexp.years.Prefernottodisclose.perc.Text}{Zero}
\definevar{demo.all.progexp.years.>5Years.text}{59}
\definevar{demo.all.progexp.years.>5Years.Text}{59}
\definevar{demo.all.progexp.years.2-5Years.text}{36}
\definevar{demo.all.progexp.years.2-5Years.Text}{36}
\definevar{demo.all.progexp.years.1-2Years.text}{ten}
\definevar{demo.all.progexp.years.1-2Years.Text}{Ten}
\definevar{demo.all.progexp.years.<1Year.text}{four}
\definevar{demo.all.progexp.years.<1Year.Text}{Four}
\definevar{demo.all.progexp.years.>5Years.perc.text}{zero}
\definevar{demo.all.progexp.years.>5Years.perc.Text}{Zero}
\definevar{demo.all.progexp.years.2-5Years.perc.text}{zero}
\definevar{demo.all.progexp.years.2-5Years.perc.Text}{Zero}
\definevar{demo.all.progexp.years.1-2Years.perc.text}{zero}
\definevar{demo.all.progexp.years.1-2Years.perc.Text}{Zero}
\definevar{demo.all.progexp.years.<1Year.perc.text}{zero}
\definevar{demo.all.progexp.years.<1Year.perc.Text}{Zero}
\definevar{demo.upwork.progexp.years.Prefernottodisclose.text}{zero}
\definevar{demo.upwork.progexp.years.Prefernottodisclose.Text}{Zero}
\definevar{demo.upwork.progexp.years.Prefernottodisclose.perc.text}{zero}
\definevar{demo.upwork.progexp.years.Prefernottodisclose.perc.Text}{Zero}
\definevar{demo.upwork.progexp.years.2-5Years.text}{21}
\definevar{demo.upwork.progexp.years.2-5Years.Text}{21}
\definevar{demo.upwork.progexp.years.>5Years.text}{18}
\definevar{demo.upwork.progexp.years.>5Years.Text}{18}
\definevar{demo.upwork.progexp.years.1-2Years.text}{eight}
\definevar{demo.upwork.progexp.years.1-2Years.Text}{Eight}
\definevar{demo.upwork.progexp.years.<1Year.text}{three}
\definevar{demo.upwork.progexp.years.<1Year.Text}{Three}
\definevar{demo.upwork.progexp.years.2-5Years.perc.text}{zero}
\definevar{demo.upwork.progexp.years.2-5Years.perc.Text}{Zero}
\definevar{demo.upwork.progexp.years.>5Years.perc.text}{zero}
\definevar{demo.upwork.progexp.years.>5Years.perc.Text}{Zero}
\definevar{demo.upwork.progexp.years.1-2Years.perc.text}{zero}
\definevar{demo.upwork.progexp.years.1-2Years.perc.Text}{Zero}
\definevar{demo.upwork.progexp.years.<1Year.perc.text}{zero}
\definevar{demo.upwork.progexp.years.<1Year.perc.Text}{Zero}
\definevar{demo.github.progexp.years.Prefernottodisclose.text}{zero}
\definevar{demo.github.progexp.years.Prefernottodisclose.Text}{Zero}
\definevar{demo.github.progexp.years.Prefernottodisclose.perc.text}{zero}
\definevar{demo.github.progexp.years.Prefernottodisclose.perc.Text}{Zero}
\definevar{demo.github.progexp.years.>5Years.text}{41}
\definevar{demo.github.progexp.years.>5Years.Text}{41}
\definevar{demo.github.progexp.years.2-5Years.text}{15}
\definevar{demo.github.progexp.years.2-5Years.Text}{15}
\definevar{demo.github.progexp.years.1-2Years.text}{two}
\definevar{demo.github.progexp.years.1-2Years.Text}{Two}
\definevar{demo.github.progexp.years.<1Year.text}{one}
\definevar{demo.github.progexp.years.<1Year.Text}{One}
\definevar{demo.github.progexp.years.>5Years.perc.text}{zero}
\definevar{demo.github.progexp.years.>5Years.perc.Text}{Zero}
\definevar{demo.github.progexp.years.2-5Years.perc.text}{zero}
\definevar{demo.github.progexp.years.2-5Years.perc.Text}{Zero}
\definevar{demo.github.progexp.years.1-2Years.perc.text}{zero}
\definevar{demo.github.progexp.years.1-2Years.perc.Text}{Zero}
\definevar{demo.github.progexp.years.<1Year.perc.text}{zero}
\definevar{demo.github.progexp.years.<1Year.perc.Text}{Zero}
\definevar{demo.all.progexp.cloudengineer.Prefernottodisclose.text}{31}
\definevar{demo.all.progexp.cloudengineer.Prefernottodisclose.Text}{31}
\definevar{demo.all.progexp.cloudengineer.Prefernottodisclose.perc.text}{zero}
\definevar{demo.all.progexp.cloudengineer.Prefernottodisclose.perc.Text}{Zero}
\definevar{demo.all.progexp.cloudengineer.2-5Years.text}{38}
\definevar{demo.all.progexp.cloudengineer.2-5Years.Text}{38}
\definevar{demo.all.progexp.cloudengineer.1-2Years.text}{15}
\definevar{demo.all.progexp.cloudengineer.1-2Years.Text}{15}
\definevar{demo.all.progexp.cloudengineer.>5Years.text}{14}
\definevar{demo.all.progexp.cloudengineer.>5Years.Text}{14}
\definevar{demo.all.progexp.cloudengineer.<1Year.text}{eleven}
\definevar{demo.all.progexp.cloudengineer.<1Year.Text}{Eleven}
\definevar{demo.all.progexp.cloudengineer.2-5Years.perc.text}{zero}
\definevar{demo.all.progexp.cloudengineer.2-5Years.perc.Text}{Zero}
\definevar{demo.all.progexp.cloudengineer.1-2Years.perc.text}{zero}
\definevar{demo.all.progexp.cloudengineer.1-2Years.perc.Text}{Zero}
\definevar{demo.all.progexp.cloudengineer.>5Years.perc.text}{zero}
\definevar{demo.all.progexp.cloudengineer.>5Years.perc.Text}{Zero}
\definevar{demo.all.progexp.cloudengineer.<1Year.perc.text}{zero}
\definevar{demo.all.progexp.cloudengineer.<1Year.perc.Text}{Zero}
\definevar{demo.upwork.progexp.cloudengineer.Prefernottodisclose.text}{zero}
\definevar{demo.upwork.progexp.cloudengineer.Prefernottodisclose.Text}{Zero}
\definevar{demo.upwork.progexp.cloudengineer.Prefernottodisclose.perc.text}{zero}
\definevar{demo.upwork.progexp.cloudengineer.Prefernottodisclose.perc.Text}{Zero}
\definevar{demo.upwork.progexp.cloudengineer.2-5Years.text}{29}
\definevar{demo.upwork.progexp.cloudengineer.2-5Years.Text}{29}
\definevar{demo.upwork.progexp.cloudengineer.1-2Years.text}{12}
\definevar{demo.upwork.progexp.cloudengineer.1-2Years.Text}{12}
\definevar{demo.upwork.progexp.cloudengineer.<1Year.text}{five}
\definevar{demo.upwork.progexp.cloudengineer.<1Year.Text}{Five}
\definevar{demo.upwork.progexp.cloudengineer.>5Years.text}{four}
\definevar{demo.upwork.progexp.cloudengineer.>5Years.Text}{Four}
\definevar{demo.upwork.progexp.cloudengineer.2-5Years.perc.text}{zero}
\definevar{demo.upwork.progexp.cloudengineer.2-5Years.perc.Text}{Zero}
\definevar{demo.upwork.progexp.cloudengineer.1-2Years.perc.text}{zero}
\definevar{demo.upwork.progexp.cloudengineer.1-2Years.perc.Text}{Zero}
\definevar{demo.upwork.progexp.cloudengineer.<1Year.perc.text}{zero}
\definevar{demo.upwork.progexp.cloudengineer.<1Year.perc.Text}{Zero}
\definevar{demo.upwork.progexp.cloudengineer.>5Years.perc.text}{zero}
\definevar{demo.upwork.progexp.cloudengineer.>5Years.perc.Text}{Zero}
\definevar{demo.github.progexp.cloudengineer.Prefernottodisclose.text}{31}
\definevar{demo.github.progexp.cloudengineer.Prefernottodisclose.Text}{31}
\definevar{demo.github.progexp.cloudengineer.Prefernottodisclose.perc.text}{zero}
\definevar{demo.github.progexp.cloudengineer.Prefernottodisclose.perc.Text}{Zero}
\definevar{demo.github.progexp.cloudengineer.>5Years.text}{ten}
\definevar{demo.github.progexp.cloudengineer.>5Years.Text}{Ten}
\definevar{demo.github.progexp.cloudengineer.2-5Years.text}{nine}
\definevar{demo.github.progexp.cloudengineer.2-5Years.Text}{Nine}
\definevar{demo.github.progexp.cloudengineer.<1Year.text}{six}
\definevar{demo.github.progexp.cloudengineer.<1Year.Text}{Six}
\definevar{demo.github.progexp.cloudengineer.1-2Years.text}{three}
\definevar{demo.github.progexp.cloudengineer.1-2Years.Text}{Three}
\definevar{demo.github.progexp.cloudengineer.>5Years.perc.text}{zero}
\definevar{demo.github.progexp.cloudengineer.>5Years.perc.Text}{Zero}
\definevar{demo.github.progexp.cloudengineer.2-5Years.perc.text}{zero}
\definevar{demo.github.progexp.cloudengineer.2-5Years.perc.Text}{Zero}
\definevar{demo.github.progexp.cloudengineer.<1Year.perc.text}{zero}
\definevar{demo.github.progexp.cloudengineer.<1Year.perc.Text}{Zero}
\definevar{demo.github.progexp.cloudengineer.1-2Years.perc.text}{zero}
\definevar{demo.github.progexp.cloudengineer.1-2Years.perc.Text}{Zero}
\definevar{demo.all.progexp.projects.Prefernottodisclose.text}{zero}
\definevar{demo.all.progexp.projects.Prefernottodisclose.Text}{Zero}
\definevar{demo.all.progexp.projects.Prefernottodisclose.perc.text}{zero}
\definevar{demo.all.progexp.projects.Prefernottodisclose.perc.Text}{Zero}
\definevar{demo.all.progexp.projects.1-5Projects.text}{33}
\definevar{demo.all.progexp.projects.1-5Projects.Text}{33}
\definevar{demo.all.progexp.projects.5-10Projects.text}{31}
\definevar{demo.all.progexp.projects.5-10Projects.Text}{31}
\definevar{demo.all.progexp.projects.10-25Projects.text}{23}
\definevar{demo.all.progexp.projects.10-25Projects.Text}{23}
\definevar{demo.all.progexp.projects.≥25Projects.text}{22}
\definevar{demo.all.progexp.projects.≥25Projects.Text}{22}
\definevar{demo.all.progexp.projects.1-5Projects.perc.text}{zero}
\definevar{demo.all.progexp.projects.1-5Projects.perc.Text}{Zero}
\definevar{demo.all.progexp.projects.5-10Projects.perc.text}{zero}
\definevar{demo.all.progexp.projects.5-10Projects.perc.Text}{Zero}
\definevar{demo.all.progexp.projects.10-25Projects.perc.text}{zero}
\definevar{demo.all.progexp.projects.10-25Projects.perc.Text}{Zero}
\definevar{demo.all.progexp.projects.≥25Projects.perc.text}{zero}
\definevar{demo.all.progexp.projects.≥25Projects.perc.Text}{Zero}
\definevar{demo.upwork.progexp.projects.Prefernottodisclose.text}{zero}
\definevar{demo.upwork.progexp.projects.Prefernottodisclose.Text}{Zero}
\definevar{demo.upwork.progexp.projects.Prefernottodisclose.perc.text}{zero}
\definevar{demo.upwork.progexp.projects.Prefernottodisclose.perc.Text}{Zero}
\definevar{demo.upwork.progexp.projects.5-10Projects.text}{19}
\definevar{demo.upwork.progexp.projects.5-10Projects.Text}{19}
\definevar{demo.upwork.progexp.projects.1-5Projects.text}{17}
\definevar{demo.upwork.progexp.projects.1-5Projects.Text}{17}
\definevar{demo.upwork.progexp.projects.10-25Projects.text}{ten}
\definevar{demo.upwork.progexp.projects.10-25Projects.Text}{Ten}
\definevar{demo.upwork.progexp.projects.≥25Projects.text}{four}
\definevar{demo.upwork.progexp.projects.≥25Projects.Text}{Four}
\definevar{demo.upwork.progexp.projects.5-10Projects.perc.text}{zero}
\definevar{demo.upwork.progexp.projects.5-10Projects.perc.Text}{Zero}
\definevar{demo.upwork.progexp.projects.1-5Projects.perc.text}{zero}
\definevar{demo.upwork.progexp.projects.1-5Projects.perc.Text}{Zero}
\definevar{demo.upwork.progexp.projects.10-25Projects.perc.text}{zero}
\definevar{demo.upwork.progexp.projects.10-25Projects.perc.Text}{Zero}
\definevar{demo.upwork.progexp.projects.≥25Projects.perc.text}{zero}
\definevar{demo.upwork.progexp.projects.≥25Projects.perc.Text}{Zero}
\definevar{demo.github.progexp.projects.Prefernottodisclose.text}{zero}
\definevar{demo.github.progexp.projects.Prefernottodisclose.Text}{Zero}
\definevar{demo.github.progexp.projects.Prefernottodisclose.perc.text}{zero}
\definevar{demo.github.progexp.projects.Prefernottodisclose.perc.Text}{Zero}
\definevar{demo.github.progexp.projects.≥25Projects.text}{18}
\definevar{demo.github.progexp.projects.≥25Projects.Text}{18}
\definevar{demo.github.progexp.projects.1-5Projects.text}{16}
\definevar{demo.github.progexp.projects.1-5Projects.Text}{16}
\definevar{demo.github.progexp.projects.10-25Projects.text}{13}
\definevar{demo.github.progexp.projects.10-25Projects.Text}{13}
\definevar{demo.github.progexp.projects.5-10Projects.text}{12}
\definevar{demo.github.progexp.projects.5-10Projects.Text}{12}
\definevar{demo.github.progexp.projects.≥25Projects.perc.text}{zero}
\definevar{demo.github.progexp.projects.≥25Projects.perc.Text}{Zero}
\definevar{demo.github.progexp.projects.1-5Projects.perc.text}{zero}
\definevar{demo.github.progexp.projects.1-5Projects.perc.Text}{Zero}
\definevar{demo.github.progexp.projects.10-25Projects.perc.text}{zero}
\definevar{demo.github.progexp.projects.10-25Projects.perc.Text}{Zero}
\definevar{demo.github.progexp.projects.5-10Projects.perc.text}{zero}
\definevar{demo.github.progexp.projects.5-10Projects.perc.Text}{Zero}
\definevar{demo.all.progexp.learning.Prefernottodisclose.text}{zero}
\definevar{demo.all.progexp.learning.Prefernottodisclose.Text}{Zero}
\definevar{demo.all.progexp.learning.Prefernottodisclose.perc.text}{zero}
\definevar{demo.all.progexp.learning.Prefernottodisclose.perc.Text}{Zero}
\definevar{demo.all.progexp.learning.Self-taught.text}{102}
\definevar{demo.all.progexp.learning.Self-taught.Text}{102}
\definevar{demo.all.progexp.learning.College/University.text}{64}
\definevar{demo.all.progexp.learning.College/University.Text}{64}
\definevar{demo.all.progexp.learning.On-the-jobtraining.text}{61}
\definevar{demo.all.progexp.learning.On-the-jobtraining.Text}{61}
\definevar{demo.all.progexp.learning.Onlineclass.text}{47}
\definevar{demo.all.progexp.learning.Onlineclass.Text}{47}
\definevar{demo.all.progexp.learning.Codingcamp.text}{14}
\definevar{demo.all.progexp.learning.Codingcamp.Text}{14}
\definevar{demo.all.progexp.learning.Other.text}{five}
\definevar{demo.all.progexp.learning.Other.Text}{Five}
\definevar{demo.all.progexp.learning.Self-taught.perc.text}{zero}
\definevar{demo.all.progexp.learning.Self-taught.perc.Text}{Zero}
\definevar{demo.all.progexp.learning.College/University.perc.text}{zero}
\definevar{demo.all.progexp.learning.College/University.perc.Text}{Zero}
\definevar{demo.all.progexp.learning.On-the-jobtraining.perc.text}{zero}
\definevar{demo.all.progexp.learning.On-the-jobtraining.perc.Text}{Zero}
\definevar{demo.all.progexp.learning.Onlineclass.perc.text}{zero}
\definevar{demo.all.progexp.learning.Onlineclass.perc.Text}{Zero}
\definevar{demo.all.progexp.learning.Codingcamp.perc.text}{zero}
\definevar{demo.all.progexp.learning.Codingcamp.perc.Text}{Zero}
\definevar{demo.all.progexp.learning.Other.perc.text}{zero}
\definevar{demo.all.progexp.learning.Other.perc.Text}{Zero}
\definevar{demo.upwork.progexp.learning.Prefernottodisclose.text}{zero}
\definevar{demo.upwork.progexp.learning.Prefernottodisclose.Text}{Zero}
\definevar{demo.upwork.progexp.learning.Prefernottodisclose.perc.text}{zero}
\definevar{demo.upwork.progexp.learning.Prefernottodisclose.perc.Text}{Zero}
\definevar{demo.upwork.progexp.learning.Self-taught.text}{46}
\definevar{demo.upwork.progexp.learning.Self-taught.Text}{46}
\definevar{demo.upwork.progexp.learning.On-the-jobtraining.text}{36}
\definevar{demo.upwork.progexp.learning.On-the-jobtraining.Text}{36}
\definevar{demo.upwork.progexp.learning.Onlineclass.text}{30}
\definevar{demo.upwork.progexp.learning.Onlineclass.Text}{30}
\definevar{demo.upwork.progexp.learning.College/University.text}{27}
\definevar{demo.upwork.progexp.learning.College/University.Text}{27}
\definevar{demo.upwork.progexp.learning.Codingcamp.text}{nine}
\definevar{demo.upwork.progexp.learning.Codingcamp.Text}{Nine}
\definevar{demo.upwork.progexp.learning.Other.text}{three}
\definevar{demo.upwork.progexp.learning.Other.Text}{Three}
\definevar{demo.upwork.progexp.learning.Self-taught.perc.text}{zero}
\definevar{demo.upwork.progexp.learning.Self-taught.perc.Text}{Zero}
\definevar{demo.upwork.progexp.learning.On-the-jobtraining.perc.text}{zero}
\definevar{demo.upwork.progexp.learning.On-the-jobtraining.perc.Text}{Zero}
\definevar{demo.upwork.progexp.learning.Onlineclass.perc.text}{zero}
\definevar{demo.upwork.progexp.learning.Onlineclass.perc.Text}{Zero}
\definevar{demo.upwork.progexp.learning.College/University.perc.text}{zero}
\definevar{demo.upwork.progexp.learning.College/University.perc.Text}{Zero}
\definevar{demo.upwork.progexp.learning.Codingcamp.perc.text}{zero}
\definevar{demo.upwork.progexp.learning.Codingcamp.perc.Text}{Zero}
\definevar{demo.upwork.progexp.learning.Other.perc.text}{zero}
\definevar{demo.upwork.progexp.learning.Other.perc.Text}{Zero}
\definevar{demo.github.progexp.learning.Prefernottodisclose.text}{zero}
\definevar{demo.github.progexp.learning.Prefernottodisclose.Text}{Zero}
\definevar{demo.github.progexp.learning.Prefernottodisclose.perc.text}{zero}
\definevar{demo.github.progexp.learning.Prefernottodisclose.perc.Text}{Zero}
\definevar{demo.github.progexp.learning.Self-taught.text}{56}
\definevar{demo.github.progexp.learning.Self-taught.Text}{56}
\definevar{demo.github.progexp.learning.College/University.text}{37}
\definevar{demo.github.progexp.learning.College/University.Text}{37}
\definevar{demo.github.progexp.learning.On-the-jobtraining.text}{25}
\definevar{demo.github.progexp.learning.On-the-jobtraining.Text}{25}
\definevar{demo.github.progexp.learning.Onlineclass.text}{17}
\definevar{demo.github.progexp.learning.Onlineclass.Text}{17}
\definevar{demo.github.progexp.learning.Codingcamp.text}{five}
\definevar{demo.github.progexp.learning.Codingcamp.Text}{Five}
\definevar{demo.github.progexp.learning.Other.text}{two}
\definevar{demo.github.progexp.learning.Other.Text}{Two}
\definevar{demo.github.progexp.learning.Self-taught.perc.text}{zero}
\definevar{demo.github.progexp.learning.Self-taught.perc.Text}{Zero}
\definevar{demo.github.progexp.learning.College/University.perc.text}{zero}
\definevar{demo.github.progexp.learning.College/University.perc.Text}{Zero}
\definevar{demo.github.progexp.learning.On-the-jobtraining.perc.text}{zero}
\definevar{demo.github.progexp.learning.On-the-jobtraining.perc.Text}{Zero}
\definevar{demo.github.progexp.learning.Onlineclass.perc.text}{zero}
\definevar{demo.github.progexp.learning.Onlineclass.perc.Text}{Zero}
\definevar{demo.github.progexp.learning.Codingcamp.perc.text}{zero}
\definevar{demo.github.progexp.learning.Codingcamp.perc.Text}{Zero}
\definevar{demo.github.progexp.learning.Other.perc.text}{zero}
\definevar{demo.github.progexp.learning.Other.perc.Text}{Zero}
\definevar{demo.interviews.progexp.learning.Prefernottodisclose.text}{zero}
\definevar{demo.interviews.progexp.learning.Prefernottodisclose.Text}{Zero}
\definevar{demo.interviews.progexp.learning.Prefernottodisclose.perc.text}{zero}
\definevar{demo.interviews.progexp.learning.Prefernottodisclose.perc.Text}{Zero}
\definevar{demo.interviews.progexp.learning.Self-taught.text}{12}
\definevar{demo.interviews.progexp.learning.Self-taught.Text}{12}
\definevar{demo.interviews.progexp.learning.College/University.text}{ten}
\definevar{demo.interviews.progexp.learning.College/University.Text}{Ten}
\definevar{demo.interviews.progexp.learning.On-the-jobtraining.text}{eight}
\definevar{demo.interviews.progexp.learning.On-the-jobtraining.Text}{Eight}
\definevar{demo.interviews.progexp.learning.Onlineclass.text}{five}
\definevar{demo.interviews.progexp.learning.Onlineclass.Text}{Five}
\definevar{demo.interviews.progexp.learning.Other.text}{one}
\definevar{demo.interviews.progexp.learning.Other.Text}{One}
\definevar{demo.interviews.progexp.learning.Self-taught.perc.text}{zero}
\definevar{demo.interviews.progexp.learning.Self-taught.perc.Text}{Zero}
\definevar{demo.interviews.progexp.learning.College/University.perc.text}{zero}
\definevar{demo.interviews.progexp.learning.College/University.perc.Text}{Zero}
\definevar{demo.interviews.progexp.learning.On-the-jobtraining.perc.text}{zero}
\definevar{demo.interviews.progexp.learning.On-the-jobtraining.perc.Text}{Zero}
\definevar{demo.interviews.progexp.learning.Onlineclass.perc.text}{zero}
\definevar{demo.interviews.progexp.learning.Onlineclass.perc.Text}{Zero}
\definevar{demo.interviews.progexp.learning.Other.perc.text}{zero}
\definevar{demo.interviews.progexp.learning.Other.perc.Text}{Zero}
\definevar{demo.all.secexp.responsible.Prefernottodisclose.text}{zero}
\definevar{demo.all.secexp.responsible.Prefernottodisclose.Text}{Zero}
\definevar{demo.all.secexp.responsible.Prefernottodisclose.perc.text}{zero}
\definevar{demo.all.secexp.responsible.Prefernottodisclose.perc.Text}{Zero}
\definevar{demo.all.secexp.responsible.Agree.text}{41}
\definevar{demo.all.secexp.responsible.Agree.Text}{41}
\definevar{demo.all.secexp.responsible.StronglyAgree.text}{38}
\definevar{demo.all.secexp.responsible.StronglyAgree.Text}{38}
\definevar{demo.all.secexp.responsible.Neutral.text}{18}
\definevar{demo.all.secexp.responsible.Neutral.Text}{18}
\definevar{demo.all.secexp.responsible.Stronglydisagree.text}{six}
\definevar{demo.all.secexp.responsible.Stronglydisagree.Text}{Six}
\definevar{demo.all.secexp.responsible.Disagree.text}{six}
\definevar{demo.all.secexp.responsible.Disagree.Text}{Six}
\definevar{demo.all.secexp.responsible.Agree.perc.text}{zero}
\definevar{demo.all.secexp.responsible.Agree.perc.Text}{Zero}
\definevar{demo.all.secexp.responsible.StronglyAgree.perc.text}{zero}
\definevar{demo.all.secexp.responsible.StronglyAgree.perc.Text}{Zero}
\definevar{demo.all.secexp.responsible.Neutral.perc.text}{zero}
\definevar{demo.all.secexp.responsible.Neutral.perc.Text}{Zero}
\definevar{demo.all.secexp.responsible.Stronglydisagree.perc.text}{zero}
\definevar{demo.all.secexp.responsible.Stronglydisagree.perc.Text}{Zero}
\definevar{demo.all.secexp.responsible.Disagree.perc.text}{zero}
\definevar{demo.all.secexp.responsible.Disagree.perc.Text}{Zero}
\definevar{demo.all.secexp.responsible.AgreeOrStronglyAgree.text}{79}
\definevar{demo.all.secexp.responsible.AgreeOrStronglyAgree.Text}{79}
\definevar{demo.all.secexp.responsible.AgreeOrStronglyAgree.perc.text}{zero}
\definevar{demo.all.secexp.responsible.AgreeOrStronglyAgree.perc.Text}{Zero}
\definevar{demo.all.secexp.takecare.Prefernottodisclose.text}{zero}
\definevar{demo.all.secexp.takecare.Prefernottodisclose.Text}{Zero}
\definevar{demo.all.secexp.takecare.Prefernottodisclose.perc.text}{zero}
\definevar{demo.all.secexp.takecare.Prefernottodisclose.perc.Text}{Zero}
\definevar{demo.all.secexp.takecare.Agree.text}{51}
\definevar{demo.all.secexp.takecare.Agree.Text}{51}
\definevar{demo.all.secexp.takecare.StronglyAgree.text}{46}
\definevar{demo.all.secexp.takecare.StronglyAgree.Text}{46}
\definevar{demo.all.secexp.takecare.Neutral.text}{six}
\definevar{demo.all.secexp.takecare.Neutral.Text}{Six}
\definevar{demo.all.secexp.takecare.Disagree.text}{three}
\definevar{demo.all.secexp.takecare.Disagree.Text}{Three}
\definevar{demo.all.secexp.takecare.Stronglydisagree.text}{three}
\definevar{demo.all.secexp.takecare.Stronglydisagree.Text}{Three}
\definevar{demo.all.secexp.takecare.Agree.perc.text}{zero}
\definevar{demo.all.secexp.takecare.Agree.perc.Text}{Zero}
\definevar{demo.all.secexp.takecare.StronglyAgree.perc.text}{zero}
\definevar{demo.all.secexp.takecare.StronglyAgree.perc.Text}{Zero}
\definevar{demo.all.secexp.takecare.Neutral.perc.text}{zero}
\definevar{demo.all.secexp.takecare.Neutral.perc.Text}{Zero}
\definevar{demo.all.secexp.takecare.Disagree.perc.text}{zero}
\definevar{demo.all.secexp.takecare.Disagree.perc.Text}{Zero}
\definevar{demo.all.secexp.takecare.Stronglydisagree.perc.text}{zero}
\definevar{demo.all.secexp.takecare.Stronglydisagree.perc.Text}{Zero}
\definevar{demo.all.secexp.takecare.AgreeOrStronglyAgree.text}{97}
\definevar{demo.all.secexp.takecare.AgreeOrStronglyAgree.Text}{97}
\definevar{demo.all.secexp.takecare.AgreeOrStronglyAgree.perc.text}{zero}
\definevar{demo.all.secexp.takecare.AgreeOrStronglyAgree.perc.Text}{Zero}
\definevar{demo.all.secexp.confident.Prefernottodisclose.text}{zero}
\definevar{demo.all.secexp.confident.Prefernottodisclose.Text}{Zero}
\definevar{demo.all.secexp.confident.Prefernottodisclose.perc.text}{zero}
\definevar{demo.all.secexp.confident.Prefernottodisclose.perc.Text}{Zero}
\definevar{demo.all.secexp.confident.Agree.text}{40}
\definevar{demo.all.secexp.confident.Agree.Text}{40}
\definevar{demo.all.secexp.confident.Neutral.text}{30}
\definevar{demo.all.secexp.confident.Neutral.Text}{30}
\definevar{demo.all.secexp.confident.StronglyAgree.text}{19}
\definevar{demo.all.secexp.confident.StronglyAgree.Text}{19}
\definevar{demo.all.secexp.confident.Disagree.text}{16}
\definevar{demo.all.secexp.confident.Disagree.Text}{16}
\definevar{demo.all.secexp.confident.Stronglydisagree.text}{four}
\definevar{demo.all.secexp.confident.Stronglydisagree.Text}{Four}
\definevar{demo.all.secexp.confident.Agree.perc.text}{zero}
\definevar{demo.all.secexp.confident.Agree.perc.Text}{Zero}
\definevar{demo.all.secexp.confident.Neutral.perc.text}{zero}
\definevar{demo.all.secexp.confident.Neutral.perc.Text}{Zero}
\definevar{demo.all.secexp.confident.StronglyAgree.perc.text}{zero}
\definevar{demo.all.secexp.confident.StronglyAgree.perc.Text}{Zero}
\definevar{demo.all.secexp.confident.Disagree.perc.text}{zero}
\definevar{demo.all.secexp.confident.Disagree.perc.Text}{Zero}
\definevar{demo.all.secexp.confident.Stronglydisagree.perc.text}{zero}
\definevar{demo.all.secexp.confident.Stronglydisagree.perc.Text}{Zero}
\definevar{demo.all.secexp.confident.AgreeOrStronglyAgree.text}{59}
\definevar{demo.all.secexp.confident.AgreeOrStronglyAgree.Text}{59}
\definevar{demo.all.secexp.confident.AgreeOrStronglyAgree.perc.text}{zero}
\definevar{demo.all.secexp.confident.AgreeOrStronglyAgree.perc.Text}{Zero}
\definevar{demo.upwork.secexp.responsible.Prefernottodisclose.text}{zero}
\definevar{demo.upwork.secexp.responsible.Prefernottodisclose.Text}{Zero}
\definevar{demo.upwork.secexp.responsible.Prefernottodisclose.perc.text}{zero}
\definevar{demo.upwork.secexp.responsible.Prefernottodisclose.perc.Text}{Zero}
\definevar{demo.upwork.secexp.responsible.Agree.text}{20}
\definevar{demo.upwork.secexp.responsible.Agree.Text}{20}
\definevar{demo.upwork.secexp.responsible.StronglyAgree.text}{16}
\definevar{demo.upwork.secexp.responsible.StronglyAgree.Text}{16}
\definevar{demo.upwork.secexp.responsible.Neutral.text}{eleven}
\definevar{demo.upwork.secexp.responsible.Neutral.Text}{Eleven}
\definevar{demo.upwork.secexp.responsible.Stronglydisagree.text}{two}
\definevar{demo.upwork.secexp.responsible.Stronglydisagree.Text}{Two}
\definevar{demo.upwork.secexp.responsible.Disagree.text}{one}
\definevar{demo.upwork.secexp.responsible.Disagree.Text}{One}
\definevar{demo.upwork.secexp.responsible.Agree.perc.text}{zero}
\definevar{demo.upwork.secexp.responsible.Agree.perc.Text}{Zero}
\definevar{demo.upwork.secexp.responsible.StronglyAgree.perc.text}{zero}
\definevar{demo.upwork.secexp.responsible.StronglyAgree.perc.Text}{Zero}
\definevar{demo.upwork.secexp.responsible.Neutral.perc.text}{zero}
\definevar{demo.upwork.secexp.responsible.Neutral.perc.Text}{Zero}
\definevar{demo.upwork.secexp.responsible.Stronglydisagree.perc.text}{zero}
\definevar{demo.upwork.secexp.responsible.Stronglydisagree.perc.Text}{Zero}
\definevar{demo.upwork.secexp.responsible.Disagree.perc.text}{zero}
\definevar{demo.upwork.secexp.responsible.Disagree.perc.Text}{Zero}
\definevar{demo.upwork.secexp.responsible.AgreeOrStronglyAgree.text}{36}
\definevar{demo.upwork.secexp.responsible.AgreeOrStronglyAgree.Text}{36}
\definevar{demo.upwork.secexp.responsible.AgreeOrStronglyAgree.perc.text}{zero}
\definevar{demo.upwork.secexp.responsible.AgreeOrStronglyAgree.perc.Text}{Zero}
\definevar{demo.upwork.secexp.takecare.Prefernottodisclose.text}{zero}
\definevar{demo.upwork.secexp.takecare.Prefernottodisclose.Text}{Zero}
\definevar{demo.upwork.secexp.takecare.Prefernottodisclose.perc.text}{zero}
\definevar{demo.upwork.secexp.takecare.Prefernottodisclose.perc.Text}{Zero}
\definevar{demo.upwork.secexp.takecare.Agree.text}{26}
\definevar{demo.upwork.secexp.takecare.Agree.Text}{26}
\definevar{demo.upwork.secexp.takecare.StronglyAgree.text}{19}
\definevar{demo.upwork.secexp.takecare.StronglyAgree.Text}{19}
\definevar{demo.upwork.secexp.takecare.Neutral.text}{two}
\definevar{demo.upwork.secexp.takecare.Neutral.Text}{Two}
\definevar{demo.upwork.secexp.takecare.Stronglydisagree.text}{two}
\definevar{demo.upwork.secexp.takecare.Stronglydisagree.Text}{Two}
\definevar{demo.upwork.secexp.takecare.Disagree.text}{one}
\definevar{demo.upwork.secexp.takecare.Disagree.Text}{One}
\definevar{demo.upwork.secexp.takecare.Agree.perc.text}{zero}
\definevar{demo.upwork.secexp.takecare.Agree.perc.Text}{Zero}
\definevar{demo.upwork.secexp.takecare.StronglyAgree.perc.text}{zero}
\definevar{demo.upwork.secexp.takecare.StronglyAgree.perc.Text}{Zero}
\definevar{demo.upwork.secexp.takecare.Neutral.perc.text}{zero}
\definevar{demo.upwork.secexp.takecare.Neutral.perc.Text}{Zero}
\definevar{demo.upwork.secexp.takecare.Stronglydisagree.perc.text}{zero}
\definevar{demo.upwork.secexp.takecare.Stronglydisagree.perc.Text}{Zero}
\definevar{demo.upwork.secexp.takecare.Disagree.perc.text}{zero}
\definevar{demo.upwork.secexp.takecare.Disagree.perc.Text}{Zero}
\definevar{demo.upwork.secexp.takecare.AgreeOrStronglyAgree.text}{45}
\definevar{demo.upwork.secexp.takecare.AgreeOrStronglyAgree.Text}{45}
\definevar{demo.upwork.secexp.takecare.AgreeOrStronglyAgree.perc.text}{zero}
\definevar{demo.upwork.secexp.takecare.AgreeOrStronglyAgree.perc.Text}{Zero}
\definevar{demo.upwork.secexp.confident.Prefernottodisclose.text}{zero}
\definevar{demo.upwork.secexp.confident.Prefernottodisclose.Text}{Zero}
\definevar{demo.upwork.secexp.confident.Prefernottodisclose.perc.text}{zero}
\definevar{demo.upwork.secexp.confident.Prefernottodisclose.perc.Text}{Zero}
\definevar{demo.upwork.secexp.confident.Agree.text}{24}
\definevar{demo.upwork.secexp.confident.Agree.Text}{24}
\definevar{demo.upwork.secexp.confident.Neutral.text}{13}
\definevar{demo.upwork.secexp.confident.Neutral.Text}{13}
\definevar{demo.upwork.secexp.confident.StronglyAgree.text}{eight}
\definevar{demo.upwork.secexp.confident.StronglyAgree.Text}{Eight}
\definevar{demo.upwork.secexp.confident.Disagree.text}{four}
\definevar{demo.upwork.secexp.confident.Disagree.Text}{Four}
\definevar{demo.upwork.secexp.confident.Stronglydisagree.text}{one}
\definevar{demo.upwork.secexp.confident.Stronglydisagree.Text}{One}
\definevar{demo.upwork.secexp.confident.Agree.perc.text}{zero}
\definevar{demo.upwork.secexp.confident.Agree.perc.Text}{Zero}
\definevar{demo.upwork.secexp.confident.Neutral.perc.text}{zero}
\definevar{demo.upwork.secexp.confident.Neutral.perc.Text}{Zero}
\definevar{demo.upwork.secexp.confident.StronglyAgree.perc.text}{zero}
\definevar{demo.upwork.secexp.confident.StronglyAgree.perc.Text}{Zero}
\definevar{demo.upwork.secexp.confident.Disagree.perc.text}{zero}
\definevar{demo.upwork.secexp.confident.Disagree.perc.Text}{Zero}
\definevar{demo.upwork.secexp.confident.Stronglydisagree.perc.text}{zero}
\definevar{demo.upwork.secexp.confident.Stronglydisagree.perc.Text}{Zero}
\definevar{demo.upwork.secexp.confident.AgreeOrStronglyAgree.text}{32}
\definevar{demo.upwork.secexp.confident.AgreeOrStronglyAgree.Text}{32}
\definevar{demo.upwork.secexp.confident.AgreeOrStronglyAgree.perc.text}{zero}
\definevar{demo.upwork.secexp.confident.AgreeOrStronglyAgree.perc.Text}{Zero}
\definevar{demo.github.secexp.responsible.Prefernottodisclose.text}{zero}
\definevar{demo.github.secexp.responsible.Prefernottodisclose.Text}{Zero}
\definevar{demo.github.secexp.responsible.Prefernottodisclose.perc.text}{zero}
\definevar{demo.github.secexp.responsible.Prefernottodisclose.perc.Text}{Zero}
\definevar{demo.github.secexp.responsible.StronglyAgree.text}{22}
\definevar{demo.github.secexp.responsible.StronglyAgree.Text}{22}
\definevar{demo.github.secexp.responsible.Agree.text}{21}
\definevar{demo.github.secexp.responsible.Agree.Text}{21}
\definevar{demo.github.secexp.responsible.Neutral.text}{seven}
\definevar{demo.github.secexp.responsible.Neutral.Text}{Seven}
\definevar{demo.github.secexp.responsible.Disagree.text}{five}
\definevar{demo.github.secexp.responsible.Disagree.Text}{Five}
\definevar{demo.github.secexp.responsible.Stronglydisagree.text}{four}
\definevar{demo.github.secexp.responsible.Stronglydisagree.Text}{Four}
\definevar{demo.github.secexp.responsible.StronglyAgree.perc.text}{zero}
\definevar{demo.github.secexp.responsible.StronglyAgree.perc.Text}{Zero}
\definevar{demo.github.secexp.responsible.Agree.perc.text}{zero}
\definevar{demo.github.secexp.responsible.Agree.perc.Text}{Zero}
\definevar{demo.github.secexp.responsible.Neutral.perc.text}{zero}
\definevar{demo.github.secexp.responsible.Neutral.perc.Text}{Zero}
\definevar{demo.github.secexp.responsible.Disagree.perc.text}{zero}
\definevar{demo.github.secexp.responsible.Disagree.perc.Text}{Zero}
\definevar{demo.github.secexp.responsible.Stronglydisagree.perc.text}{zero}
\definevar{demo.github.secexp.responsible.Stronglydisagree.perc.Text}{Zero}
\definevar{demo.github.secexp.responsible.AgreeOrStronglyAgree.text}{43}
\definevar{demo.github.secexp.responsible.AgreeOrStronglyAgree.Text}{43}
\definevar{demo.github.secexp.responsible.AgreeOrStronglyAgree.perc.text}{zero}
\definevar{demo.github.secexp.responsible.AgreeOrStronglyAgree.perc.Text}{Zero}
\definevar{demo.github.secexp.takecare.Prefernottodisclose.text}{zero}
\definevar{demo.github.secexp.takecare.Prefernottodisclose.Text}{Zero}
\definevar{demo.github.secexp.takecare.Prefernottodisclose.perc.text}{zero}
\definevar{demo.github.secexp.takecare.Prefernottodisclose.perc.Text}{Zero}
\definevar{demo.github.secexp.takecare.StronglyAgree.text}{27}
\definevar{demo.github.secexp.takecare.StronglyAgree.Text}{27}
\definevar{demo.github.secexp.takecare.Agree.text}{25}
\definevar{demo.github.secexp.takecare.Agree.Text}{25}
\definevar{demo.github.secexp.takecare.Neutral.text}{four}
\definevar{demo.github.secexp.takecare.Neutral.Text}{Four}
\definevar{demo.github.secexp.takecare.Disagree.text}{two}
\definevar{demo.github.secexp.takecare.Disagree.Text}{Two}
\definevar{demo.github.secexp.takecare.Stronglydisagree.text}{one}
\definevar{demo.github.secexp.takecare.Stronglydisagree.Text}{One}
\definevar{demo.github.secexp.takecare.StronglyAgree.perc.text}{zero}
\definevar{demo.github.secexp.takecare.StronglyAgree.perc.Text}{Zero}
\definevar{demo.github.secexp.takecare.Agree.perc.text}{zero}
\definevar{demo.github.secexp.takecare.Agree.perc.Text}{Zero}
\definevar{demo.github.secexp.takecare.Neutral.perc.text}{zero}
\definevar{demo.github.secexp.takecare.Neutral.perc.Text}{Zero}
\definevar{demo.github.secexp.takecare.Disagree.perc.text}{zero}
\definevar{demo.github.secexp.takecare.Disagree.perc.Text}{Zero}
\definevar{demo.github.secexp.takecare.Stronglydisagree.perc.text}{zero}
\definevar{demo.github.secexp.takecare.Stronglydisagree.perc.Text}{Zero}
\definevar{demo.github.secexp.takecare.AgreeOrStronglyAgree.text}{52}
\definevar{demo.github.secexp.takecare.AgreeOrStronglyAgree.Text}{52}
\definevar{demo.github.secexp.takecare.AgreeOrStronglyAgree.perc.text}{zero}
\definevar{demo.github.secexp.takecare.AgreeOrStronglyAgree.perc.Text}{Zero}
\definevar{demo.github.secexp.confident.Prefernottodisclose.text}{zero}
\definevar{demo.github.secexp.confident.Prefernottodisclose.Text}{Zero}
\definevar{demo.github.secexp.confident.Prefernottodisclose.perc.text}{zero}
\definevar{demo.github.secexp.confident.Prefernottodisclose.perc.Text}{Zero}
\definevar{demo.github.secexp.confident.Neutral.text}{17}
\definevar{demo.github.secexp.confident.Neutral.Text}{17}
\definevar{demo.github.secexp.confident.Agree.text}{16}
\definevar{demo.github.secexp.confident.Agree.Text}{16}
\definevar{demo.github.secexp.confident.Disagree.text}{12}
\definevar{demo.github.secexp.confident.Disagree.Text}{12}
\definevar{demo.github.secexp.confident.StronglyAgree.text}{eleven}
\definevar{demo.github.secexp.confident.StronglyAgree.Text}{Eleven}
\definevar{demo.github.secexp.confident.Stronglydisagree.text}{three}
\definevar{demo.github.secexp.confident.Stronglydisagree.Text}{Three}
\definevar{demo.github.secexp.confident.Neutral.perc.text}{zero}
\definevar{demo.github.secexp.confident.Neutral.perc.Text}{Zero}
\definevar{demo.github.secexp.confident.Agree.perc.text}{zero}
\definevar{demo.github.secexp.confident.Agree.perc.Text}{Zero}
\definevar{demo.github.secexp.confident.Disagree.perc.text}{zero}
\definevar{demo.github.secexp.confident.Disagree.perc.Text}{Zero}
\definevar{demo.github.secexp.confident.StronglyAgree.perc.text}{zero}
\definevar{demo.github.secexp.confident.StronglyAgree.perc.Text}{Zero}
\definevar{demo.github.secexp.confident.Stronglydisagree.perc.text}{zero}
\definevar{demo.github.secexp.confident.Stronglydisagree.perc.Text}{Zero}
\definevar{demo.github.secexp.confident.AgreeOrStronglyAgree.text}{27}
\definevar{demo.github.secexp.confident.AgreeOrStronglyAgree.Text}{27}
\definevar{demo.github.secexp.confident.AgreeOrStronglyAgree.perc.text}{zero}
\definevar{demo.github.secexp.confident.AgreeOrStronglyAgree.perc.Text}{Zero}
\definevar{demo.all.secexp.training.Prefernottodisclose.text}{zero}
\definevar{demo.all.secexp.training.Prefernottodisclose.Text}{Zero}
\definevar{demo.all.secexp.training.Prefernottodisclose.perc.text}{zero}
\definevar{demo.all.secexp.training.Prefernottodisclose.perc.Text}{Zero}
\definevar{demo.all.secexp.training.Self-taught.text}{86}
\definevar{demo.all.secexp.training.Self-taught.Text}{86}
\definevar{demo.all.secexp.training.On-the-jobtraining.text}{55}
\definevar{demo.all.secexp.training.On-the-jobtraining.Text}{55}
\definevar{demo.all.secexp.training.Onlineclass.text}{39}
\definevar{demo.all.secexp.training.Onlineclass.Text}{39}
\definevar{demo.all.secexp.training.College/University.text}{24}
\definevar{demo.all.secexp.training.College/University.Text}{24}
\definevar{demo.all.secexp.training.None.text}{17}
\definevar{demo.all.secexp.training.None.Text}{17}
\definevar{demo.all.secexp.training.Other.text}{seven}
\definevar{demo.all.secexp.training.Other.Text}{Seven}
\definevar{demo.all.secexp.training.Codingcamp.text}{six}
\definevar{demo.all.secexp.training.Codingcamp.Text}{Six}
\definevar{demo.all.secexp.training.Self-taught.perc.text}{zero}
\definevar{demo.all.secexp.training.Self-taught.perc.Text}{Zero}
\definevar{demo.all.secexp.training.On-the-jobtraining.perc.text}{zero}
\definevar{demo.all.secexp.training.On-the-jobtraining.perc.Text}{Zero}
\definevar{demo.all.secexp.training.Onlineclass.perc.text}{zero}
\definevar{demo.all.secexp.training.Onlineclass.perc.Text}{Zero}
\definevar{demo.all.secexp.training.College/University.perc.text}{zero}
\definevar{demo.all.secexp.training.College/University.perc.Text}{Zero}
\definevar{demo.all.secexp.training.None.perc.text}{zero}
\definevar{demo.all.secexp.training.None.perc.Text}{Zero}
\definevar{demo.all.secexp.training.Other.perc.text}{zero}
\definevar{demo.all.secexp.training.Other.perc.Text}{Zero}
\definevar{demo.all.secexp.training.Codingcamp.perc.text}{zero}
\definevar{demo.all.secexp.training.Codingcamp.perc.Text}{Zero}
\definevar{demo.upwork.secexp.training.Prefernottodisclose.text}{zero}
\definevar{demo.upwork.secexp.training.Prefernottodisclose.Text}{Zero}
\definevar{demo.upwork.secexp.training.Prefernottodisclose.perc.text}{zero}
\definevar{demo.upwork.secexp.training.Prefernottodisclose.perc.Text}{Zero}
\definevar{demo.upwork.secexp.training.On-the-jobtraining.text}{38}
\definevar{demo.upwork.secexp.training.On-the-jobtraining.Text}{38}
\definevar{demo.upwork.secexp.training.Self-taught.text}{36}
\definevar{demo.upwork.secexp.training.Self-taught.Text}{36}
\definevar{demo.upwork.secexp.training.Onlineclass.text}{27}
\definevar{demo.upwork.secexp.training.Onlineclass.Text}{27}
\definevar{demo.upwork.secexp.training.College/University.text}{16}
\definevar{demo.upwork.secexp.training.College/University.Text}{16}
\definevar{demo.upwork.secexp.training.Codingcamp.text}{five}
\definevar{demo.upwork.secexp.training.Codingcamp.Text}{Five}
\definevar{demo.upwork.secexp.training.None.text}{four}
\definevar{demo.upwork.secexp.training.None.Text}{Four}
\definevar{demo.upwork.secexp.training.Other.text}{four}
\definevar{demo.upwork.secexp.training.Other.Text}{Four}
\definevar{demo.upwork.secexp.training.On-the-jobtraining.perc.text}{zero}
\definevar{demo.upwork.secexp.training.On-the-jobtraining.perc.Text}{Zero}
\definevar{demo.upwork.secexp.training.Self-taught.perc.text}{zero}
\definevar{demo.upwork.secexp.training.Self-taught.perc.Text}{Zero}
\definevar{demo.upwork.secexp.training.Onlineclass.perc.text}{zero}
\definevar{demo.upwork.secexp.training.Onlineclass.perc.Text}{Zero}
\definevar{demo.upwork.secexp.training.College/University.perc.text}{zero}
\definevar{demo.upwork.secexp.training.College/University.perc.Text}{Zero}
\definevar{demo.upwork.secexp.training.Codingcamp.perc.text}{zero}
\definevar{demo.upwork.secexp.training.Codingcamp.perc.Text}{Zero}
\definevar{demo.upwork.secexp.training.None.perc.text}{zero}
\definevar{demo.upwork.secexp.training.None.perc.Text}{Zero}
\definevar{demo.upwork.secexp.training.Other.perc.text}{zero}
\definevar{demo.upwork.secexp.training.Other.perc.Text}{Zero}
\definevar{demo.github.secexp.training.Prefernottodisclose.text}{zero}
\definevar{demo.github.secexp.training.Prefernottodisclose.Text}{Zero}
\definevar{demo.github.secexp.training.Prefernottodisclose.perc.text}{zero}
\definevar{demo.github.secexp.training.Prefernottodisclose.perc.Text}{Zero}
\definevar{demo.github.secexp.training.Self-taught.text}{50}
\definevar{demo.github.secexp.training.Self-taught.Text}{50}
\definevar{demo.github.secexp.training.On-the-jobtraining.text}{17}
\definevar{demo.github.secexp.training.On-the-jobtraining.Text}{17}
\definevar{demo.github.secexp.training.None.text}{13}
\definevar{demo.github.secexp.training.None.Text}{13}
\definevar{demo.github.secexp.training.Onlineclass.text}{12}
\definevar{demo.github.secexp.training.Onlineclass.Text}{12}
\definevar{demo.github.secexp.training.College/University.text}{eight}
\definevar{demo.github.secexp.training.College/University.Text}{Eight}
\definevar{demo.github.secexp.training.Other.text}{three}
\definevar{demo.github.secexp.training.Other.Text}{Three}
\definevar{demo.github.secexp.training.Codingcamp.text}{one}
\definevar{demo.github.secexp.training.Codingcamp.Text}{One}
\definevar{demo.github.secexp.training.Self-taught.perc.text}{zero}
\definevar{demo.github.secexp.training.Self-taught.perc.Text}{Zero}
\definevar{demo.github.secexp.training.On-the-jobtraining.perc.text}{zero}
\definevar{demo.github.secexp.training.On-the-jobtraining.perc.Text}{Zero}
\definevar{demo.github.secexp.training.None.perc.text}{zero}
\definevar{demo.github.secexp.training.None.perc.Text}{Zero}
\definevar{demo.github.secexp.training.Onlineclass.perc.text}{zero}
\definevar{demo.github.secexp.training.Onlineclass.perc.Text}{Zero}
\definevar{demo.github.secexp.training.College/University.perc.text}{zero}
\definevar{demo.github.secexp.training.College/University.perc.Text}{Zero}
\definevar{demo.github.secexp.training.Other.perc.text}{zero}
\definevar{demo.github.secexp.training.Other.perc.Text}{Zero}
\definevar{demo.github.secexp.training.Codingcamp.perc.text}{zero}
\definevar{demo.github.secexp.training.Codingcamp.perc.Text}{Zero}

\definevar{repotype.Prefernottodisclose}{0}
\definevar{repotype.Prefernottodisclose.perc}{0.0\%}
\definevar{repotype.Publicrepository(thirdpartyprovider)}{95}
\definevar{repotype.Privaterepository(thirdpartyprovider)}{77}
\definevar{repotype.Privaterepository(onlocaldevice)}{55}
\definevar{repotype.Workonlocalproject(withoutarepository)}{45}
\definevar{repotype.Privaterepository(self-hostedownserver)}{39}
\definevar{repotype.Publicrepository(self-hostedownserver)}{26}
\definevar{repotype.Publicrepository(thirdpartyprovider).perc}{87.2\%}
\definevar{repotype.Privaterepository(thirdpartyprovider).perc}{70.6\%}
\definevar{repotype.Privaterepository(onlocaldevice).perc}{50.5\%}
\definevar{repotype.Workonlocalproject(withoutarepository).perc}{41.3\%}
\definevar{repotype.Privaterepository(self-hostedownserver).perc}{35.8\%}
\definevar{repotype.Publicrepository(self-hostedownserver).perc}{23.9\%}
\definevar{vcs.Prefernottodisclose}{0}
\definevar{vcs.Prefernottodisclose.perc}{0.0\%}
\definevar{vcs.Git}{109}
\definevar{vcs.Subversion}{12}
\definevar{vcs.CVS}{5}
\definevar{vcs.MSTeamFoundationVC}{4}
\definevar{vcs.Perforce}{2}
\definevar{vcs.Mercurial}{2}
\definevar{vcs.Git.perc}{100.0\%}
\definevar{vcs.Subversion.perc}{11.0\%}
\definevar{vcs.CVS.perc}{4.6\%}
\definevar{vcs.MSTeamFoundationVC.perc}{3.7\%}
\definevar{vcs.Perforce.perc}{1.8\%}
\definevar{vcs.Mercurial.perc}{1.8\%}
\definevar{platform.Prefernottodisclose}{5}
\definevar{platform.Prefernottodisclose.perc}{4.6\%}
\definevar{platform.GitHub}{99}
\definevar{platform.GitLab}{48}
\definevar{platform.Bitbucket}{36}
\definevar{platform.AzureDevOpsServer}{18}
\definevar{platform.AWSCodeCommit}{12}
\definevar{platform.Sourceforge}{3}
\definevar{platform.Gitea}{3}
\definevar{platform.GoogleCloudSourceRepositories}{2}
\definevar{platform.Gitee}{1}
\definevar{platform.MicrosoftTFS}{1}
\definevar{platform.PerforceClient/Server}{1}
\definevar{platform.GeneralFileSync}{1}
\definevar{platform.pagure.io}{1}
\definevar{platform.Sourcehut}{1}
\definevar{platform.GitHub.perc}{90.8\%}
\definevar{platform.GitLab.perc}{44.0\%}
\definevar{platform.Bitbucket.perc}{33.0\%}
\definevar{platform.AzureDevOpsServer.perc}{16.5\%}
\definevar{platform.AWSCodeCommit.perc}{11.0\%}
\definevar{platform.Sourceforge.perc}{2.8\%}
\definevar{platform.Gitea.perc}{2.8\%}
\definevar{platform.GoogleCloudSourceRepositories.perc}{1.8\%}
\definevar{platform.Gitee.perc}{0.9\%}
\definevar{platform.MicrosoftTFS.perc}{0.9\%}
\definevar{platform.PerforceClient/Server.perc}{0.9\%}
\definevar{platform.GeneralFileSync.perc}{0.9\%}
\definevar{platform.pagure.io.perc}{0.9\%}
\definevar{platform.Sourcehut.perc}{0.9\%}
\definevar{secrets.past.n/a}{56}
\definevar{secrets.past.n/a.perc}{51.4\%}
\definevar{secrets.past.APIKeys}{35}
\definevar{secrets.past.LoginCredentials}{35}
\definevar{secrets.past.Auth/AccessTokens}{33}
\definevar{secrets.past.PersonalCryptographicKeys}{22}
\definevar{secrets.past.ServicePasswords}{17}
\definevar{secrets.past.CI/CDKeys}{14}
\definevar{secrets.past.Other}{2}
\definevar{secrets.past.APIKeys.perc}{32.1\%}
\definevar{secrets.past.LoginCredentials.perc}{32.1\%}
\definevar{secrets.past.Auth/AccessTokens.perc}{30.3\%}
\definevar{secrets.past.PersonalCryptographicKeys.perc}{20.2\%}
\definevar{secrets.past.ServicePasswords.perc}{15.6\%}
\definevar{secrets.past.CI/CDKeys.perc}{12.8\%}
\definevar{secrets.past.Other.perc}{1.8\%}
\definevar{secrets.current.n/a}{0}
\definevar{secrets.current.n/a.perc}{0.0\%}
\definevar{secrets.current.Auth/AccessTokens}{52}
\definevar{secrets.current.LoginCredentials}{51}
\definevar{secrets.current.APIKeys}{45}
\definevar{secrets.current.PersonalCryptographicKeys}{30}
\definevar{secrets.current.ServerCertificates}{22}
\definevar{secrets.current.ServicePasswords}{22}
\definevar{secrets.current.CI/CDKeys}{20}
\definevar{secrets.current.None}{17}
\definevar{secrets.current.PersonalCertificates}{7}
\definevar{secrets.current.Other}{1}
\definevar{secrets.current.Auth/AccessTokens.perc}{47.7\%}
\definevar{secrets.current.LoginCredentials.perc}{46.8\%}
\definevar{secrets.current.APIKeys.perc}{41.3\%}
\definevar{secrets.current.PersonalCryptographicKeys.perc}{27.5\%}
\definevar{secrets.current.ServerCertificates.perc}{20.2\%}
\definevar{secrets.current.ServicePasswords.perc}{20.2\%}
\definevar{secrets.current.CI/CDKeys.perc}{18.3\%}
\definevar{secrets.current.None.perc}{15.6\%}
\definevar{secrets.current.PersonalCertificates.perc}{6.4\%}
\definevar{secrets.current.Other.perc}{0.9\%}
\definevar{threat-model.detailed.APIKeys.Teammembers.perc}{20.8\%}
\definevar{threat-model.detailed.APIKeys.Management.perc}{7.1\%}
\definevar{threat-model.detailed.APIKeys.Membersofotherteams.perc}{4.8\%}
\definevar{threat-model.detailed.APIKeys.Public.perc}{1.2\%}
\definevar{threat-model.detailed.Auth/AccessTokens.Teammembers.perc}{20.7\%}
\definevar{threat-model.detailed.Auth/AccessTokens.Management.perc}{11.9\%}
\definevar{threat-model.detailed.Auth/AccessTokens.Membersofotherteams.perc}{3.1\%}
\definevar{threat-model.detailed.Auth/AccessTokens.Public.perc}{1.6\%}
\definevar{threat-model.detailed.CI/CDKeys.Teammembers.perc}{19.7\%}
\definevar{threat-model.detailed.CI/CDKeys.Management.perc}{5.6\%}
\definevar{threat-model.detailed.CI/CDKeys.Membersofotherteams.perc}{1.4\%}
\definevar{threat-model.detailed.LoginCredentials.Teammembers.perc}{19.4\%}
\definevar{threat-model.detailed.LoginCredentials.Management.perc}{11.3\%}
\definevar{threat-model.detailed.LoginCredentials.Membersofotherteams.perc}{7.0\%}
\definevar{threat-model.detailed.LoginCredentials.Public.perc}{1.6\%}
\definevar{threat-model.detailed.Other.Management.perc}{25.0\%}
\definevar{threat-model.detailed.Other.Membersofotherteams.perc}{25.0\%}
\definevar{threat-model.detailed.PersonalCertificates.Management.perc}{11.1\%}
\definevar{threat-model.detailed.PersonalCertificates.Membersofotherteams.perc}{7.4\%}
\definevar{threat-model.detailed.PersonalCertificates.Teammembers.perc}{7.4\%}
\definevar{threat-model.detailed.PersonalCryptographicKeys.Teammembers.perc}{9.6\%}
\definevar{threat-model.detailed.PersonalCryptographicKeys.Management.perc}{7.0\%}
\definevar{threat-model.detailed.PersonalCryptographicKeys.Membersofotherteams.perc}{1.8\%}
\definevar{threat-model.detailed.ServerCertificates.Teammembers.perc}{16.0\%}
\definevar{threat-model.detailed.ServerCertificates.Management.perc}{9.9\%}
\definevar{threat-model.detailed.ServerCertificates.Membersofotherteams.perc}{6.2\%}
\definevar{threat-model.detailed.ServicePasswords.Teammembers.perc}{25.0\%}
\definevar{threat-model.detailed.ServicePasswords.Management.perc}{12.5\%}
\definevar{threat-model.detailed.ServicePasswords.Membersofotherteams.perc}{5.0\%}
\definevar{threat-model.overview.Teammembers.Hadaccess}{156}
\definevar{threat-model.overview.Teammembers.Hadaccess.perc}{70,5\%}
\definevar{threat-model.overview.Teammembers.Hadnoaccess}{49}
\definevar{threat-model.overview.Teammembers.Hadnoaccess.perc}{22.2\%}
\definevar{threat-model.overview.Teammembers.Idon'tknow}{2}
\definevar{threat-model.overview.Teammembers.Idon'tknow.perc}{0.9\%}
\definevar{threat-model.overview.Teammembers.Prefernottodisclose}{14}
\definevar{threat-model.overview.Teammembers.Prefernottodisclose.perc}{6.3\%}
\definevar{threat-model.overview.Management.Hadaccess}{79}
\definevar{threat-model.overview.Management.Hadaccess.perc}{35.7\%}
\definevar{threat-model.overview.Management.Hadnoaccess}{116}
\definevar{threat-model.overview.Management.Hadnoaccess.perc}{52.5\%}
\definevar{threat-model.overview.Management.Idon'tknow}{10}
\definevar{threat-model.overview.Management.Idon'tknow.perc}{45.2\%}
\definevar{threat-model.overview.Management.Prefernottodisclose}{16}
\definevar{threat-model.overview.Management.Prefernottodisclose.perc}{7.2}
\definevar{threat-model.overview.Membersofotherteams.Hadaccess}{34}
\definevar{threat-model.overview.Membersofotherteams.Hadaccess.perc}{15.8\%}
\definevar{threat-model.overview.Membersofotherteams.Hadnoaccess}{170}
\definevar{threat-model.overview.Membersofotherteams.Hadnoaccess.perc}{76.9\%}
\definevar{threat-model.overview.Membersofotherteams.Idon'tknow}{4}
\definevar{threat-model.overview.Membersofotherteams.Idon'tknow.perc}{1.8\%}
\definevar{threat-model.overview.Membersofotherteams.Prefernottodisclose}{12}
\definevar{threat-model.overview.Membersofotherteams.Prefernottodisclose.perc}{5.4\%}
\definevar{threat-model.overview.Public.Hadaccess}{8}
\definevar{threat-model.overview.Public.Hadaccess.perc}{3.6\%}
\definevar{threat-model.overview.Public.Hadnoaccess}{203}
\definevar{threat-model.overview.Public.Hadnoaccess.perc}{91.9\%}
\definevar{threat-model.overview.Public.Idon'tknow}{3}
\definevar{threat-model.overview.Public.Idon'tknow.perc}{1.4\%}
\definevar{threat-model.overview.Public.Prefernottodisclose}{7}
\definevar{threat-model.overview.Public.Prefernottodisclose.perc}{3.2\%}
\definevar{secretDecision.n/a}{17}
\definevar{secretDecision.n/a.perc}{15.6\%}
\definevar{secretDecision.themselves}{44}
\definevar{secretDecision.involved}{18}
\definevar{secretDecision.nan}{17}
\definevar{secretDecision.notInvolved}{15}
\definevar{secretDecision.joint}{15}
\definevar{secretDecision.themselves.perc}{40.4\%}
\definevar{secretDecision.involved.perc}{16.5\%}
\definevar{secretDecision.nan.perc}{15.6\%}
\definevar{secretDecision.notInvolved.perc}{13.8\%}
\definevar{secretDecision.joint.perc}{13.8\%}
\definevar{repotype.Prefernottodisclose.text}{zero}
\definevar{repotype.Prefernottodisclose.Text}{Zero}
\definevar{repotype.Prefernottodisclose.perc.text}{zero}
\definevar{repotype.Prefernottodisclose.perc.Text}{Zero}
\definevar{repotype.Publicrepository(thirdpartyprovider).text}{95}
\definevar{repotype.Publicrepository(thirdpartyprovider).Text}{95}
\definevar{repotype.Privaterepository(thirdpartyprovider).text}{77}
\definevar{repotype.Privaterepository(thirdpartyprovider).Text}{77}
\definevar{repotype.Privaterepository(onlocaldevice).text}{55}
\definevar{repotype.Privaterepository(onlocaldevice).Text}{55}
\definevar{repotype.Workonlocalproject(withoutarepository).text}{45}
\definevar{repotype.Workonlocalproject(withoutarepository).Text}{45}
\definevar{repotype.Privaterepository(self-hostedownserver).text}{39}
\definevar{repotype.Privaterepository(self-hostedownserver).Text}{39}
\definevar{repotype.Publicrepository(self-hostedownserver).text}{26}
\definevar{repotype.Publicrepository(self-hostedownserver).Text}{26}
\definevar{repotype.Publicrepository(thirdpartyprovider).perc.text}{zero}
\definevar{repotype.Publicrepository(thirdpartyprovider).perc.Text}{Zero}
\definevar{repotype.Privaterepository(thirdpartyprovider).perc.text}{zero}
\definevar{repotype.Privaterepository(thirdpartyprovider).perc.Text}{Zero}
\definevar{repotype.Privaterepository(onlocaldevice).perc.text}{zero}
\definevar{repotype.Privaterepository(onlocaldevice).perc.Text}{Zero}
\definevar{repotype.Workonlocalproject(withoutarepository).perc.text}{zero}
\definevar{repotype.Workonlocalproject(withoutarepository).perc.Text}{Zero}
\definevar{repotype.Privaterepository(self-hostedownserver).perc.text}{zero}
\definevar{repotype.Privaterepository(self-hostedownserver).perc.Text}{Zero}
\definevar{repotype.Publicrepository(self-hostedownserver).perc.text}{zero}
\definevar{repotype.Publicrepository(self-hostedownserver).perc.Text}{Zero}
\definevar{vcs.Prefernottodisclose.text}{zero}
\definevar{vcs.Prefernottodisclose.Text}{Zero}
\definevar{vcs.Prefernottodisclose.perc.text}{zero}
\definevar{vcs.Prefernottodisclose.perc.Text}{Zero}
\definevar{vcs.Git.text}{109}
\definevar{vcs.Git.Text}{109}
\definevar{vcs.Subversion.text}{12}
\definevar{vcs.Subversion.Text}{12}
\definevar{vcs.CVS.text}{five}
\definevar{vcs.CVS.Text}{Five}
\definevar{vcs.MSTeamFoundationVC.text}{four}
\definevar{vcs.MSTeamFoundationVC.Text}{Four}
\definevar{vcs.Perforce.text}{two}
\definevar{vcs.Perforce.Text}{Two}
\definevar{vcs.Mercurial.text}{two}
\definevar{vcs.Mercurial.Text}{Two}
\definevar{vcs.Git.perc.text}{one}
\definevar{vcs.Git.perc.Text}{One}
\definevar{vcs.Subversion.perc.text}{zero}
\definevar{vcs.Subversion.perc.Text}{Zero}
\definevar{vcs.CVS.perc.text}{zero}
\definevar{vcs.CVS.perc.Text}{Zero}
\definevar{vcs.MSTeamFoundationVC.perc.text}{zero}
\definevar{vcs.MSTeamFoundationVC.perc.Text}{Zero}
\definevar{vcs.Perforce.perc.text}{zero}
\definevar{vcs.Perforce.perc.Text}{Zero}
\definevar{vcs.Mercurial.perc.text}{zero}
\definevar{vcs.Mercurial.perc.Text}{Zero}
\definevar{platform.Prefernottodisclose.text}{five}
\definevar{platform.Prefernottodisclose.Text}{Five}
\definevar{platform.Prefernottodisclose.perc.text}{zero}
\definevar{platform.Prefernottodisclose.perc.Text}{Zero}
\definevar{platform.GitHub.text}{99}
\definevar{platform.GitHub.Text}{99}
\definevar{platform.GitLab.text}{48}
\definevar{platform.GitLab.Text}{48}
\definevar{platform.Bitbucket.text}{36}
\definevar{platform.Bitbucket.Text}{36}
\definevar{platform.AzureDevOpsServer.text}{18}
\definevar{platform.AzureDevOpsServer.Text}{18}
\definevar{platform.AWSCodeCommit.text}{12}
\definevar{platform.AWSCodeCommit.Text}{12}
\definevar{platform.Sourceforge.text}{three}
\definevar{platform.Sourceforge.Text}{Three}
\definevar{platform.Gitea.text}{three}
\definevar{platform.Gitea.Text}{Three}
\definevar{platform.GoogleCloudSourceRepositories.text}{two}
\definevar{platform.GoogleCloudSourceRepositories.Text}{Two}
\definevar{platform.Gitee.text}{one}
\definevar{platform.Gitee.Text}{One}
\definevar{platform.MicrosoftTFS.text}{one}
\definevar{platform.MicrosoftTFS.Text}{One}
\definevar{platform.PerforceClient/Server.text}{one}
\definevar{platform.PerforceClient/Server.Text}{One}
\definevar{platform.GeneralFileSync.text}{one}
\definevar{platform.GeneralFileSync.Text}{One}
\definevar{platform.pagure.io.text}{one}
\definevar{platform.pagure.io.Text}{One}
\definevar{platform.Sourcehut.text}{one}
\definevar{platform.Sourcehut.Text}{One}
\definevar{platform.GitHub.perc.text}{zero}
\definevar{platform.GitHub.perc.Text}{Zero}
\definevar{platform.GitLab.perc.text}{zero}
\definevar{platform.GitLab.perc.Text}{Zero}
\definevar{platform.Bitbucket.perc.text}{zero}
\definevar{platform.Bitbucket.perc.Text}{Zero}
\definevar{platform.AzureDevOpsServer.perc.text}{zero}
\definevar{platform.AzureDevOpsServer.perc.Text}{Zero}
\definevar{platform.AWSCodeCommit.perc.text}{zero}
\definevar{platform.AWSCodeCommit.perc.Text}{Zero}
\definevar{platform.Sourceforge.perc.text}{zero}
\definevar{platform.Sourceforge.perc.Text}{Zero}
\definevar{platform.Gitea.perc.text}{zero}
\definevar{platform.Gitea.perc.Text}{Zero}
\definevar{platform.GoogleCloudSourceRepositories.perc.text}{zero}
\definevar{platform.GoogleCloudSourceRepositories.perc.Text}{Zero}
\definevar{platform.Gitee.perc.text}{zero}
\definevar{platform.Gitee.perc.Text}{Zero}
\definevar{platform.MicrosoftTFS.perc.text}{zero}
\definevar{platform.MicrosoftTFS.perc.Text}{Zero}
\definevar{platform.PerforceClient/Server.perc.text}{zero}
\definevar{platform.PerforceClient/Server.perc.Text}{Zero}
\definevar{platform.GeneralFileSync.perc.text}{zero}
\definevar{platform.GeneralFileSync.perc.Text}{Zero}
\definevar{platform.pagure.io.perc.text}{zero}
\definevar{platform.pagure.io.perc.Text}{Zero}
\definevar{platform.Sourcehut.perc.text}{zero}
\definevar{platform.Sourcehut.perc.Text}{Zero}
\definevar{secrets.past.n/a.text}{56}
\definevar{secrets.past.n/a.Text}{56}
\definevar{secrets.past.n/a.perc.text}{zero}
\definevar{secrets.past.n/a.perc.Text}{Zero}
\definevar{secrets.past.APIKeys.text}{35}
\definevar{secrets.past.APIKeys.Text}{35}
\definevar{secrets.past.LoginCredentials.text}{35}
\definevar{secrets.past.LoginCredentials.Text}{35}
\definevar{secrets.past.Auth/AccessTokens.text}{33}
\definevar{secrets.past.Auth/AccessTokens.Text}{33}
\definevar{secrets.past.PersonalCryptographicKeys.text}{22}
\definevar{secrets.past.PersonalCryptographicKeys.Text}{22}
\definevar{secrets.past.ServicePasswords.text}{17}
\definevar{secrets.past.ServicePasswords.Text}{17}
\definevar{secrets.past.CI/CDKeys.text}{14}
\definevar{secrets.past.CI/CDKeys.Text}{14}
\definevar{secrets.past.Other.text}{two}
\definevar{secrets.past.Other.Text}{Two}
\definevar{secrets.past.APIKeys.perc.text}{zero}
\definevar{secrets.past.APIKeys.perc.Text}{Zero}
\definevar{secrets.past.LoginCredentials.perc.text}{zero}
\definevar{secrets.past.LoginCredentials.perc.Text}{Zero}
\definevar{secrets.past.Auth/AccessTokens.perc.text}{zero}
\definevar{secrets.past.Auth/AccessTokens.perc.Text}{Zero}
\definevar{secrets.past.PersonalCryptographicKeys.perc.text}{zero}
\definevar{secrets.past.PersonalCryptographicKeys.perc.Text}{Zero}
\definevar{secrets.past.ServicePasswords.perc.text}{zero}
\definevar{secrets.past.ServicePasswords.perc.Text}{Zero}
\definevar{secrets.past.CI/CDKeys.perc.text}{zero}
\definevar{secrets.past.CI/CDKeys.perc.Text}{Zero}
\definevar{secrets.past.Other.perc.text}{zero}
\definevar{secrets.past.Other.perc.Text}{Zero}
\definevar{secrets.current.n/a.text}{zero}
\definevar{secrets.current.n/a.Text}{Zero}
\definevar{secrets.current.n/a.perc.text}{zero}
\definevar{secrets.current.n/a.perc.Text}{Zero}
\definevar{secrets.current.Auth/AccessTokens.text}{52}
\definevar{secrets.current.Auth/AccessTokens.Text}{52}
\definevar{secrets.current.LoginCredentials.text}{51}
\definevar{secrets.current.LoginCredentials.Text}{51}
\definevar{secrets.current.APIKeys.text}{45}
\definevar{secrets.current.APIKeys.Text}{45}
\definevar{secrets.current.PersonalCryptographicKeys.text}{30}
\definevar{secrets.current.PersonalCryptographicKeys.Text}{30}
\definevar{secrets.current.ServerCertificates.text}{22}
\definevar{secrets.current.ServerCertificates.Text}{22}
\definevar{secrets.current.ServicePasswords.text}{22}
\definevar{secrets.current.ServicePasswords.Text}{22}
\definevar{secrets.current.CI/CDKeys.text}{20}
\definevar{secrets.current.CI/CDKeys.Text}{20}
\definevar{secrets.current.None.text}{17}
\definevar{secrets.current.None.Text}{17}
\definevar{secrets.current.PersonalCertificates.text}{seven}
\definevar{secrets.current.PersonalCertificates.Text}{Seven}
\definevar{secrets.current.Other.text}{one}
\definevar{secrets.current.Other.Text}{One}
\definevar{secrets.current.Auth/AccessTokens.perc.text}{zero}
\definevar{secrets.current.Auth/AccessTokens.perc.Text}{Zero}
\definevar{secrets.current.LoginCredentials.perc.text}{zero}
\definevar{secrets.current.LoginCredentials.perc.Text}{Zero}
\definevar{secrets.current.APIKeys.perc.text}{zero}
\definevar{secrets.current.APIKeys.perc.Text}{Zero}
\definevar{secrets.current.PersonalCryptographicKeys.perc.text}{zero}
\definevar{secrets.current.PersonalCryptographicKeys.perc.Text}{Zero}
\definevar{secrets.current.ServerCertificates.perc.text}{zero}
\definevar{secrets.current.ServerCertificates.perc.Text}{Zero}
\definevar{secrets.current.ServicePasswords.perc.text}{zero}
\definevar{secrets.current.ServicePasswords.perc.Text}{Zero}
\definevar{secrets.current.CI/CDKeys.perc.text}{zero}
\definevar{secrets.current.CI/CDKeys.perc.Text}{Zero}
\definevar{secrets.current.None.perc.text}{zero}
\definevar{secrets.current.None.perc.Text}{Zero}
\definevar{secrets.current.PersonalCertificates.perc.text}{zero}
\definevar{secrets.current.PersonalCertificates.perc.Text}{Zero}
\definevar{secrets.current.Other.perc.text}{zero}
\definevar{secrets.current.Other.perc.Text}{Zero}
\definevar{threat-model.detailed.APIKeys.Teammembers.perc.text}{zero}
\definevar{threat-model.detailed.APIKeys.Teammembers.perc.Text}{Zero}
\definevar{threat-model.detailed.APIKeys.Management.perc.text}{zero}
\definevar{threat-model.detailed.APIKeys.Management.perc.Text}{Zero}
\definevar{threat-model.detailed.APIKeys.Membersofotherteams.perc.text}{zero}
\definevar{threat-model.detailed.APIKeys.Membersofotherteams.perc.Text}{Zero}
\definevar{threat-model.detailed.APIKeys.Public.perc.text}{zero}
\definevar{threat-model.detailed.APIKeys.Public.perc.Text}{Zero}
\definevar{threat-model.detailed.Auth/AccessTokens.Teammembers.perc.text}{zero}
\definevar{threat-model.detailed.Auth/AccessTokens.Teammembers.perc.Text}{Zero}
\definevar{threat-model.detailed.Auth/AccessTokens.Management.perc.text}{zero}
\definevar{threat-model.detailed.Auth/AccessTokens.Management.perc.Text}{Zero}
\definevar{threat-model.detailed.Auth/AccessTokens.Membersofotherteams.perc.text}{zero}
\definevar{threat-model.detailed.Auth/AccessTokens.Membersofotherteams.perc.Text}{Zero}
\definevar{threat-model.detailed.Auth/AccessTokens.Public.perc.text}{zero}
\definevar{threat-model.detailed.Auth/AccessTokens.Public.perc.Text}{Zero}
\definevar{threat-model.detailed.CI/CDKeys.Teammembers.perc.text}{zero}
\definevar{threat-model.detailed.CI/CDKeys.Teammembers.perc.Text}{Zero}
\definevar{threat-model.detailed.CI/CDKeys.Management.perc.text}{zero}
\definevar{threat-model.detailed.CI/CDKeys.Management.perc.Text}{Zero}
\definevar{threat-model.detailed.CI/CDKeys.Membersofotherteams.perc.text}{zero}
\definevar{threat-model.detailed.CI/CDKeys.Membersofotherteams.perc.Text}{Zero}
\definevar{threat-model.detailed.LoginCredentials.Teammembers.perc.text}{zero}
\definevar{threat-model.detailed.LoginCredentials.Teammembers.perc.Text}{Zero}
\definevar{threat-model.detailed.LoginCredentials.Management.perc.text}{zero}
\definevar{threat-model.detailed.LoginCredentials.Management.perc.Text}{Zero}
\definevar{threat-model.detailed.LoginCredentials.Membersofotherteams.perc.text}{zero}
\definevar{threat-model.detailed.LoginCredentials.Membersofotherteams.perc.Text}{Zero}
\definevar{threat-model.detailed.LoginCredentials.Public.perc.text}{zero}
\definevar{threat-model.detailed.LoginCredentials.Public.perc.Text}{Zero}
\definevar{threat-model.detailed.Other.Management.perc.text}{zero}
\definevar{threat-model.detailed.Other.Management.perc.Text}{Zero}
\definevar{threat-model.detailed.Other.Membersofotherteams.perc.text}{zero}
\definevar{threat-model.detailed.Other.Membersofotherteams.perc.Text}{Zero}
\definevar{threat-model.detailed.PersonalCertificates.Management.perc.text}{zero}
\definevar{threat-model.detailed.PersonalCertificates.Management.perc.Text}{Zero}
\definevar{threat-model.detailed.PersonalCertificates.Membersofotherteams.perc.text}{zero}
\definevar{threat-model.detailed.PersonalCertificates.Membersofotherteams.perc.Text}{Zero}
\definevar{threat-model.detailed.PersonalCertificates.Teammembers.perc.text}{zero}
\definevar{threat-model.detailed.PersonalCertificates.Teammembers.perc.Text}{Zero}
\definevar{threat-model.detailed.PersonalCryptographicKeys.Teammembers.perc.text}{zero}
\definevar{threat-model.detailed.PersonalCryptographicKeys.Teammembers.perc.Text}{Zero}
\definevar{threat-model.detailed.PersonalCryptographicKeys.Management.perc.text}{zero}
\definevar{threat-model.detailed.PersonalCryptographicKeys.Management.perc.Text}{Zero}
\definevar{threat-model.detailed.PersonalCryptographicKeys.Membersofotherteams.perc.text}{zero}
\definevar{threat-model.detailed.PersonalCryptographicKeys.Membersofotherteams.perc.Text}{Zero}
\definevar{threat-model.detailed.ServerCertificates.Teammembers.perc.text}{zero}
\definevar{threat-model.detailed.ServerCertificates.Teammembers.perc.Text}{Zero}
\definevar{threat-model.detailed.ServerCertificates.Management.perc.text}{zero}
\definevar{threat-model.detailed.ServerCertificates.Management.perc.Text}{Zero}
\definevar{threat-model.detailed.ServerCertificates.Membersofotherteams.perc.text}{zero}
\definevar{threat-model.detailed.ServerCertificates.Membersofotherteams.perc.Text}{Zero}
\definevar{threat-model.detailed.ServicePasswords.Teammembers.perc.text}{zero}
\definevar{threat-model.detailed.ServicePasswords.Teammembers.perc.Text}{Zero}
\definevar{threat-model.detailed.ServicePasswords.Management.perc.text}{zero}
\definevar{threat-model.detailed.ServicePasswords.Management.perc.Text}{Zero}
\definevar{threat-model.detailed.ServicePasswords.Membersofotherteams.perc.text}{zero}
\definevar{threat-model.detailed.ServicePasswords.Membersofotherteams.perc.Text}{Zero}
\definevar{threat-model.overview.Teammembers.Hadaccess.text}{171}
\definevar{threat-model.overview.Teammembers.Hadaccess.Text}{171}
\definevar{threat-model.overview.Teammembers.Hadnoaccess.text}{62}
\definevar{threat-model.overview.Teammembers.Hadnoaccess.Text}{62}
\definevar{threat-model.overview.Teammembers.Idon'tknow.text}{two}
\definevar{threat-model.overview.Teammembers.Idon'tknow.Text}{Two}
\definevar{threat-model.overview.Teammembers.Prefernottodisclose.text}{15}
\definevar{threat-model.overview.Teammembers.Prefernottodisclose.Text}{15}
\definevar{threat-model.overview.Management.Hadaccess.text}{90}
\definevar{threat-model.overview.Management.Hadaccess.Text}{90}
\definevar{threat-model.overview.Management.Hadnoaccess.text}{129}
\definevar{threat-model.overview.Management.Hadnoaccess.Text}{129}
\definevar{threat-model.overview.Management.Idon'tknow.text}{13}
\definevar{threat-model.overview.Management.Idon'tknow.Text}{13}
\definevar{threat-model.overview.Management.Prefernottodisclose.text}{18}
\definevar{threat-model.overview.Management.Prefernottodisclose.Text}{18}
\definevar{threat-model.overview.Membersofotherteams.Hadaccess.text}{42}
\definevar{threat-model.overview.Membersofotherteams.Hadaccess.Text}{42}
\definevar{threat-model.overview.Membersofotherteams.Hadnoaccess.text}{191}
\definevar{threat-model.overview.Membersofotherteams.Hadnoaccess.Text}{191}
\definevar{threat-model.overview.Membersofotherteams.Idon'tknow.text}{four}
\definevar{threat-model.overview.Membersofotherteams.Idon'tknow.Text}{Four}
\definevar{threat-model.overview.Membersofotherteams.Prefernottodisclose.text}{13}
\definevar{threat-model.overview.Membersofotherteams.Prefernottodisclose.Text}{13}
\definevar{threat-model.overview.Public.Hadaccess.text}{eight}
\definevar{threat-model.overview.Public.Hadaccess.Text}{Eight}
\definevar{threat-model.overview.Public.Hadnoaccess.text}{231}
\definevar{threat-model.overview.Public.Hadnoaccess.Text}{231}
\definevar{threat-model.overview.Public.Idon'tknow.text}{three}
\definevar{threat-model.overview.Public.Idon'tknow.Text}{Three}
\definevar{threat-model.overview.Public.Prefernottodisclose.text}{eight}
\definevar{threat-model.overview.Public.Prefernottodisclose.Text}{Eight}
\definevar{secretDecision.n/a.text}{17}
\definevar{secretDecision.n/a.Text}{17}
\definevar{secretDecision.n/a.perc.text}{zero}
\definevar{secretDecision.n/a.perc.Text}{Zero}
\definevar{secretDecision.themselves.text}{44}
\definevar{secretDecision.themselves.Text}{44}
\definevar{secretDecision.involved.text}{18}
\definevar{secretDecision.involved.Text}{18}
\definevar{secretDecision.nan.text}{17}
\definevar{secretDecision.nan.Text}{17}
\definevar{secretDecision.notInvolved.text}{15}
\definevar{secretDecision.notInvolved.Text}{15}
\definevar{secretDecision.joint.text}{15}
\definevar{secretDecision.joint.Text}{15}
\definevar{secretDecision.themselves.perc.text}{zero}
\definevar{secretDecision.themselves.perc.Text}{Zero}
\definevar{secretDecision.involved.perc.text}{zero}
\definevar{secretDecision.involved.perc.Text}{Zero}
\definevar{secretDecision.nan.perc.text}{zero}
\definevar{secretDecision.nan.perc.Text}{Zero}
\definevar{secretDecision.notInvolved.perc.text}{zero}
\definevar{secretDecision.notInvolved.perc.Text}{Zero}
\definevar{secretDecision.joint.perc.text}{zero}
\definevar{secretDecision.joint.perc.Text}{Zero}

\definevar{leak.prevalence.Prefernottodisclose}{0}
\definevar{leak.prevalence.Prefernottodisclose.perc}{0.0\%}
\definevar{leak.prevalence.Agree}{37}
\definevar{leak.prevalence.Neutral}{22}
\definevar{leak.prevalence.Disagree}{21}
\definevar{leak.prevalence.Stronglyagree}{14}
\definevar{leak.prevalence.Stronglydisagree}{7}
\definevar{leak.prevalence.Agree.perc}{33.9\%}
\definevar{leak.prevalence.Neutral.perc}{20.2\%}
\definevar{leak.prevalence.Disagree.perc}{19.3\%}
\definevar{leak.prevalence.Stronglyagree.perc}{12.8\%}
\definevar{leak.prevalence.Stronglydisagree.perc}{6.4\%}
\definevar{leak.prevalence.AgreeOrStronglyAgree}{51}
\definevar{leak.prevalence.Notexperienced}{57}
\definevar{leak.prevalence.Knowpeople}{42}
\definevar{leak.prevalence.Experienced}{33}
\definevar{leak.prevalence.Notexperienced.perc}{52.3\%}
\definevar{leak.prevalence.Knowpeople.perc}{38.5\%}
\definevar{leak.prevalence.Experienced.perc}{30.3\%}
\definevar{leak.prevalence.Prefernottodisclose.text}{zero}
\definevar{leak.prevalence.Prefernottodisclose.Text}{Zero}
\definevar{leak.prevalence.Prefernottodisclose.perc.text}{zero}
\definevar{leak.prevalence.Prefernottodisclose.perc.Text}{Zero}
\definevar{leak.prevalence.Agree.text}{37}
\definevar{leak.prevalence.Agree.Text}{37}
\definevar{leak.prevalence.Neutral.text}{22}
\definevar{leak.prevalence.Neutral.Text}{22}
\definevar{leak.prevalence.Disagree.text}{21}
\definevar{leak.prevalence.Disagree.Text}{21}
\definevar{leak.prevalence.Stronglyagree.text}{14}
\definevar{leak.prevalence.Stronglyagree.Text}{14}
\definevar{leak.prevalence.Stronglydisagree.text}{seven}
\definevar{leak.prevalence.Stronglydisagree.Text}{Seven}
\definevar{leak.prevalence.Agree.perc.text}{zero}
\definevar{leak.prevalence.Agree.perc.Text}{Zero}
\definevar{leak.prevalence.Neutral.perc.text}{zero}
\definevar{leak.prevalence.Neutral.perc.Text}{Zero}
\definevar{leak.prevalence.Disagree.perc.text}{zero}
\definevar{leak.prevalence.Disagree.perc.Text}{Zero}
\definevar{leak.prevalence.Stronglyagree.perc.text}{zero}
\definevar{leak.prevalence.Stronglyagree.perc.Text}{Zero}
\definevar{leak.prevalence.Stronglydisagree.perc.text}{zero}
\definevar{leak.prevalence.Stronglydisagree.perc.Text}{Zero}
\definevar{leak.prevalence.AgreeOrStronglyAgree.text}{51}
\definevar{leak.prevalence.AgreeOrStronglyAgree.Text}{51}
\definevar{leak.prevalence.Notexperienced.text}{57}
\definevar{leak.prevalence.Notexperienced.Text}{57}
\definevar{leak.prevalence.Knowpeople.text}{42}
\definevar{leak.prevalence.Knowpeople.Text}{42}
\definevar{leak.prevalence.Experienced.text}{33}
\definevar{leak.prevalence.Experienced.Text}{33}
\definevar{leak.prevalence.Notexperienced.perc.text}{zero}
\definevar{leak.prevalence.Notexperienced.perc.Text}{Zero}
\definevar{leak.prevalence.Knowpeople.perc.text}{zero}
\definevar{leak.prevalence.Knowpeople.perc.Text}{Zero}
\definevar{leak.prevalence.Experienced.perc.text}{zero}
\definevar{leak.prevalence.Experienced.perc.Text}{Zero}

\definevar{pas.failed}{8}
\definevar{pas.failed.perc}{7.3\%}
\definevar{pas.failed.text}{eight}
\definevar{pas.failed.Text}{Eight}
\definevar{pas.failed.perc.text}{zero}
\definevar{pas.failed.perc.Text}{Zero}
\definevar{coding.prevention.other}{8}
\definevar{coding.prevention.other.perc}{7.3\%}
\definevar{coding.remediation.accessmanagement}{6}
\definevar{coding.remediation.accessmanagement.perc}{5.5\%}
\definevar{coding.remediation.removalfromsourcecode}{12}
\definevar{coding.remediation.removalfromsourcecode.perc}{11.0\%}
\definevar{coding.remediation.serveroperations}{2}
\definevar{coding.remediation.serveroperations.perc}{1.8\%}
\definevar{coding.prevention.encryptedsecrets}{30}
\definevar{coding.prevention.encryptedsecrets.perc}{27.5\%}
\definevar{coding.prevention.monitoring}{16}
\definevar{coding.prevention.monitoring.perc}{14.7\%}
\definevar{coding.prevention.restrictaccess}{19}
\definevar{coding.prevention.restrictaccess.perc}{17.4\%}
\definevar{coding.prevention.externalizesecrets}{60}
\definevar{coding.prevention.externalizesecrets.perc}{55.0\%}
\definevar{coding.remediation.reneworrevokesecret}{59}
\definevar{coding.remediation.reneworrevokesecret.perc}{54.1\%}
\definevar{coding.prevention.rotation}{6}
\definevar{coding.prevention.rotation.perc}{5.5\%}
\definevar{coding.prevention.education&awareness}{9}
\definevar{coding.prevention.education&awareness.perc}{8.3\%}
\definevar{coding.prevention.codeandsecretreviews}{4}
\definevar{coding.prevention.codeandsecretreviews.perc}{3.7\%}
\definevar{coding.remediation.cleanupvcshistory}{19}
\definevar{coding.remediation.cleanupvcshistory.perc}{17.4\%}
\definevar{coding.prevention.blocksecrets}{32}
\definevar{coding.prevention.blocksecrets.perc}{29.4\%}
\definevar{coding.remediation.analyzeleak}{17}
\definevar{coding.remediation.analyzeleak.perc}{15.6\%}
\definevar{coding.remediation.systemicconsequences}{3}
\definevar{coding.remediation.systemicconsequences.perc}{2.8\%}
\definevar{coding.remediation.retractrepository}{5}
\definevar{coding.remediation.retractrepository.perc}{4.6\%}
\definevar{coding.remediation.notifyconcernedroles}{8}
\definevar{coding.remediation.notifyconcernedroles.perc}{7.3\%}
\definevar{coding.prevention.other.text}{eight}
\definevar{coding.prevention.other.Text}{Eight}
\definevar{coding.prevention.other.perc.text}{zero}
\definevar{coding.prevention.other.perc.Text}{Zero}
\definevar{coding.remediation.accessmanagement.text}{six}
\definevar{coding.remediation.accessmanagement.Text}{Six}
\definevar{coding.remediation.accessmanagement.perc.text}{zero}
\definevar{coding.remediation.accessmanagement.perc.Text}{Zero}
\definevar{coding.remediation.removalfromsourcecode.text}{12}
\definevar{coding.remediation.removalfromsourcecode.Text}{12}
\definevar{coding.remediation.removalfromsourcecode.perc.text}{zero}
\definevar{coding.remediation.removalfromsourcecode.perc.Text}{Zero}
\definevar{coding.remediation.serveroperations.text}{two}
\definevar{coding.remediation.serveroperations.Text}{Two}
\definevar{coding.remediation.serveroperations.perc.text}{zero}
\definevar{coding.remediation.serveroperations.perc.Text}{Zero}
\definevar{coding.prevention.encryptedsecrets.text}{30}
\definevar{coding.prevention.encryptedsecrets.Text}{30}
\definevar{coding.prevention.encryptedsecrets.perc.text}{zero}
\definevar{coding.prevention.encryptedsecrets.perc.Text}{Zero}
\definevar{coding.prevention.monitoring.text}{16}
\definevar{coding.prevention.monitoring.Text}{16}
\definevar{coding.prevention.monitoring.perc.text}{zero}
\definevar{coding.prevention.monitoring.perc.Text}{Zero}
\definevar{coding.prevention.restrictaccess.text}{19}
\definevar{coding.prevention.restrictaccess.Text}{19}
\definevar{coding.prevention.restrictaccess.perc.text}{zero}
\definevar{coding.prevention.restrictaccess.perc.Text}{Zero}
\definevar{coding.prevention.externalizesecrets.text}{60}
\definevar{coding.prevention.externalizesecrets.Text}{60}
\definevar{coding.prevention.externalizesecrets.perc.text}{zero}
\definevar{coding.prevention.externalizesecrets.perc.Text}{Zero}
\definevar{coding.remediation.reneworrevokesecret.text}{59}
\definevar{coding.remediation.reneworrevokesecret.Text}{59}
\definevar{coding.remediation.reneworrevokesecret.perc.text}{zero}
\definevar{coding.remediation.reneworrevokesecret.perc.Text}{Zero}
\definevar{coding.prevention.rotation.text}{six}
\definevar{coding.prevention.rotation.Text}{Six}
\definevar{coding.prevention.rotation.perc.text}{zero}
\definevar{coding.prevention.rotation.perc.Text}{Zero}
\definevar{coding.prevention.education&awareness.text}{nine}
\definevar{coding.prevention.education&awareness.Text}{Nine}
\definevar{coding.prevention.education&awareness.perc.text}{zero}
\definevar{coding.prevention.education&awareness.perc.Text}{Zero}
\definevar{coding.prevention.codeandsecretreviews.text}{four}
\definevar{coding.prevention.codeandsecretreviews.Text}{Four}
\definevar{coding.prevention.codeandsecretreviews.perc.text}{zero}
\definevar{coding.prevention.codeandsecretreviews.perc.Text}{Zero}
\definevar{coding.remediation.cleanupvcshistory.text}{19}
\definevar{coding.remediation.cleanupvcshistory.Text}{19}
\definevar{coding.remediation.cleanupvcshistory.perc.text}{zero}
\definevar{coding.remediation.cleanupvcshistory.perc.Text}{Zero}
\definevar{coding.prevention.blocksecrets.text}{32}
\definevar{coding.prevention.blocksecrets.Text}{32}
\definevar{coding.prevention.blocksecrets.perc.text}{zero}
\definevar{coding.prevention.blocksecrets.perc.Text}{Zero}
\definevar{coding.remediation.analyzeleak.text}{17}
\definevar{coding.remediation.analyzeleak.Text}{17}
\definevar{coding.remediation.analyzeleak.perc.text}{zero}
\definevar{coding.remediation.analyzeleak.perc.Text}{Zero}
\definevar{coding.remediation.systemicconsequences.text}{three}
\definevar{coding.remediation.systemicconsequences.Text}{Three}
\definevar{coding.remediation.systemicconsequences.perc.text}{zero}
\definevar{coding.remediation.systemicconsequences.perc.Text}{Zero}
\definevar{coding.remediation.retractrepository.text}{five}
\definevar{coding.remediation.retractrepository.Text}{Five}
\definevar{coding.remediation.retractrepository.perc.text}{zero}
\definevar{coding.remediation.retractrepository.perc.Text}{Zero}
\definevar{coding.remediation.notifyconcernedroles.text}{eight}
\definevar{coding.remediation.notifyconcernedroles.Text}{Eight}
\definevar{coding.remediation.notifyconcernedroles.perc.text}{zero}
\definevar{coding.remediation.notifyconcernedroles.perc.Text}{Zero}
\definevar{analysis.prevention.notapplicable}{28}
\definevar{analysis.prevention.notapplicable.perc}{56.0\%}
\definevar{analysis.prevention.restrictaccess}{4}
\definevar{analysis.prevention.restrictaccess.perc}{8.0\%}
\definevar{analysis.prevention.monitoring}{11}
\definevar{analysis.prevention.monitoring.perc}{22.0\%}
\definevar{analysis.prevention.externalizesecrets}{9}
\definevar{analysis.prevention.externalizesecrets.perc}{18.0\%}
\definevar{analysis.prevention.rotation}{1}
\definevar{analysis.prevention.rotation.perc}{2.0\%}
\definevar{analysis.prevention.blocksecrets}{12}
\definevar{analysis.prevention.blocksecrets.perc}{24.0\%}
\definevar{analysis.prevention.encryptedsecrets}{9}
\definevar{analysis.prevention.encryptedsecrets.perc}{18.0\%}
\definevar{analysis.prevention.codeandsecretreviews}{0.0}
\definevar{analysis.prevention.codeandsecretreviews.perc}{0.0\%}
\definevar{analysis.prevention.education&awareness}{1}
\definevar{analysis.prevention.education&awareness.perc}{2.0\%}
\definevar{analysis.prevention.other}{1}
\definevar{analysis.prevention.other.perc}{2.0\%}
\definevar{analysis.prevention.misconception}{nan}
\definevar{analysis.prevention.misconception.perc}{nan\%}
\definevar{analysis.remediation.notapplicable}{43}
\definevar{analysis.remediation.notapplicable.perc}{86.0\%}
\definevar{analysis.remediation.cleanupvcshistory}{2}
\definevar{analysis.remediation.cleanupvcshistory.perc}{4.0\%}
\definevar{analysis.remediation.reneworrevokesecret}{7}
\definevar{analysis.remediation.reneworrevokesecret.perc}{14.0\%}
\definevar{analysis.remediation.removalfromsourcecode}{0.0}
\definevar{analysis.remediation.removalfromsourcecode.perc}{0.0\%}
\definevar{analysis.remediation.serveroperations}{0.0}
\definevar{analysis.remediation.serveroperations.perc}{0.0\%}
\definevar{analysis.remediation.retractrepository}{0.0}
\definevar{analysis.remediation.retractrepository.perc}{0.0\%}
\definevar{analysis.remediation.analyzeleak}{1}
\definevar{analysis.remediation.analyzeleak.perc}{2.0\%}
\definevar{analysis.remediation.systemicconsequences}{0.0}
\definevar{analysis.remediation.systemicconsequences.perc}{0.0\%}
\definevar{analysis.remediation.accessmanagement}{1}
\definevar{analysis.remediation.accessmanagement.perc}{2.0\%}
\definevar{analysis.remediation.notifyconcernedroles}{0.0}
\definevar{analysis.remediation.notifyconcernedroles.perc}{0.0\%}
\definevar{analysis.prevention.notapplicable.text}{28}
\definevar{analysis.prevention.notapplicable.Text}{28}
\definevar{analysis.prevention.notapplicable.perc.text}{zero}
\definevar{analysis.prevention.notapplicable.perc.Text}{Zero}
\definevar{analysis.prevention.restrictaccess.text}{four}
\definevar{analysis.prevention.restrictaccess.Text}{Four}
\definevar{analysis.prevention.restrictaccess.perc.text}{zero}
\definevar{analysis.prevention.restrictaccess.perc.Text}{Zero}
\definevar{analysis.prevention.monitoring.text}{eleven}
\definevar{analysis.prevention.monitoring.Text}{Eleven}
\definevar{analysis.prevention.monitoring.perc.text}{zero}
\definevar{analysis.prevention.monitoring.perc.Text}{Zero}
\definevar{analysis.prevention.externalizesecrets.text}{nine}
\definevar{analysis.prevention.externalizesecrets.Text}{Nine}
\definevar{analysis.prevention.externalizesecrets.perc.text}{zero}
\definevar{analysis.prevention.externalizesecrets.perc.Text}{Zero}
\definevar{analysis.prevention.rotation.text}{one}
\definevar{analysis.prevention.rotation.Text}{One}
\definevar{analysis.prevention.rotation.perc.text}{zero}
\definevar{analysis.prevention.rotation.perc.Text}{Zero}
\definevar{analysis.prevention.blocksecrets.text}{12}
\definevar{analysis.prevention.blocksecrets.Text}{12}
\definevar{analysis.prevention.blocksecrets.perc.text}{zero}
\definevar{analysis.prevention.blocksecrets.perc.Text}{Zero}
\definevar{analysis.prevention.encryptedsecrets.text}{nine}
\definevar{analysis.prevention.encryptedsecrets.Text}{Nine}
\definevar{analysis.prevention.encryptedsecrets.perc.text}{zero}
\definevar{analysis.prevention.encryptedsecrets.perc.Text}{Zero}
\definevar{analysis.prevention.codeandsecretreviews.text}{zero}
\definevar{analysis.prevention.codeandsecretreviews.Text}{Zero}
\definevar{analysis.prevention.codeandsecretreviews.perc.text}{zero}
\definevar{analysis.prevention.codeandsecretreviews.perc.Text}{Zero}
\definevar{analysis.prevention.education&awareness.text}{one}
\definevar{analysis.prevention.education&awareness.Text}{One}
\definevar{analysis.prevention.education&awareness.perc.text}{zero}
\definevar{analysis.prevention.education&awareness.perc.Text}{Zero}
\definevar{analysis.prevention.other.text}{one}
\definevar{analysis.prevention.other.Text}{One}
\definevar{analysis.prevention.other.perc.text}{zero}
\definevar{analysis.prevention.other.perc.Text}{Zero}
\definevar{analysis.remediation.notapplicable.text}{43}
\definevar{analysis.remediation.notapplicable.Text}{43}
\definevar{analysis.remediation.notapplicable.perc.text}{zero}
\definevar{analysis.remediation.notapplicable.perc.Text}{Zero}
\definevar{analysis.remediation.cleanupvcshistory.text}{two}
\definevar{analysis.remediation.cleanupvcshistory.Text}{Two}
\definevar{analysis.remediation.cleanupvcshistory.perc.text}{zero}
\definevar{analysis.remediation.cleanupvcshistory.perc.Text}{Zero}
\definevar{analysis.remediation.reneworrevokesecret.text}{seven}
\definevar{analysis.remediation.reneworrevokesecret.Text}{Seven}
\definevar{analysis.remediation.reneworrevokesecret.perc.text}{zero}
\definevar{analysis.remediation.reneworrevokesecret.perc.Text}{Zero}
\definevar{analysis.remediation.removalfromsourcecode.text}{zero}
\definevar{analysis.remediation.removalfromsourcecode.Text}{Zero}
\definevar{analysis.remediation.removalfromsourcecode.perc.text}{zero}
\definevar{analysis.remediation.removalfromsourcecode.perc.Text}{Zero}
\definevar{analysis.remediation.serveroperations.text}{zero}
\definevar{analysis.remediation.serveroperations.Text}{Zero}
\definevar{analysis.remediation.serveroperations.perc.text}{zero}
\definevar{analysis.remediation.serveroperations.perc.Text}{Zero}
\definevar{analysis.remediation.retractrepository.text}{zero}
\definevar{analysis.remediation.retractrepository.Text}{Zero}
\definevar{analysis.remediation.retractrepository.perc.text}{zero}
\definevar{analysis.remediation.retractrepository.perc.Text}{Zero}
\definevar{analysis.remediation.analyzeleak.text}{one}
\definevar{analysis.remediation.analyzeleak.Text}{One}
\definevar{analysis.remediation.analyzeleak.perc.text}{zero}
\definevar{analysis.remediation.analyzeleak.perc.Text}{Zero}
\definevar{analysis.remediation.systemicconsequences.text}{zero}
\definevar{analysis.remediation.systemicconsequences.Text}{Zero}
\definevar{analysis.remediation.systemicconsequences.perc.text}{zero}
\definevar{analysis.remediation.systemicconsequences.perc.Text}{Zero}
\definevar{analysis.remediation.accessmanagement.text}{one}
\definevar{analysis.remediation.accessmanagement.Text}{One}
\definevar{analysis.remediation.accessmanagement.perc.text}{zero}
\definevar{analysis.remediation.accessmanagement.perc.Text}{Zero}
\definevar{analysis.remediation.notifyconcernedroles.text}{zero}
\definevar{analysis.remediation.notifyconcernedroles.Text}{Zero}
\definevar{analysis.remediation.notifyconcernedroles.perc.text}{zero}
\definevar{analysis.remediation.notifyconcernedroles.perc.Text}{Zero}

\definevar{guides}{100}
\definevar{guides.text}{100}
\definevar{guides.Text}{100}
\definevar{guides.each}{50}
\definevar{guides.each.text}{50}
\definevar{guides.each.Text}{50}

\definevar{guides.queries.distinct}{17}
\definevar{guides.queries.distinct.text}{17}
\definevar{guides.queries.distinct.Text}{17}

\definevar{survey.participants.compensation.upwork}{\$25}
\definevar{survey.time}{20} %
\definevar{survey.time.screening}{5} %
\definevar{survey.time.screening.text}{five}
\definevar{survey.time.screening.Text}{Five}

\definevar{survey.participants.piloting.total.text}{11}
\definevar{survey.participants.piloting.cognitive.text}{four}
\definevar{survey.participants.piloting.cognitive.Text}{Four}
\definevar{survey.participants.piloting.unsupervised.text}{seven}
\definevar{survey.participants.piloting.unsupervised.Text}{Seven}

\definevar{survey.failed}{7}
\definevar{survey.failed.text}{seven}
\definevar{survey.failed.Text}{Seven}
\definevar{survey.failed.perc}{6.4\%}

\definevar{github.mails.total}{103,646}
\definevar{github.mails.written}{5,310}

\definevar{demo.upwork.gender.Non-binary.perc}{0.0\%}
\definevar{demo.upwork.countries.Germany.perc}{0.0\%}
\definevar{demo.github.countries.Kenya.perc}{0.0\%}
\definevar{demo.upwork.edu.Highschoolorequivalent.perc}{0.0\%}
\definevar{demo.upwork.employment.Retired.perc}{0.0\%}
\definevar{demo.interviews.gender.Woman.perc}{0.0\%}
\definevar{demo.interviews.progexp.learning.Codingcamp.perc}{0.0\%}
\definevar{demo.interviews.countries.Germany.perc}{0.0\%}
\definevar{demo.interviews.countries.other.perc}{42.9\%}

\begin{document}

\date{}

\title{Pushed by Accident: A Mixed-Methods Study on Strategies\\ of Handling Secrets in Source Code Repositories}

\author{
{\rm Alexander Krause}\,\orcidlink{0000-0003-2993-2568}\textsuperscript{$\mathcal{C}$}
\and
{\rm Jan H.\ Klemmer}\,\orcidlink{0000-0002-6994-7206}\textsuperscript{$\ast$}
\and
{\rm Nicolas Huaman}\,\orcidlink{0000-0003-2733-5073}\textsuperscript{$\ast$}
\and
{\rm Dominik Wermke}\textsuperscript{$\mathcal{C}$}
\and
{\rm Yasemin Acar}\,\orcidlink{0000-0001-7167-7383}\textsuperscript{$\dagger$, $\ddagger$}
\and
{\rm Sascha Fahl}\,\orcidlink{0000-0002-5644-3316}\textsuperscript{$\mathcal{C}$}\\[-12pt]
\and
\textsuperscript{$\mathcal{C}$}CISPA Helmholtz Center for Information Security, Germany, \\
\hypersetup{hidelinks}\texttt{\{\href{mailto:alexander.krause@cispa.de}{alexander.krause},\href{mailto:dominik.wermke@cispa.de}{dominik.wermke},\href{mailto:sascha.fahl@cispa.de}{sascha.fahl}\}@cispa.de}\\
\textsuperscript{$\ast$}Leibniz University Hannover, Germany, \hypersetup{hidelinks}\texttt{\{\href{mailto:klemmer@sec.uni-hannover.de}{klemmer},\href{mailto:stransky@sec.uni-hannover.de}{huaman}\}@sec.uni-hannover.de}\\
\textsuperscript{$\dagger$}Paderborn University, Germany, \hypersetup{hidelinks}\href{mailto:yasemin.acar@uni-paderborn.de}{\texttt{yasemin.acar@uni-paderborn.de}}\\
\textsuperscript{$\ddagger$}The George Washington University,  USA \\
}

\def\plainauthor{Alexander Krause, Jan H. Klemmer, Nicolas Huaman, Dominik Wermke, Yasemin Acar, Sascha Fahl}

\maketitle

\thispagestyle{plain}
\pagestyle{plain}

\begin{abstract}
\Glsentrylongpl{vcs} for source code, such as Git, are key tools in modern software development. Many developers use services like GitHub or GitLab for collaborative software development. Many software projects include code secrets such as API keys or passwords that need to be managed securely.
Previous research and blog posts found that developers struggle with secure code secret management and accidentally leaked code secrets to public Git repositories. Leaking code secrets to the public can have disastrous consequences, such as abusing services and systems or making sensitive user data available to attackers.
In a mixed-methods study, we surveyed \var{survey.participants.total} developers with version control system experience. Additionally, %
we conducted 14 in-depth semi-structured interviews with developers who experienced secret leakage in the past.
\var{leak.prevalence.Experienced.perc} of our participants encountered code secret leaks in the past. Most of them face several challenges with secret leakage prevention and remediation. 
Based on our findings, we discuss challenges, such as estimating the risks of leaked secrets, and the needs of developers in remediating and preventing code secret leaks, such as low adoption requirements. We conclude with recommendations for developers and source code platform providers to reduce the risk of secret leakage.

\end{abstract}

\section{Introduction}\label{sec:intro}
\Glspl{vcs} are an essential technology for collaborative software development.
Git~\cite{git}, a fundamental tool to orchestrate collaborative development, has been voted as the most common tool in the recent Stack Overflow Developer Survey~\cite{2022StackOverflowSurvey} with 93.4\% of participants specifying to use this tool in their development workflow.
Git-based code repository platforms (\eg{} GitHub~\cite{github-2} and GitLab~\cite{gitlab}) aim to ease sharing, reviewing, and contributing to software projects.
In modern development pipelines, software is commonly directly built, tested, and deployed within and from these code repositories.
To deploy software on server infrastructure, automate interactions with third-party services, or handle authentication, developers need to provide secrets, \eg{} credentials, authentication tokens, or secret encryption keys.
However, these secrets must be protected from being leaked accidentally into the public codebase.
Unfortunately, this is no straightforward task. Recent work by Meli et al.~\cite{meli2019bad} found that on GitHub, the most popular code sharing platform\footnote{According to the Tranco list~\cite{LePochat2019} generated on May~25, 2023, available at \url{https://tranco-list.eu/list/JX43Y}.}, thousands of automatically detectable secrets are leaked daily. 

A leaked secret can have a significant impact depending on the type of secret and how long it takes for the secret owner to revoke it after noticing its leak. In some cases, a leak can be highly critical, as in the case of Toyota: a hard-coded credential for accessing a data server was publicly pushed to GitHub in 2017. It allowed attackers to control the Toyota T-Connect accounts for 296,019~customers~\cite{toyotabreach}. T-Connect provides features like remote starting, in-car Wi-Fi, digital key access, full control over dashboard-provided metrics, and a direct line to the My Toyota service app. After more than five years, Toyota invalidated the key. Such a long time could mean multiple malicious actors have already gained access.
\revision{GitHub recently leaked a private SSH host key of their production Git servers in a public repository and replaced them to prevent abuse. This incident illustrates the complexity of secret management, even for large companies experienced with code secrets and public source code repositories~\cite{githubhostkeys}.}

\revision{There is anecdotal knowledge~\cite{anecdotalknowledge-spectralops,endpointprotector,owasp, gitguardian} on secret leaks through source code repositories. However, there has been little prior research trying to understand better the reasons for and experiences with code secret leakage in source code repositories. To address this gap, to the best of our knowledge we are the first to conduct a mixed-methods study, including an online survey and semi-structured in-depth interviews with experienced developers. We investigate the following research questions:}

\revision{
\boldparagraph{RQ1}\textit{How widespread is code secret leakage among developers?}
Leaking secrets and access tokens in source code poses a potentially serious security threat. We asked 109~developers how often they encountered secret leaks in the past.}
\boldparagraph{RQ2}\textit{What are secret leakage prevention approaches, and what are developers experiences?}
Depending on the scenario and context, prevention approaches can differ widely.
We surveyed and interviewed developers to reveal prevention approaches and the experiences, challenges, and needs that developers have when using them.
\boldparagraph{RQ3}\textit{What are developers' experiences with code secret leakage incidents?}
Little is known about developers' experiences when remediating code secret leaks.
We interviewed  developers on their latest and most impactful secret leaks to learn from their experiences, how they recognized a leak, and their consequences.
\boldparagraph{RQ4}\textit{What are developers' experiences with code secret remediation techniques and tools?}
Remediating code secret leakage can be challenging.
We examine deployed remediation approaches, developers' experiences with these approaches, and their requirements for approaches.

\vspace{2pt plus 1pt minus 1pt} %

Overall, we conducted a survey with \var{survey.participants.upwork}~freelancers from Upwork and \var{survey.participants.github}~developers from GitHub. \revision{For in-depth insights in developers' experiences with code secret leakage incidents and the approaches they use to prevent and remediate leaks, we interviewed another 14~developers who experienced code secret leakage.}
We make the following key contributions:

\boldparagraph{Identifying 18 Secret Leakage Prevention and Remediation Approaches}
We present survey results with freelancers and GitHub developers, investigating their approaches and experiences with code secret prevention and remediation (Section~\ref{sec:study-survey}). We discovered 18 approaches to prevent and remediate code secret leakage (Table~\ref{tab:findings}). \var{leak.prevalence.Experienced.perc} of our participants reported first-hand experience with secret leakage in their projects. 

\boldparagraph{Identifying Challenges Developers Face with Secret Leakage Prevention and Remediation Approaches}
\revision{In addition to the survey, we interviewed GitHub developers who experienced code secret leakage to gather more qualitative insights (Section~\ref{sec:study-interviews}).} 
We report on how they detected the leaks, their experiences with code secret leak incidents, their approaches to preventing leaks, and their techniques and tools for secret leakage remediation.
We identified several challenges with common remediation and prevention approaches and motivations to use them.

\boldparagraph{Providing Recommendations to Reduce the Risk of Secret Leakage}
Based on our findings, we provide recommendations for future research, software developers, and collaborative source code platforms to prevent and remediate code secret leakage in Section~\ref{sec:discussion}.

\boldparagraph{Providing a Full Replication Package}
To support future research, a full replication package is available in line with the effort to support replication of our work, containing all study materials in the \availability{} (after Section~\ref{sec:conclusion}). %

\section{Related Work}\label{sec:related-work}
We present and discuss related work in two key areas: previous research on secret leakage in source code repositories and \revisionII{developers' secure development approaches and practices}.

\subsection{Secret Leakage in Code Repositories} 
In recent years, researchers made efforts to measure secret leakage in source code repositories. In addition, we discuss secret detection and methods to improve its accuracy.

\boldparagraph{Measurement Studies}
Sinha et al.\ discussed different approaches on how to detect, prevent, and fix code secret leakage in source code repositories~\cite{sinha2015detecting}.
In 2019, Meli et al.\ presented a large-scale measurement study on secret leakage in public GitHub repositories, finding more than 100,000~repositories with leaked secrets. They demonstrated and evaluated approaches to detect secrets on GitHub. The authors examined potential root causes, including the developer experience and practice of storing secret information in repositories~\cite{meli2019bad}.

\boldparagraph{Secret Leakage Detection}
Secret scanners produce a high rate of false-positive results, which is a major problem because developers have to review them manually.
Meli et al.\ used past secret detection strategies; while they avoided high false-positive results. They combined multiple methods to detect potential secrets and secret evaluation~\cite{meli2019bad}.
However, automatic detection of secret leaks can be challenging. 
In 2020, Saha et al.\ applied machine learning to reduce the false-positive rate of secret scanners~\cite{saha2020reducing}.
Similarly, Lounici et al.\ developed and evaluated machine learning classifiers to reduce false-positives~\cite{lounici2021detection}.
Recently, Kall and Trabelsi proposed and evaluated an approach to improve the detection of leaked credentials in source code repositories~\cite{kall2021federated}.
In 2022, Feng et al.\ developed an automated approach to effectively detect password leakage from public repositories~\cite{10.1145/3510003.3510150}.
In 2022, Rahman et al.\ investigated human factors on code secret leakage detection tool warnings~\cite{rahman2022secret}.
In 2022, Basak et al.\ conducted a gray literature review to identify developer and organizational practices~\cite{9973029}.

\revision{The previous research on code secret leakage focused on assessing its prevalence and identifying code secret leaks after an incident occurred. We investigate developers' experiences with code secret leakage incidents and explore their strategies for preventing and remediating them by conducting two developer studies.}

\revision{\subsection{Exploring Secure Development Approaches and Practices}
Researchers have extensively studied secure software development practices. This section describes these research efforts dedicated to improving secure development methodologies, offering insights and guidance to empower developers in achieving secure software.

Assal et al.\ interviewed 13~developers to investigate their real-life software security practices during each stage of the development lifecycle. Real-life security practices differ markedly from best practices identified in the literature~\cite{10.5555/3291228.3291251}.
Assal et al.\ continued by conducting an online survey with 123~software developers to explore the interplay between developers and software security processes. They found that security vulnerabilities often result from a lack of organizational or process support~\cite{10.1145/3290605.3300519}.
In 2017, Haney et al.\ surveyed 121~representatives from organizations that work in cryptographic development. They characterized the cryptographic practices, types of resources, and standards used by cryptographic developers. They found that participants used cryptography for a variety of purposes, with the majority relying on generally accepted, standards-based implementations as guides~\cite{8228643}.
In 2018, the researchers conducted 21~in-depth interviews with highly experienced individuals from organizations that employ cryptographic implementations to gain a more profound understanding of their cryptographic development practices. They demonstrate a strong organizational security culture that guides the careful selection of resources and informs formal, rigorous development and testing practices~\cite{219400}.
In 2020, Votipka et al.\ qualitatively analyzed BIBIFI~\cite{ruef2016build} submissions on how and why programmers make security errors. They found that most vulnerabilities result from misconceptions. Furthermore, they suggest APIs should be simpler and more precisely documented, including multiple use cases and edge cases~\cite{247694}.
\revisionII{In 2020, Palombo et al.\ presented an ethnographic study of secure software development processes in a software company finding that sometimes vulnerabilities were ignored or even consciously introduced to fix other issues~\cite{10.5555/3488905.3488917}.}
In 2022, Wermke et al.\ interviewed 27~open source developers to investigate their security and trust practices. \revisionII{They found that open source projects are highly diverse in deployed security measures, trust processes, and their underlying motivations~\cite{9833686}.}
Naiakshina et al.\ qualitatively analyzed security problems when implementing secure password storage. The authors found conflicting advice to be an obstacle that developers struggle with~\cite{Naiakshina:2017}.
\revisionII{Lopez et al.\ analyzed security-related conversations on Stack Overflow and found that developers use online environments to actively connect, exchange information, and provide assistance, despite concerns about their reliability as security information sources.~\cite{lopez2019Anatomy}.}
Recently, Fischer et al.\ conducted a study analyzing the effect of Google Search on security in software development. Besides insecure resources among search results, they demonstrated that re-ranking search results significantly improved security. %

While previous research explored secure development practices and approaches in general, in companies, and open source communities, focusing on security vulnerabilities and implementations, we extend this by focusing on the practices developers use to prevent and remediate code secret leakage at code repository platforms.
}

\section{Methodology}\label{sec:studies}
\begin{figure}[t]
    \centering
    \includegraphics[width=0.8\linewidth]{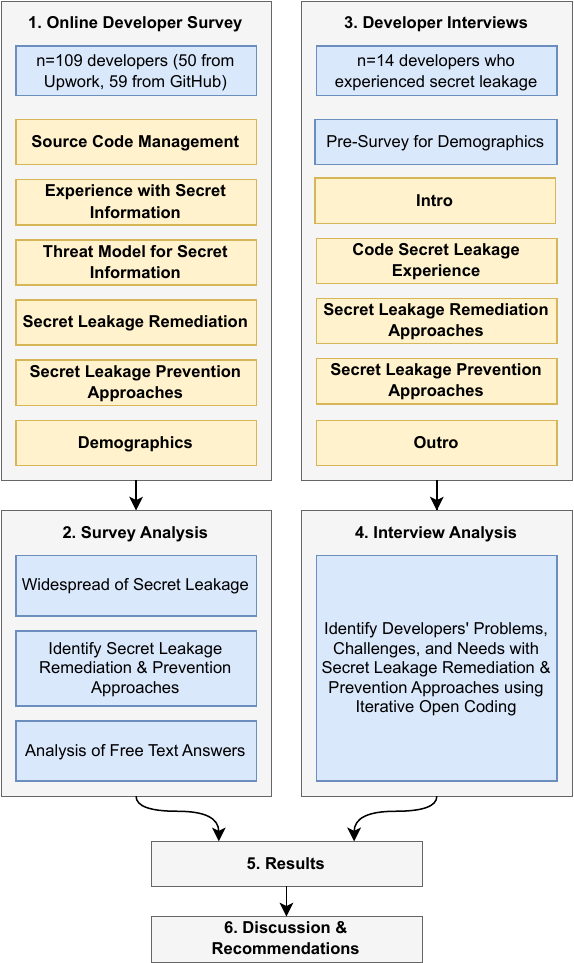}
    \caption{\revision{Study overview, showing methodology in blue and the content of the survey/interview sections in yellow.}}
    \label{fig:studyoverview}
\end{figure}

In this section, we explain the methodology of our studies. An overview is depicted in Figure~\ref{fig:studyoverview}. After a detailed description of the online survey with Upwork and GitHub developers, we describe our interview study with developers from GitHub, in which we gained further qualitative insights into code secret incidents and their remediation and prevention approaches.

\subsection{Online Survey with Developers}
\label{sec:study-survey}
\revision{Below, we provide details on the approach and structure of the survey which we conducted with \var{survey.participants.total.text}~developers. We detail the analysis of both qualitative and quantitative data points.}

\subsubsection{Survey Procedure}
Between September 6 and October 26, 2021, we conducted an online survey with \var{survey.participants.upwork}~freelancers from Upwork and \var{survey.participants.github}~developers from GitHub. We used Qualtrics~\cite{qualtrics} to provide the survey and collect the respondent's data.

\boldparagraph{Questionnaire Development}
First, we conducted an exploratory analysis of online guides, reviewing 100 web pages for an impression of what kind of information for code secret leakage prevention and remediation is provided online (\cf{}~Appendix~\ref{sec:guide-analysis}). We developed our questionnaire based on this exploratory analysis, previous research (\cf{}~Section~\ref{sec:related-work}), and the research questions. In addition, we established additional areas of interest for our survey based on participants' input during pre-testing.
\revision{In both our survey and interview studies, we provided explanations for the terms \emph{code secret}, \emph{code secret leakage}, and \emph{code secret handling approaches} for a shared understanding of the terms with participants. These explanations emerged from the exploratory analysis and were validated for understanding in survey pilots.}

\boldparagraph{Piloting}
Before conducting the survey, we piloted our questionnaire with \var{survey.participants.piloting.total.text}~participants.
First, we conducted cognitive walkthroughs~\cite{presser2004methods} with \var{survey.participants.piloting.cognitive.text} usable security researchers. Subsequently, \var{survey.participants.piloting.unsupervised.text} participants completed the survey unsupervised, with an additional feedback text box on each page of the questionnaire. The results were used to verify and improve survey clarity and flow.

\subsubsection{Recruitment and Inclusion Criteria}
\label{sec:survey-recruitment}
\revision{We used two different recruitment strategies to obtain a diverse sample of developers: We recruited Upwork freelancers and GitHub developers.}
We include the recruitment material in our replication package (\cf{}~\availability). %

\boldparagraph{Upwork}
\revision{
On the freelancer platform Upwork~\cite{upwork}, we published several identical postings for developers from September~8 through October~23, 2021. Freelancers applied by writing a short application. This included answering screening questions, which we used to accept only participants that met our inclusion criteria. We accepted freelancers who had worked with \glspl{vcs} and related platforms to manage their source code. They were also required to have collaborated with other people in the past. These criteria ensured they potentially had to handle secrets on source code platforms.}
We compensated the freelancers with \var{survey.participants.compensation.upwork} to offer a competitive reward (\$1/minute), so that professionals would also have an incentive to participate~\cite{upwork-costs}.

\boldparagraph{GitHub}
\revision{
Following security research from 2021 and 2022~\cite{conf/oakland/wermke22, 10.1145/3491102.3501822}, we implemented the following procedure to recruit developers from GitHub.
From October~18 to October~20, 2021, we invited \var{github.mails.written} GitHub users, which had an activity on September~1, 2021. We also verified that all recruited developers were also active contributors in general by manually analyzing their GitHub commits. We only invited developers who had published their contact email addresses on their profiles, and stopped inviting developers once we reached saturation. We refrained from contacting developers who did not want to be contacted, contacted developers only once, and offered to add them to an ignore list\footnote{1.4\% of the users we invited made use of this option.}. We received no complaints from any of the invited developers (\cf{}~Section~\ref{sec:ethics}).}

\subsubsection{Survey Structure}
Below, we outline topics and questions covered in the questionnaire (\cf{}~Figure~\ref{fig:studyoverview}).
\revision{The full survey including the consent form can be found in our replication package (\cf{}~\availability).}

\boldparagraph{Source Code Management}
We asked questions regarding what types of source code repositories participants used (\eg{} local only, remote platform, self- or third-party-hosted), and about usage of different \glspl{vcs} and platforms to gather a general understanding of respondents source code management.

\boldparagraph{Experience with Secret Information}
This block of questions aims to understand what type of different secrets participants encountered. We asked these questions for all past projects and the most recent project.

\boldparagraph{Threat Model for Secret Information}
To understand participants' threat model, we asked participants which other stakeholders had access to which secret information and who made this decision in their most recent project. 
We followed up with brief definitions of \emph{code secret} and \emph{code secret leakage} to ensure a shared understanding for the following questions.
This question block concludes with two Likert scale questions regarding the perceived prevalence of code secret leakage and potential consequences.%

\boldparagraph{Secret Leakage Remediation}
Next, we asked questions regarding whether the participants experienced code secret leakage themselves or knew someone who did. To investigate current secret leakage remediation practices, participants were prompted to describe (based on the previous question's answer) how they remediated code secret leakage themselves, how others did, or how they hypothetically would.

\boldparagraph{Code Secret Handling Approaches}
Another important aspect, covered in this question block, are approaches to prevent code secret leakage in general. Again, we defined \emph{code secret handling approach} (all approaches to avoid code secret leakage).
To gather prevalent practices, we asked participants in free text questions which approaches they have heard of, which ones they used, and why.
Additionally, we asked participants for any issues they had, with which approaches, why they failed with an approach, and what they wanted to use an approach for.
Finally, we asked whether and how the participants helped co-workers handle secrets.

\boldparagraph{Demographics}
The concluding block was about demographics. This included standard questions (\eg{} age, gender, country, education, employment status), but also specialized questions to investigate the development (\eg{} years in development, number of projects) and security experience.

\subsubsection{Analysis}
\label{sec:study-survey-analysis}
Our analysis is a mix of quantitative and qualitative evaluation.
We report various counts and percentages of single- and multiple-choice questions in text and figures in the survey results (\cf{}~Section~\ref{sec:study-results-survey}).
For the free-text questions, two researchers used an iterative open-coding approach to extract the used approaches on code secret prevention and remediation (Q11--15)~\cite{charmaz2014constructing, strauss1997grounded, corbin1990grounded}. 
To prevent mislabeling, the researchers first coded ten answers together and discussed problematic codes for each question. Subsequently, two researchers coded the answers independently, followed by a third researcher independently reviewing the coding. \revision{Finally, they resolved any coding conflicts in a consensus discussion or introduced new codes if necessary~\cite{IRR}. All previous answers were re-coded if new codes were introduced.}

\subsection{Interviews with Developers}
\label{sec:study-interviews}
To enrich and deepen the survey insights, we decided to complement those with qualitative insights from semi-structured interviews~($n=\var{interviews.participants.total}$). The interview results cover the reasons, experiences, and processes for the prevention and remediation approaches we collected in our survey (\cf{} Section~\ref{sec:study-results-interviews}). We reached saturation within the high-level codes of our codebook. The average interview duration was 32 minutes (median: 32.5~minutes).

\subsubsection{Interview Procedure}
All \var{interviews.participants.total}~interviews were conducted in June and July 2022. We utilized a setup with two interviewers. A main interviewer held the conversation with the interviewee and asked the questions according to the interview guide. A so-called shadow interviewer was present to listen and note what questions were asked, and to make sure none were forgotten. At the end of the interview, the shadow interviewer also had the chance to ask questions to follow up on interesting aspects that emerged. %
All interviews were conducted in English and remotely via a GDPR-compliant conference tool. We recorded each interview to create a transcript later on, after which we destroyed the recordings. All transcripts have been manually checked and corrected by us for possible errors.

\boldparagraph{Pre-Questionnaire}
Before the actual interview, each participant had to fill a short pre-questionnaire. This had multiple purposes. (1)~We screened participants by only accepting those who experienced secret leakage. (2)~We explained the purpose of the study and obtained consent for participation, our data handling, and recording. (3)~Finally, we asked several demographics and background questions, which also helped the interviewer to prepare for the interview.

\boldparagraph{Piloting}
We iteratively tested and improved the initial version of the interview guide. This mainly included three cognitive walkthroughs with usable security researchers. We used this interview simulations to obtain feedback on question clarity, completeness, and to generally improve the interview guide with the interviewed researchers' experience. After each interview, we tweaked question clarity to ensure a good interview flow. This was followed by two pilot interviews with developers from Upwork.

\subsubsection{Recruitment and Inclusion Criteria}

As the goal of this study is to investigate secret leakage prevention as well as remediation, we decided to only interview developers who experienced secret leakage and therefore can report on remediation and past incidents. This was the only eligibility criteria to participate in an interview.
Two developers from Upwork who also participated in our survey and stated that they experienced code secret leakage were invited and compensated with \$60.
\revision{Apart from the two piloting interviews with Upwork users, we recruited the remaining twelve participants from GitHub -- following the same approach as for the online survey (\cf{} Section~\ref{sec:survey-recruitment}). Due to institutional restrictions, we could not compensate GitHub interview participants directly. Instead, we could offer these participants to sponsor a GitHub project of their choice with \$60.%
}

\subsubsection{Interview Structure and Interview Guide}
In the following, we describe the interview structure and questions. We outline all sections, each containing top-level questions and corresponding follow-up questions. The full interview guide can be found in Appendix~\ref{sec:interview-appendix}.

\boldparagraph{Introduction}
Each interview started with greeting the participant, explaining the interview's purpose and procedure, and obtaining consent from the participants. We underlined that we are only interested in personal opinions and experiences and not judging their case of secret leakage. The participants could skip questions anytime.

\boldparagraph{Code Secrets}
In the first section, we talked about code secrets and established a shared understanding of what a code secret is. Moreover, we asked participants about their broader experiences with code secrets. This included where they came into touch with code secrets and their experiences regarding sensitivity, and code secret access.

\boldparagraph{Secret Leakage and Remediation Approaches}
We continued with the code secret leakage incident that the participant had experienced. First, we established a uniform understanding of what a secret leak is. Then, we queried participants about their most impactful or latest (depending on what they remember best) secret leak. To get detailed insights, we asked for reasons why the leak occurred, consequences, and changes to secret handling due to the incident.
Related to the leak, we asked for its remediation, including experiences, and challenges, involved individuals/teams, and consulted resources. 

\boldparagraph{Secret Leakage Prevention Approaches}
The third section was about preventive measures against secret leakage. We asked questions on approaches that participants have used, including their experiences, understanding of the approach, and any challenges.
We also asked for approaches they tried to use but failed with and the reason for this, as well as approaches they know of but do not use.
If a participant had not taken any prevention approach, we instead asked for potential reasons. The section concludes with an open question on wishes and improvements for future prevention approaches.

\boldparagraph{Outro \& Debriefing}
After we asked all the questions, we held a debriefing with the participants to clarify any remaining questions, to give participants an opportunity to add something we might not have specifically asked for, and to gather feedback for the interview.

\subsubsection{Analysis and Coding}
\label{sec:study-analysis}
We used an iterative open-coding approach to analyze all interview transcripts~\cite{charmaz2014constructing, strauss1997grounded, corbin1990grounded}. First, two researchers developed an initial codebook based on their interview impressions and the interview guide.
Afterward, the same two researchers coded the interviews in multiple rounds. After each iteration, they resolved conflicts by consensus discussion or by introducing new sub-codes. We continued iterative coding until no new codes and themes emerged~\cite{birks15, cathy13}.
\revision{We do not report inter-coder agreement. We resolved each conflict immediately when it emerged (resulting in a hypothetical final agreement of 100\%)~\cite{IRR, 9833686, conf/oakland/wermke23, conf/soups/huaman22, conf/oakland/gutfleisch22}. The final codebook is part of our replication package (\cf{}~\availability).}

\subsection{Participant Demographics}
\label{sec:demographics}
We hired \var{survey.participants.total.text}~respondents for the survey, and recruited \var{interviews.participants.total.text}~participants for the interviews. 
Overall, our survey respondents and interview participants were predominantly male, with roughly 10\% female, and were about 30~years old on average.
Country-wise, we have a highly diverse sample of survey respondents from more than \var{demo.all.countries.distinct.text}~distinct  countries, including the U.S., India, and Germany as the top three, as well as Canada, the UK, Russia, Pakistan, Portugal, the Netherlands, Mexico, Australia, Egypt, Brazil, and Indonesia.
Interview participants were from nine distinct countries, including the U.S., India, and Pakistan as the top three, as well as Canada, Belarus, Italy, Kenya, and Brazil.
While most survey respondents were full-time (\var{demo.all.employment.Employedfull-time}, \var{demo.all.employment.Employedfull-time.perc}) or part-time employees (\var{demo.all.employment.Employedpart-time}, \var{demo.all.employment.Employedpart-time.perc}), some participants reported to be self-employed/freelancers (\var{demo.all.employment.Self-employed/Freelancer}, \var{demo.all.employment.Self-employed/Freelancer.perc}), or students (\var{demo.all.employment.Student}, \var{demo.all.employment.Student.perc}). 
Overall, the respondents and participants were highly experienced, with the majority of survey respondents (\var{demo.all.progexp.years.>5Years}, \var{demo.all.progexp.years.>5Years.perc}) having developed software for more than five years. 
About 85\% of the survey respondents and 95\% of the interview participants said they taught themselves how to program, often in addition to other ways of learning, \eg{} at college or university, on the job, or in online classes.
In total, the demographics in terms of gender, age, top three countries, and education are comparable to those of the 2022 Stack Overflow developer survey~\cite{2022StackOverflowSurvey}.
Table~\ref{tab:demographics} provides the detailed overview of survey respondents' and interview participants' demographics.
\begin{table}[t]
    \centering
    \scriptsize
    \renewcommand{\arraystretch}{1.05}
    \setlength{\tabcolsep}{1\tabcolsep}
    \setlength{\defaultaddspace}{0.3\defaultaddspace} %
    \caption{Selected participant demographics from both the survey and interviews. We omit ``Other'' and ``Prefer not to disclose'' answers for space reasons.}
    \label{tab:demographics}
    \begin{threeparttable}
        \begin{tabular}{lrrrr}
            \toprule
             & \multicolumn{3}{c}{\textbf{Survey}} & \multirow{2}{*}{\textbf{Interviews}} \\
            \cmidrule(lr){2-4} 
             & \textbf{Upwork} & \textbf{GitHub} & \textbf{Combined} &  \\
            \midrule
            
            \textbf{Participants:} & \\
            \hspace{2mm}Started & \var{survey.participants.upwork.started} & \var{survey.participants.github.started} & \var{survey.participants.total.started} & n/a \\
            \hspace{2mm}Finished & \var{survey.participants.upwork.finished} & \var{survey.participants.github.finished} & \var{survey.participants.total.finished} & n/a \\
            \hspace{2mm}Valid/Total ($n=$) & \var{survey.participants.upwork} & \var{survey.participants.github} & \var{survey.participants.total} & \var{interviews.participants.total} \\
            
            \textbf{Gender:} & \\
            \hspace{2mm}Male & \var{demo.upwork.gender.Man.perc} & \var{demo.github.gender.Man.perc} & \var{demo.all.gender.Man.perc} & \var{demo.interviews.gender.Man.perc} \\
            \hspace{2mm}Female & \var{demo.upwork.gender.Woman.perc} & \var{demo.github.gender.Woman.perc} & \var{demo.all.gender.Woman.perc} & \var{demo.interviews.gender.Woman.perc}  \\
            \hspace{2mm}Non-Binary & \var{demo.upwork.gender.Non-binary.perc} & \var{demo.github.gender.Non-binary.perc} & \var{demo.all.gender.Non-binary.perc} & \var{demo.interviews.gender.Non-binary.perc} \\
            
            \textbf{Age [years]:} \\
            \hspace{2mm}Median & \var{demo.upwork.age.50percentile} & \var{demo.github.age.50percentile} & \var{demo.all.age.50percentile} & \var{demo.interviews.age.50percentile} \\
            \hspace{2mm}Mean & \var{demo.upwork.age.mean} & \var{demo.github.age.mean} & \var{demo.all.age.mean} & \var{demo.interviews.age.mean} \\

            \multicolumn{2}{l}{\textbf{Country of Residence:}} \\
            \hspace{2mm}U.S. & \var{demo.upwork.countries.UnitedStatesofAmerica.perc} & \var{demo.github.countries.UnitedStatesofAmerica.perc} & \var{demo.all.countries.UnitedStatesofAmerica.perc} & \var{demo.interviews.countries.UnitedStatesofAmerica.perc} \\
            \hspace{2mm}India & \var{demo.upwork.countries.India.perc} & \var{demo.github.countries.India.perc} & \var{demo.all.countries.India.perc} & \var{demo.interviews.countries.India.perc} \\
            \hspace{2mm}Germany & \var{demo.upwork.countries.Germany.perc} & \var{demo.github.countries.Germany.perc} & \var{demo.all.countries.Germany.perc} & \var{demo.interviews.countries.Germany.perc} \\
            \hspace{2mm}Pakistan & \var{demo.upwork.countries.Pakistan.perc} & \var{demo.github.countries.Pakistan.perc} & \var{demo.all.countries.Pakistan.perc} & \var{demo.interviews.countries.Pakistan.perc} \\
            \hspace{2mm}Other & \var{demo.upwork.countries.other.perc} & \var{demo.github.countries.other.perc} & \var{demo.all.countries.other.perc} & \var{demo.interviews.countries.other.perc} \\

            \multicolumn{3}{l}{\textbf{Development/Programming Education:\tnote{1}}} \\
            \hspace{2mm}Self-taught & \var{demo.upwork.progexp.learning.Self-taught.perc} & \var{demo.github.progexp.learning.Self-taught.perc} & \var{demo.all.progexp.learning.Self-taught.perc} & \var{demo.interviews.progexp.learning.Self-taught.perc} \\
            \hspace{2mm}College/University & \var{demo.upwork.progexp.learning.College/University.perc} & \var{demo.github.progexp.learning.College/University.perc} & \var{demo.all.progexp.learning.College/University.perc} & \var{demo.interviews.progexp.learning.College/University.perc} \\
            \hspace{2mm}On-the-job training & \var{demo.upwork.progexp.learning.On-the-jobtraining.perc} & \var{demo.github.progexp.learning.On-the-jobtraining.perc} & \var{demo.all.progexp.learning.On-the-jobtraining.perc} & \var{demo.interviews.progexp.learning.On-the-jobtraining.perc} \\
            \hspace{2mm}Online class & \var{demo.upwork.progexp.learning.Onlineclass.perc} & \var{demo.github.progexp.learning.Onlineclass.perc} & \var{demo.all.progexp.learning.Onlineclass.perc} & \var{demo.interviews.progexp.learning.Onlineclass.perc} \\
            \hspace{2mm}Coding camp & \var{demo.upwork.progexp.learning.Codingcamp.perc} & \var{demo.github.progexp.learning.Codingcamp.perc} & \var{demo.all.progexp.learning.Codingcamp.perc} & \var{demo.interviews.progexp.learning.Codingcamp.perc} \\

            \bottomrule
        \end{tabular}
        \begin{tablenotes}
            \item [1] Multiple answers allowed; may not sum to 100\%.
        \end{tablenotes}
    \end{threeparttable}
\end{table}

\subsection{Ethics \& Data Protection}
\label{sec:ethics}\label{sec:data-protection}
The research in this paper was approved by one of our institutions' \gls{erb} (IRB equivalent).
Overall, we adhere to the principles for ethical research outlined in the \emph{Menlo Report}~\cite{Kenneally:2012:menlo}.
In addition, we handle all data according to the strict \gls{gdpr} laws of the European Union (EU). Furthermore, all data is stored in de-identified form, so there is no link between the participants' survey or interview answers and their identity.
Before participating in our studies, we encouraged potential participants to familiarize themselves with consent and data handling information on the study website. 
We obtained informed consent from all participants for participation in the study and having their interview’s audio recorded and transcribed. Before, during, and after the interview, (potential) participants were able to contact us at listed contact addresses for any questions or additional information. Reporting security incidents such as leaked secrets might be very sensitive for participants. Therefore, we always offered the option to skip a question or select ``Prefer not to answer''.

 \revision{Recruiting developers from GitHub has been common in the past~\cite{Acar:2017, Gorski:2018, 9833686}. \revisionII{However, a change in GitHub's terms of service prohibits contacting their users for research purposes~\cite[\S~7]{github-tos}. Therefore, we suggest researchers avoid this recruitment procedure in the future.}

We compensated all Upworkers who completed the survey with \var{survey.participants.compensation.upwork}.
Due to institutional restrictions, we could not compensate survey participants recruited from GitHub. When conducting the interview study later, we solved this issue and could offer compensation to interview participants recruited from GitHub.
}
\subsection{Limitations}\label{sec:limitations}
Our work includes some limitations typical for these survey and interview studies and should be interpreted in context. In general, self-report studies may suffer from several biases, including over- and under-reporting, sample bias, self-selection bias, and social desirability bias~\cite{redmiles2018asking, Acar:2017, peixoto_understanding_2020, senarath_embedprivacy_2018, mhaidli_ad_networks_2019}. Developers who agreed to speak with us could be more (or less) security-conscious than those who declined.

Furthermore, focusing our recruitment on Upwork and GitHub participants might introduce a sampling bias by excluding developers active on other platforms. We chose GitHub due to its high popularity compared to other platforms like GitLab. Kaur et al.\ compared different freelancer recruitment channels for studies with developers~\cite{conf/usenix/kaur22}, and Upwork appears to be the best choice for our study. It offers recruitment for freelancers worldwide, allowing us to gather a more complete picture than country-specific platforms. Our demographics also reflect this diversity (Section~\ref{sec:demographics}).

\section{Survey on Secret Management}
\label{sec:study-results-survey}
\revision{In this section, we report on the findings based on all \var{survey.participants.total}~valid survey responses (\var{survey.participants.total.started}~respondents started, one respondent tried to scam us by copying answers from websites). This includes an overview of 18~prevention and remediation approaches that survey respondents used. 
We include survey respondents’ quotes as transcribed, with minor grammatical corrections and omissions marked by brackets (``\textelp{}''). Survey  respondents are numbered with a leading \textit{S} (\eg{} S4).}
Below, we report on the most relevant and interesting survey questions and responses. A full list of counts and codes for all questions is provided as part of our replication package (\cf{}~\availability). %
For the interview study that we conducted with developers who experienced secret leakage, we report on their experiences in Section~\ref{sec:study-results-interviews}.

\subsection{Version Control Systems and Platforms}
In the survey, we asked respondents for the different \glspl{vcs} and corresponding platforms they used within the last 12~months.
All \var{vcs.Git.text}~respondents reported using Git, which was by far the most popular \gls{vcs}. In addition to Git, \var{vcs.Subversion.text}~respondents (\var{vcs.Subversion.perc}) used Subversion, being the second most popular \gls{vcs}. Others mentioned \gls{cvs} (\var{vcs.CVS}, \var{vcs.CVS.perc}), \gls{tfvc} (\var{vcs.MSTeamFoundationVC}, \var{vcs.MSTeamFoundationVC.perc}), Perforce (\var{vcs.Perforce}, \var{vcs.Perforce.perc}), and Mercurial (\var{vcs.Mercurial}, \var{vcs.Mercurial.perc}).

\begin{figure}[t]
    \centering
    \includegraphics[width=0.68\linewidth]{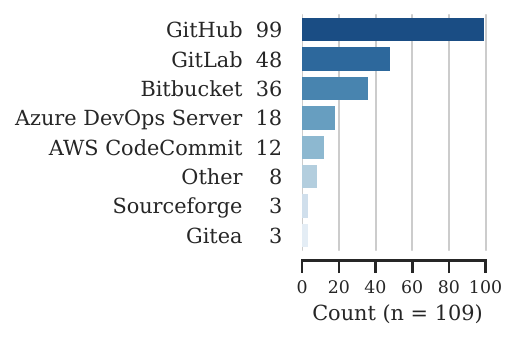}
    \caption{Platform usage reported by the survey  respondents. We allowed multiple answers. }
    \label{fig:platforms}
\end{figure}

Closely related to the high prevalence of Git, both GitHub and GitLab were the platforms most often reported by \var{platform.GitHub.text}~survey respondents (\var{platform.GitHub.perc}) and \var{platform.GitLab.text}~respondents (\var{platform.GitLab.perc}), respectively.
Besides these public services mainly known for open-source software development, respondents reported services targeted at companies that offer commercial features for private source code repositories. 
This included cloud solutions like Microsoft's Azure DevOps Server (\var{platform.AzureDevOpsServer}, \var{platform.AzureDevOpsServer.perc}), \gls{aws} CodeCommit (\var{platform.AWSCodeCommit}, \var{platform.AWSCodeCommit.perc}), and \gls{gc} Source Repositories (\var{platform.GoogleCloudSourceRepositories}, \var{platform.GoogleCloudSourceRepositories.perc}), but also self-hosted solutions like Gitea (\var{platform.Gitea}, \var{platform.Gitea.perc}).
\var{platform.GeneralFileSync.Text} respondent reported to use general file synchronization solutions instead of specialized services or hosting platforms.
Figure~\ref{fig:platforms} depicts system and service usage in detail.

Overall, hosting source code repositories using a third-party service or provider is the most prevalent for both public (\var{repotype.Publicrepository(thirdpartyprovider)}, \var{repotype.Publicrepository(thirdpartyprovider).perc}) or private (\var{repotype.Privaterepository(thirdpartyprovider)}, \var{repotype.Privaterepository(thirdpartyprovider).perc}) repositories. Contrary, self-hosting repositories were reported less than half as often and are more prevalent for private repositories (\var{repotype.Privaterepository(self-hostedownserver)}, \var{repotype.Privaterepository(self-hostedownserver).perc}) than for public ones (\var{repotype.Publicrepository(self-hostedownserver)}, \var{repotype.Publicrepository(self-hostedownserver).perc}).

\subsection{Secrets, Access, and Threat Model}
\label{sec:results-threatmodel}
\revisionII{Regarding the types of secret information, our respondents reported that they predominantly handled login credentials (\var{secrets.past.LoginCredentials}, \var{secrets.past.LoginCredentials.perc}), \gls{api} keys (\var{secrets.past.APIKeys}, \var{secrets.past.APIKeys.perc}), and authentication and access tokens (\var{secrets.past.Auth/AccessTokens}, \var{secrets.past.Auth/AccessTokens.perc}) within their past projects. This is similar for their most recent projects, as shown in Figure~\ref{fig:secret-usage}. Respondents also said that they encountered cryptographic keys, either for personal or server use, as well as special service passwords or keys used in CI/CD pipelines.

\begin{figure}[t]
    \centering
    \includegraphics[width=0.75\linewidth]{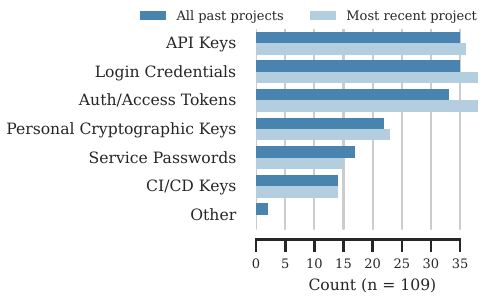}
    \caption{\revision{Secret usage reported by the survey respondents. We allowed multiple answers.}}
    \label{fig:secret-usage}
\end{figure}

Regarding secret access and sharing, the respondents reported that other members within the same team had access \var{threat-model.overview.Teammembers.Hadaccess.text}~times. Most commonly, these were the management (\var{threat-model.overview.Management.Hadaccess.text}) and members of other teams (\var{threat-model.overview.Membersofotherteams.Hadaccess.text}). The general public had access in only \var{threat-model.overview.Public.Hadaccess.text}~cases. Conversely, the public had no access in the majority of cases, which can be inferred from Table~\ref{tab:threat-model-summary} among other details.
\begin{table}[t]

\newcolumntype{R}[2]{%
    >{\adjustbox{angle=#1,lap=\width-(#2)}\bgroup}%
    l%
    <{\egroup}%
}
\newcommand*\rot{\multicolumn{1}{R{30}{1em}}}%

\centering
\scriptsize
\caption{Summary of which groups of persons had access to secrets (besides the participants themselves)\revision{; percentages normalized to 100\% for all groups of each access level.}}
\label{tab:threat-model-summary}
\setlength{\tabcolsep}{0.6\tabcolsep}
\begin{tabular}{lrrrr}
\toprule
{} &  \rot{\shortstack[l]{Other team\\members}} &  \rot{Management} &  \rot{\shortstack[l]{Members of\\other teams}} & \rot{Public} \\
\midrule
Had access             &      \revision{           \var{threat-model.overview.Teammembers.Hadaccess} (\var{threat-model.overview.Teammembers.Hadaccess.perc}) } &    \revision{      \var{threat-model.overview.Management.Hadaccess} (\var{threat-model.overview.Management.Hadaccess.perc}) }&   \revision{                   \var{threat-model.overview.Membersofotherteams.Hadaccess} (\var{threat-model.overview.Membersofotherteams.Hadaccess.perc}) } &  \revision{     \var{threat-model.overview.Public.Hadaccess} (\var{threat-model.overview.Public.Hadaccess.perc}) }\\
Had no access          &                \revision{  \var{threat-model.overview.Teammembers.Hadnoaccess} (\var{threat-model.overview.Teammembers.Hadnoaccess.perc}) }&    \revision{     \var{threat-model.overview.Management.Hadnoaccess} (\var{threat-model.overview.Management.Hadnoaccess.perc}) }& \revision{                    \var{threat-model.overview.Membersofotherteams.Hadnoaccess} (\var{threat-model.overview.Membersofotherteams.Hadnoaccess.perc}) }&  \revision{   \var{threat-model.overview.Public.Hadnoaccess} (\var{threat-model.overview.Public.Hadnoaccess.perc})} \\
I don't know           &           \revision{        \var{threat-model.overview.Teammembers.Idon'tknow} (\var{threat-model.overview.Teammembers.Idon'tknow.perc})} &       \revision{   \var{threat-model.overview.Management.Idon'tknow} (\var{threat-model.overview.Management.Idon'tknow.perc})} &          \revision{             \var{threat-model.overview.Membersofotherteams.Idon'tknow} (\var{threat-model.overview.Membersofotherteams.Idon'tknow.perc}) }&   \revision{    \var{threat-model.overview.Public.Idon'tknow} (\var{threat-model.overview.Public.Idon'tknow.perc}) }\\
Prefer not to disclose &        \revision{          \var{threat-model.overview.Teammembers.Prefernottodisclose} (\var{threat-model.overview.Teammembers.Prefernottodisclose.perc})} &   \revision{       \var{threat-model.overview.Management.Prefernottodisclose} (\var{threat-model.overview.Management.Prefernottodisclose.perc}) }&     \revision{                 \var{threat-model.overview.Membersofotherteams.Prefernottodisclose} (\var{threat-model.overview.Membersofotherteams.Prefernottodisclose.perc}) }&   \revision{      \var{threat-model.overview.Public.Prefernottodisclose} (\var{threat-model.overview.Public.Prefernottodisclose.perc})}  \\
\bottomrule
\end{tabular}
\end{table}

Most respondents shared secrets with their team members, some with their management, only a few with other teams, or with the public (\cf{}~Figure~\ref{fig:threat-model}).
Developers shared passwords for services, API keys, and authentication/access tokens more often than personal keys. 

\begin{figure}[t]
    \centering
    \includegraphics[width=0.75\linewidth]{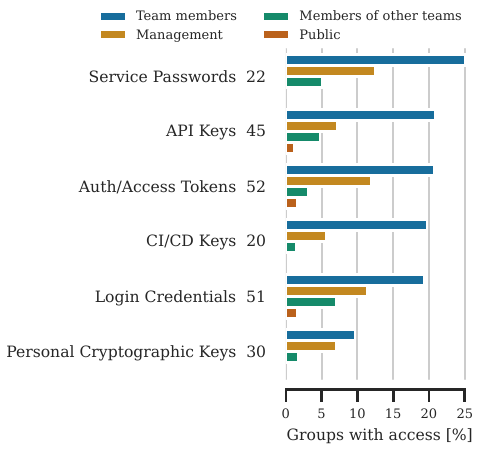}
    \caption{\revision{Respondents' reports on who had access to which secret information in their most recent project. Not considering ``I don't know'' and ``Prefer not to disclose'' answers; normalized to 100\% for all groups of each secret type.}}
    \label{fig:threat-model}
\end{figure}

Access control for secrets was most often configured by the respondents themselves, as \var{secretDecision.themselves}~respondents (\var{secretDecision.themselves.perc}) reported. Sporadically, these decisions were jointly made (\var{secretDecision.joint}, \var{secretDecision.joint.perc}), or at least with some involvement of the respondent (\var{secretDecision.involved}, \var{secretDecision.involved.perc}).
In those cases, decisions were made with architects, team leaders, management, the whole development team, or a security team.}

\revision{
\subsection{Code Secret Leakage Incidents}
\label{sec:survey-incidents}
While a third of all survey respondents (\var{leak.prevalence.Experienced.text}, \var{leak.prevalence.Experienced.perc}) reported a code secret leakage in the past, only a few 10 participants provided further information in the corresponding free text field. 
Survey respondents mainly reported on two types of incidents: secrets that were included in source code, and secrets that were placed in dedicated files, \eg{} secret files or \emph{.env} files, and had been pushed to the repo accidentally, \eg{} \blockquote[S150]{I mistakenly pushed my .env file which includes all of my API keys to the Github repository.}.

\boldparagraph{Impact of Code Secret Leakage}
Respondents faced various consequences of code secret leakage. They reported on, \eg{} delayed project schedules and service downtimes due to the code secret leak.
Furthermore, worse consequences occurred, like the leakage of confidential data, or team members that got fired because of the code secret leak, \eg{} \blockquote[S49]{There was a code secret leakage in one of the projects that I was part of. It led \textelp{} to the loss of confidential information to the public. The project team leader had to cut every team member of the project because, \textins{in his opinion, one team member was involved in the leakage.}}
}
\vspace{-0.8em}

\subsection{Prevention and Remediation Approaches}
\begin{table*}[t]
	\caption{Approaches for preventing and remediating code secret leakage as reported in the survey ($n=109$).}
	\label{tab:findings}
	\begin{threeparttable}
	\centering
	\scriptsize
	\setlength{\tabcolsep}{3pt}
\rowcolors{3}{lightgray}{}
\begin{tabularx}{\textwidth}{lp{3cm}Xrr}
\toprule
            \multicolumn{2}{l}{\textbf{Approach}} &                                                                                                                                                                                                                                                                                                           \textbf{Description} &                                          \textbf{$\#$} &                                                 \textbf{$\%$}\\
\midrule
\multicolumn{2}{l}{\textit{Prevention}}\\
            & \textbf{Externalize Secrets} &                                                                                                                       Separation of code secrets and committed code so that secrets are loaded at runtime, e.g., storing secrets on a central server or secret management system, using environment variables or files \revision{[Hashicorp Vault, Azure Key Vault, AWS Vault, *Password, KeePass, Doppler, python-decouple, GitLab CI, GitHub CI, Travis CI]*} &          \var{coding.prevention.externalizesecrets} &      \var{coding.prevention.externalizesecrets.perc} \\
            & \textbf{Block Secrets} &  Prevent code secrets to be contained in code, config, or any other files or prevent including them in publicly available source code repositories, for example, usage of \textit{.gitignore} files, minimizing secret usage in general or use none, remove secrets from version control before publishing to repository &              \var{coding.prevention.blocksecrets} &            \var{coding.prevention.blocksecrets.perc} \\
            & \textbf{Encrypted Secrets} &                                                                                                                                                                                                                                              Use encryption to store secrets securely within source code repositories \revision{[git-secret, git-crypt, SOPS, GPG, kube-seal]*} &          \var{coding.prevention.encryptedsecrets} &        \var{coding.prevention.encryptedsecrets.perc} \\
            & \textbf{Restrict access} &                                                                                                                                                                    Limit the scope of entities including systems and users with access to code secrets, e.g., by user management, policies, role-based access control &            \var{coding.prevention.restrictaccess} &          \var{coding.prevention.restrictaccess.perc} \\
            & \textbf{Monitoring} &                                                                                                                                     Regular scanning for code secrets and leaks both locally and remote e.g., using secrets scanners in CI/CD pipelines or pre-commit hooks, or review which entities have/had access \revision{[SonarQube, Checkmarx, GitGuardian, AWS Cloud Trail]*} &                \var{coding.prevention.monitoring} &              \var{coding.prevention.monitoring.perc} \\
            & \textbf{Education \& Awareness} &                                                                                                                                                                                   Raise awareness for code secret leakage and educate developers how to handle code secrets, e.g., coding guide, best practices wiki &      \var{coding.prevention.education&awareness} &     \var{coding.prevention.education&awareness.perc} \\
            & \textbf{Other} &                                                                                                                                                                                                                                   Miscellaneous other approaches named by participants, not limited to secret handling &                     \var{coding.prevention.other} &                   \var{coding.prevention.other.perc}\\
            & \textbf{Rotation} &                                                                                                                                                                                                                                                                     Use short-lived secrets, rotate them periodically \revision{[Doppler]*} &                    \var{coding.prevention.rotation} &                \var{coding.prevention.rotation.perc} \\

            & \textbf{Code \& Secret Reviews} &                                                                                                                                                                                                           Manual code reviewing which also focus on code secrets; four or more eyes principle to approve code changes &        \var{coding.prevention.codeandsecretreviews} &    \var{coding.prevention.codeandsecretreviews.perc} \\

\midrule
\multicolumn{2}{l}{\textit{Remediation}} &  &  &  \\
            & \textbf{Renew or Revoke Secret} &                                                                                                                                                                                                                                                           Invalidate leaked code secrets to prevent any future misuse \revision{[Doppler]*} &      \var{coding.remediation.reneworrevokesecret} &    \var{coding.remediation.reneworrevokesecret.perc} \\
            & \textbf{Cleanup VCS History} &                                                                                                                                                                                     Remove leaked secrets from VCSs whole history, e.g., by rewriting the history, clean caches, or reinitialize the whole repository \revision{[BFG Repo Cleaner]*}&       \var{coding.remediation.cleanupvcshistory} &      \var{coding.remediation.cleanupvcshistory.perc}\\
              & \textbf{Analyze Leak} &                                                                                                                                                         Analysis and forensics on the code secret leak to identify root causes or how the leak was exploited, e.g., by auditing logs or consulting security experts, &           \var{coding.remediation.analyzeleak} &            \var{coding.remediation.analyzeleak.perc} \\
            & \textbf{Removal from Source Code} &                                                                                                                                                                                                             Remove leaked secrets from the current code base. This doesn't include version history, caches or similar &      \var{coding.remediation.removalfromsourcecode} &  \var{coding.remediation.removalfromsourcecode.perc}\\
            & \textbf{Notify Concerned Roles} &                                                                                                                                                                                              Inform stakeholders affected or involved in the leak, e.g., security team, management, customers, providers, authorities &      \var{coding.remediation.notifyconcernedroles} &   \var{coding.remediation.notifyconcernedroles.perc} \\
            & \textbf{Access Management} &                                                                                                                                                                                                                    Re-evaluation of access control concepts and applying more restrictive access management if needed &        \var{coding.remediation.accessmanagement} &       \var{coding.remediation.accessmanagement.perc} \\
            & \textbf{Retract Repository} &                                                                                                                                                                                             Delete public repositories affected by the leak or make them private, possibly temporarily until remediation is completed &        \var{coding.remediation.retractrepository} &      \var{coding.remediation.retractrepository.perc}\\

            & \textbf{Systemic Consequences} &                                                                                                                                                                                             Applying consequences due to the secret leak, e.g., new processes, specific education, removal of team members or clients &       \var{coding.remediation.systemicconsequences} &   \var{coding.remediation.systemicconsequences.perc} \\
            & \textbf{Server Operations} &                                                                                                                                                                                      Actions taken to remediate secret leakage in running software, e.g., by backuping systems, or pruning and re-initializing servers &           \var{coding.remediation.serveroperations} &       \var{coding.remediation.serveroperations.perc} \\
            
\bottomrule

\end{tabularx}
\begin{tablenotes}
            \revision{\item [*] Tools our respondents used.}
        \end{tablenotes}
\rowcolors{50}{}{}
\end{threeparttable}

\end{table*}

Based on the survey answers, we found 18~approaches to prevent and remediate code secret leakage that participants had used before, or at least knew about. Table~\ref{tab:findings} provides an overview of both the nine prevention and nine remediation approaches, and their prevalence rates. Each approach is a theme that we extracted from all survey responses.

While most respondents reported the combined usage of multiple approaches (\eg{} externalizing secrets and keeping them out of repositories, or encrypting them within the repositories), some approaches are designed for specific use cases and occurred less frequently (\eg{} code or secret reviews). This applies to both prevention and remediation approaches.
Approaches were either technical or organizational.

\boldparagraph{Prevention Approaches}
The most reported measures were a combination of externalizing secrets (\var{coding.prevention.externalizesecrets}, \var{coding.prevention.externalizesecrets.perc}), \eg{} using dedicated config files or using environment variables, and blocking secrets from the repository itself (\var{coding.prevention.blocksecrets}, \var{coding.prevention.blocksecrets.perc}), \eg{} by using a \textit{.gitignore} file. About a quarter of the participants (\var{coding.prevention.encryptedsecrets}, \var{coding.prevention.encryptedsecrets.perc}) indicated to have stored secrets encrypted to prevent misuse if the secrets had to stay in the repository.

\boldparagraph{Remediation Approaches}
The majority of participants (\var{coding.remediation.reneworrevokesecret}, \var{coding.remediation.reneworrevokesecret.perc}) reported on renewing or at least revoking a leaked secret to prevent further misuse. In addition, they (\var{coding.remediation.cleanupvcshistory}, \var{coding.remediation.cleanupvcshistory.perc}) reported cleaning the \gls{vcs} history, or removing the secret (\var{coding.remediation.removalfromsourcecode}, \var{coding.remediation.removalfromsourcecode.perc}) from source code without cleaning the \gls{vcs} history.

\section{Interviews: Experiences with Secret Leakage}
\label{sec:study-results-interviews}
Based on the 14~developer interviews, we report in-depth qualitative findings below.
We use quantifiers to determine which qualitative findings were more relevant or minor. They do not serve as quantitative statistical.
We include interview participants’ quotes with minor grammatical corrections and omissions marked by brackets (``\textelp{}''). Interview participants are numbered with a leading \textit{I} (\eg{} I4).

\subsection{Secret Leakage Prevention Approaches}
\label{sec:study-results-interviews-prevention}
Interview participants used different approaches to prevent code secret leakage. They also elaborated on several factors on applying these measures, challenges, and lessons learned.
Most participants used approaches to block secrets from getting in their repositories: half (7) of the participants used environment variables to avoid statements of secrets in the source code, many (6) used secret managers like \textit{HashiCorp Vault}~\cite{vaultproject} or \textit{AWS Parameter Store}~\cite{awsparameterstore}, and a few (3) used \textit{.gitignore} files.
A few (3) participants reported that they educated and trained developers to raise their awareness.
Moreover, a small number (3) of participants reported that they use secret scanners and also that they use separate environments for development and production.
Notably, some (3) participants did not use prevention approaches before encountering code secret leakage.

\boldparagraph{Factors}
Participants reported on several factors that influenced the choice and usage of different prevention approaches.
Most important for all (14) participants' satisfaction with an approach was that the approach worked well (in terms of usability and actual code secret leakage prevention). Other than that, approaches have to be effective(9), efficient(5), secure(3), and compliant with company requirements(2).
Participants reported several factors that lead to a negative influence on the usage of prevention approaches. One (1) participant reported that they do not rely on third parties. In consequence, they did not integrate a third-party code secret scanner. Another (1) participant reported, that there were too many constraints for the approach to work: \blockquote[I3]{We are not going to use vault because it would be too many constraints for us.}.
At least, one (1) participant said, that the integration of the approach to their infrastructure would be too complicated.
\subsubsection{Challenges}
\label{sec:study-results-interviews-prevention-challanges}
While a third of all survey respondents (\var{leak.prevalence.Experienced.text}, \var{leak.prevalence.Experienced.perc}) reported to have had code secret leakage in the past, only a few respondents (\var{survey.failed}, \var{survey.failed.perc}) reported that they had failed to apply a code secret handling approach when using environment variables, the \texttt{git filter-branch} command, or external tools, \eg{} vaults.
We could identify further problems with prevention measures from the participants we interviewed.
\boldparagraph{Cost and Time Constraints}
Almost all interview participants complained about time and cost requirements of adopting a code secret management tool. This included the time it took to set up a method or educate all involved developers about the method. It also involved adopting the method into existing projects, often requiring refactoring work.
\boldparagraph{Documentation}
Finally, documentation was one of the more commonly cited challenges, especially \blockquote[I14]{the lack of actual documentation available \textins{for} open source software or open source approaches.}. This required special care and more time from developers during the setup of an approach, while securely deploying an approach in their infrastructure, and while using it.
\boldparagraph{Awareness and Education}
The interviewees also indicated that getting developers to work carefully and use a secret management approach was a challenge: if a tool required too many workflow changes, it would slow down development, so developers try to bypass prevention approaches, \eg{} \blockquote[I6]{Someone was doing something off the books \textelp{}: They were just creating another repository \textelp{} not within the organisation, but maybe just under a personal account or something. Those you can't really fix with tooling, at the end of the day those are just people problems \textelp{} and we can fix that through training \textelp{or} policy.}.
\boldparagraph{Maintenance} Maintaining an approach can be challenging, for example, having to update secret definitions in secret scanners. About half of our interview participants reported that maintenance proved to be challenging when adopting prevention approaches. The biggest complaint was that maintenance was too time-consuming and not user-friendly.

\subsubsection{Lessons Learned}
All in all, participants learned from their previous experiences regarding code secret leakage prevention. Many (6) participants suggested using more (automated) secret scanning, \eg{} \blockquote[I8]{We need an approach to statically analyze \textelp{code for} secret leakage. For example, we are totally missing this kind of automation in our pipeline.}. Furthermore, a few (2) participants used a prevention approach because it is cheaper compared to facing a code secret leakage incident. Moreover, a few (2) participants stated that they would like to know if the company's prevention approach would work in case of a potential leak. They felt uncomfortable because the approach was never tested.
Besides, the human error seemed to be the most relevant factor when it comes to code secret leakage. Therefore, some (8) participants suggested to focus on a better onboarding of new developers.

\subsection{Secret Leakage Incidents}
\label{sec:results-incidents}
There are many ways to leak code secrets in \glspl{vcs}. They depend on the place of a leak, the secret type, if and how a leak was recognized, and the potential consequences. This section sheds light on incidents that our participants reported. 
We learned that some participants experienced code secret leakage multiple times, \eg{} in the same or across different software projects. A few two participants stated that it happened repeatedly at a high frequency. \blockquote[I5]{\textins{Code secret leakage} happens four or five times a year, I would say.}.

\boldparagraph{Places of Leak}
Our interview participants experienced code secret leaks in various places. Most (12) participants reported a secret leak through public source code repositories, including GitHub and GitLab.
Few participants (2) reported leaks through private repositories, to which only a specific group of potentially unauthorized users had access.
 One participant (1) experienced code secret leakage through the log files of the \textit{GitHub Workflow} feature~\cite{github-workflow}. Another (1) participant reported secret leaks on platforms such as \textit{pastebin}~\cite{pastebin} or \textit{gist}~\cite{github-gists}, \eg{} \blockquote[I11]{It wasn't like `I'm going to publish this on the Internet.' They had actually copied and pasted some code to get assistance from somebody in a gist, and the gist happened to be public.}.

\boldparagraph{Type of Leak}
The type and frequency of secret leaks vastly varied between participants. All (14) participants reported secret leaks through hard-coded information in source code or configuration files of a project.
Some (5) interviewees experienced leaks through accidental commits of configuration files that included API keys, tokens, or login credentials, for example.
Other (4) participants leaked access keys and API tokens they used for authentication purposes in deployment infrastructures. %
Few (2) participants leaked \gls{aws} tokens~\cite{aws}.
Finally, few (2) interviewees leaked passwords for databases containing sensitive internal information or customer data.
One participant accidentally pushed all their project secrets to a publicly available GitHub repository, \eg{} \blockquote[I10]{\textins{I was} pushing the commits to GitHub and when I pushed the remote repository, I found that my [password manager database] has gone into GitHub without me wanting it to go to there.}.

\boldparagraph{Leak Detection}
The impact of code secret leaks highly depends on their detection. In particular, the time span until leaks are detected is important. 
Most (7) of our participants reported that GitHub notified them in case of a secret leak so that they could respond quickly. In two (2) cases, GitHub even triggered the revocation of the leaked secrets.

Although GitHub notifications provide immediate feedback in case of secret leaks, some (3) participants reported, that they discovered the secret leak only some weeks after the notification.
\blockquote[I11]{It was probably out there for a couple of weeks. So, yes, that was not amazing.}.
Furthermore, some (4) participants noticed their leak randomly while they did another task, \eg{} reading the logs to debug a program error (1). %
In addition, a few participants (2) were informed by random people that saw their secret in the repository, \eg{} \blockquote[I13]{Actually, it was someone who saw it for me because it was really recently.}.
Moreover, a participant (1) got informed by a third-party secret scanner.

\boldparagraph{Impact}
We observed two different types of consequences caused by code secret leakage. Consequences may directly affect the company or software team that is responsible for the leak. However, we also saw consequences for external stakeholders, such as clients of the developed software or the customers of that client.
Most (6) of our participants reported, that the secret leak caused additional workload for the team. This included the investigation of the leak, as well as the remediation process.
Furthermore, some (3) participants told us, that their company or their team suffered from financial damage.
A few (2) participants reported that they experienced reputational damage to their company, \eg{} \blockquote[I11]{All these issues were there, and we had to tell our clients that this happened, and we had to release it in that there was a security breach \textelp{}.}
Besides these consequences, one participant told us that their leaked secrets got used to crypto mine on their systems. %
In a few (3) cases, external stakeholders got affected since they had to renew all customer and client secrets to prevent misuse. 

\boldparagraph{Root Causes, Access, and Threat Model}
There were various reasons for code secret leakage. Most (4) participants reported that their leaks happened because developers, especially new developers that recently joined the team, were not aware of the consequences of a leak. Furthermore, most (8) participants stated that they did not use any prevention approaches before an incident happened. In summary, developers leaked code secrets because they hard-code secrets in source code, or they did not add secret-containing files to a \textit{.gitignore} file.

Access control configurations for code secrets and the threat model of the developers can affect the likelihood of code secret leakage.
Most (12) participants could describe a threat model for their use cases and included sharing secrets as a risk. Some (6) of them have placed special emphasis on confidentiality. One participant stated that code secrets were not critical to them at all.

Some developers (6) granted access to secrets on request. Few (3) participants reported that their companies or team considered all employees as trusted and granted them full access to all secrets, \eg{} \blockquote[I6]{Really just any time you ask, you'll just get access to whatever you want.}.

\subsection{Secret Leakage Remediation}
\label{sec:study-results-interviews-remediation}
All interviewees experienced code secret leakage. %
Below, we report on the developers' experiences during the remediation approaches, their challenges, and lessons learned.
Most (10) participants reported to have revoked or updated the affected code secrets after they got aware of the leakage. Some (5) immediately implemented remediation. They also analyzed root causes to prevent further future leaks. 
Few (2) participants remediated the leak by taking down the repository or making the repository private.
Moreover, few (2) reported developer education measures on handling secrets securely.
In addition, some participants (3,4) removed the leaked code secrets by rewriting git history or deleting the secret with a new commit.
Notably, a small number (3) did neither revoke nor update leaked secrets. Participants repeatedly reported misconceptions of how \textit{git} works.

While they used several approaches, the kinds of incidents can be various, including the consequences caused by the leak. Depending on the leakage and the consequences, participants chose different remediation approaches.
    
\boldparagraph{Challenges}
Overall, most (5) participants described the process of remediation as cumbersome.
Several (4) complained that they were facing an incident response process which they had never used before and was complicated to use.
Some (3) participants reported, that it was hard to estimate the consequences of their incident. Without being aware of all consequences, it was difficult to implement remediation, \eg{} 

\blockquote[I4]{It was a challenge not being able to say with 100\% certainty that these secrets had never been misused \textelp{}. There are scenarios where somebody could've \textelp{} used it in such a way that it would be very difficult to detect, and we would have missed it. I didn't like that feeling that I couldn't say with certainty there was no impact.}.
Moreover, few (2) participants complained, that no all-in-one solution exists for any kind of incident. Selecting, learning, and applying different or multiple remediation approaches would be too complex and time-consuming.

\boldparagraph{Lessons Learned}
Most (9) participants reported that their remediation process worked well in general. Of those, a few (4) would apply the same process for future incidents.
Besides that, participants requested changes for future incidents. While many (6) participants expressed the need for better tooling to faster and easier remediate an incident, a few (2) generally requested secret scanning to prevent or at least faster remediate leakage.

\section{Discussion}\label{sec:discussion}
\revision{This section discusses our findings, makes recommendations for developers and service providers, and provides ideas for future research. We base the discussion on both studies, using their individual findings to complement each other.

\boldparagraph{Encountering Code Secret Leakage}
Our first point of discussion returns to the \var{leak.prevalence.Experienced.perc} of survey respondents that experienced code secret leakage, which turns out to be highly prevailing (\cf{}~Section~\ref{sec:survey-incidents}) compared to related work. One reason for this is that previous work~\cite{meli2019bad,sinha2015detecting,saha2020reducing,lounici2021detection,kall2021federated} focused on detectable secrets, for example, \gls{api} keys, which were included in our survey.} However, we additionally asked for credentials that need to be shared, and encryption keys a program needs to access. Our participants reported relying on the externalization, blocking, and encryption of secrets. Monitoring secrets through code scanners was less commonly mentioned, which we relate to the high false positive rates~\cite{sinha2015detecting,meli2019bad} that make manual developer review necessary, \eg{} \blockquote[I13]{Most of the time, it just raises warnings about some secrets that are really supposed to be in the code and you have to manually exclude it from being scanned.}.

\boldparagraph{Usability and Adoption Aspects}  
Few participants reported significant challenges with secret handling approaches. Many participants deployed specific approaches due to good usability. At the same time, using approaches that participants used before or already knew about was a major theme. This might indicate an adoption burden that developers could not be willing to overcome, since light-weight approaches like blocking secrets via \gls{vcs} (\eg{} \texttt{.gitignore}) were adopted often while more resource-intensive approaches were not. For example, using short-living secrets that are rotated regularly, is a good fit to reduce the temporal attack surface and therefore potential secret leakage's impact – but was reported rarely. However, this would require more automation. While \emph{\gls{saas}} solutions and secret management tools like \emph{HashiCorp Vault} can provide this out-of-the-box, developers, especially in small development teams, might not use them because they require additional setup and learning (\cf{}~Section~\ref{sec:study-results-interviews-prevention}). \revision{Our findings regarding tool usage are consistent with previous research that focused on security tool adoption~\cite{10.1145/2786805.2786816, 10.1145/2531602.2531722}.
Overall, we believe that approaches need to be light-weight to be adopted, or ideally require no developer effort at all. An excellent example for the latter is GitHub's secret scanning program, which is enabled by default for all public repositories~\cite{github-secret-scanning} -- therefore driving adoption at scale.}

\boldparagraph{Secret Leakage Reporting}
\revision{Interestingly, we found a mismatch in our survey between the low number of reports with problems using a code secret handling approach (\var{survey.failed}, \var{survey.failed.perc}) in relation to the number of people that experienced leakage (\var{leak.prevalence.Experienced}, \var{leak.prevalence.Experienced.perc}).}
Developers could experience leaks in their teams, for example, through a team member accidentally leaking hard-coded data not sufficiently secured through secret management approaches.
Possible explanations include that developers do not relate the experienced code secret leakage with failed secret handling approaches. \revision{As we found in our interviews, they may also have not tried to use any approach, so they could not fail, \eg{} \blockquote[I2]{We were a startup, \textins{we didn't had any prevention approaches in place}, we took all the measures after the secret leakage.}.
Moreover, as a final factor, developers may have used an insufficient approach, so they have not failed this approach directly, but leaked a code secret nevertheless. If developers are unaware of problems or do not report them, it does not mean there are no issues.}

\revision{
\boldparagraph{Access Control Models}
Some developers reported not feeling responsible for secret leakage. Instead, other team members, other teams, management, and clients had access to secrets and caused leaks.} This is likely caused by secrets being shared as part of the development or deployment process (\cf{}~Figure~\ref{fig:threat-model}).
To better support developers with this, future research should work on secure and easy-to-use secret-sharing and management platforms providing secret transfer and revocation between involved stakeholders.

\boldparagraph{Lack of Helpful Resources}
\revision{In both the survey and the interviews, we found a lack of comprehensive resources that guide developers in the case of secret incidents. This is a burden for developers when they experience secret leakage. Although secret leakage is widespread, developers do not encounter it on a daily basis; they cannot be expected to instantly know what measures to take. Nonetheless, an incident ideally requires instant action. Therefore, we argue that easily accessible online resources are needed, and should contain actionable steps for easy, fast, and secure remediation.} We think that code hosting platforms are a place to provide such information.
This also holds for prevention approaches: developers complained about insufficient or missing documentation for deployment or usage (\cf{}~Section~\ref{sec:study-results-interviews-prevention} and \ref{sec:study-results-interviews-remediation}).

\boldparagraph{Constructive Incident Handling}
One interesting consequence of secret leakage was the firing of a full team or the complete termination of client contracts as a reaction to a secret leakage, potentially caused only by a single person. Considering our findings on the resources for secret leakage in general, we assume that this is mainly a problem with education and awareness, so \emph{systemic consequences} like these are unlikely to prevent further leakage, especially considering that team members or clients are usually replaced with new, potentially less experienced stakeholders. Furthermore, it may discourage developers from reporting secret leakage to other team members or leaders, \eg{} \blockquote[I5]{I didn't ask anyone, I knew what to do, I just responded directly.}. \revision{This can be highly problematic considering that the same participant applied insufficient remediation approaches and others could still misuse the leaked secrets.
We propose to instead use restrictive \emph{access management}, support through security teams, and education of team members and clients to prevent issues in companies~\cite{Huaman:2021:cybercrime}.}

\boldparagraph{Developer Awareness}
Developers must be more aware of the risk and consequences of code secret leakage. After studying the experiences, challenges, and needs that developers had with prevention approaches, the human factor seems to be the most relevant regarding code secret leakage prevention, \eg{} \blockquote[I2]{Even with all the technology \textelp{} to prevent secret leakage, the biggest contributor to secret leakage is the human factor, or negligence.}. \revision{Companies need to properly onboard new developers to reduce the risk of secret leakage by training them or by policies (\cf{}~Section~\ref{sec:study-results-interviews-prevention-challanges}). They also need to be trained and should be supported in developing and understanding threat models~\cite{Shostack}. Typically, companies can also strengthen their cybersecurity advocates, as recommended by Haney et al.~\cite{219408}.}

\boldparagraph{Manual vs.\ Automation}
We found that developers desire increased automation for code secret leakage prevention. However, usability and maintenance challenges make the use of automated approaches, including code secret scanners, vaults, and automated rotation services, complicated.  Manual approaches, such as restrictive access models and manual blocking of secrets in the repository, are subject to human error. \revision{Our findings illustrate the challenges of making the right trade-off decisions between security, ease of use, and maintainability of manual and automated approaches. Ultimately, the approach should be based on the developer or team's specific use case, considering their unique requirements and constraints.}

\revision{
\subsection{Recommendations for Developers}
We derive recommendations based on the approaches that we identified (\cf{} Table~\ref{tab:findings}). We discuss these recommendations for more secure and usable code secret management for developers and focus on approaches that our survey and interview participants found to be secure and usable.}

\boldparagraph{Prevention}
\revision{We suggest combining different approaches to decrease the likelihood of code secret leakage. \revisionII{First, developers should \emph{externalize secrets}, \eg{} using environment variables or tools like vaults and secret managers, and \emph{block secrets} in the repositories using, \eg{} .gitignore files.} These approaches can both prevent accidental commits. \emph{Monitoring} , \eg{} using secret scanners, especially as a pre-commit approach, can provide additional security. Some scenarios require developers to share code secrets with others. In such cases, we recommend to \emph{encrypt secrets}, so unauthorized third parties cannot access them. Tools like \eg{} git-secret, SOPS and GPP support comfortable secret encryption.}

\boldparagraph{Remediation}
\revision{Typical steps that should always be taken to remediate code secret leakage effectively are to \emph{renew or revoke secret}s that leaked to prevent further misuse of affected services, \emph{analyze leak}s to identify the root causes of the leak, and \emph{revise the access management} using those results, \eg{} apply more restrictive access management if needed. 
We also consider it essential to \emph{notify the concerned roles} (\eg{} management, security team, customers) for legal and ethical reasons, if not to get the appropriate help from security and privacy experts. 
Overall, we consider the above steps necessary because these steps will handle all consequences of a secret leak. 
The other approaches (\cf{}~Table~\ref{tab:findings}) can be used to complement the essential ones and should be considered as a second step in the circumstantial situation. \emph{Removal from source code} and \emph{cleaning up VCS history} are important steps. 
However, they cannot save a leaked secret on GitHub or similar platforms since those services are frequently crawled for archival purposes~\cite{meli2019bad}. This, and the risk of archiving of public websites emphasize the need to \emph{renew or revoke secret}s that have leaked in public spaces. \emph{Server Operations} and \emph{systemic consequences} (\eg{} introducing new processes)} depend heavily on company policies, the type of leak, and how well leakage damage can be prevented when developers can just \emph{renew or revoke secret}s that have leaked. In case developers have remediation approaches prepared, we suggest testing them to make sure they work as expected in case of a leak. %

\subsection{Recommendations for Service Providers}
Based on our findings, we discuss recommendations for improved code secret management for service providers.
\revision{\boldparagraph{Improving Online Documentation}
We identified a lack of available documentation for secret leakage countermeasures. As platforms are the central instance, those can potentially reach many developers and should therefore provide easy-to-understand, accessible, and actionable guidance on secret leakage prevention and (especially) remediation. We believe that the approaches and recommendations for developers identified in this paper could be a good starting point for providing comprehensive guidance by service providers}.

\boldparagraph{Provide and Extend Secret Scanning}
We highly appreciate the platforms' effort in deploying large-scale secret \emph{monitoring}, \eg{} GitHub~\cite{github-secret-scanning} and GitLab~\cite{gitlab-secret-scanning}, which automatically scan public repositories for secrets and notify developers in case of a leak.
In the example of GitHub, secrets with a known format are scanned using regular expressions. Those formats are supplied by several partnering service providers and limited to their \gls{api} keys and access tokens. Additionally, the tokens are checked by the partner services for validity and automatically revoked if valid~\cite{github-secret-scanning-partner-program}. We believe this to be a suitable approach as it allows fully automated and, therefore, instant remediation. This is currently limited to a set of secrets and would be better if it included more types of secrets, like SSH keys. Although those cannot be revoked automatically (there is no central service provider), at least notifying the developers would be possible.

\subsection{Outlook}
\boldparagraph{Usability of Prevention and Remediation Approaches}
To investigate and improve the  Those would have to solve a programming task applying different code secret leakage prevention approaches to measure and compare their usability, or have to remediate a given secret leak.

\boldparagraph{Improving Secret Detection and Leakage Prevention}
As discussed in related work (Section~\ref{sec:related-work}), general secret detection has high false-positive rates~\cite{sinha2015detecting,saha2020reducing,lounici2021detection,kall2021federated}. Future work should aim to improve detection accuracy so that platforms and developers can get useful secret scanners at hand.
Another aspect that needs to be researched is secret leakage prevention. One approach is to develop and evaluate \gls{api} designs that aim to prevent secret leakage by forcing a strict separation of code and data, \eg{} the secrets. This may ensure security by default.
The appropriate time for both secret scanning and prevention remains uncertain.

\revision{
\boldparagraph{Comparison of Supporting Tools}
Several tools support our identified approaches, \eg{} secret scanners or vaults to externalize secrets and enable automatic rotation.
A user study could investigate the challenges of using these tools. Additionally, how supportive they are in preventing and remediating code secret leakage could be measured.

}

\section{Conclusion}\label{sec:conclusion}
In our online survey with \var{survey.participants.total.text}~experienced software developers, we learned about their experiences with code secret leakage. We identified nine approaches developers use to prevent code secret leakage and nine approaches for code secret leakage remediation. 
\var{leak.prevalence.Experienced.perc} of the survey respondents experienced code secret leakage in the past.
In 14 in-depth, semi-structured interviews with developers who experienced code secret leakage, we identified several problems and challenges that developers face when preventing and remediating code secret leakage.
We make recommendations for both developers and service providers and outline ideas for future research based on our analysis.
Overall, we strongly recommend that developers take preventative measures before any secret is accidentally leaked and be aware of the risks associated with leaking secret information.

\section*{Acknowledgments}
We want to thank all survey participants and interviewees for supporting our research. \revision{Furthermore, we thank the anonymous reviewers and our shepherd for their constructive feedback.}
This research was funded in part by the \mbox{VolkswagenStiftung} Niedersächsisches Vorab -- ZN3695, the Deutsche Forschungsgemeinschaft (DFG, German Research Foundation) under Germany's Excellence Strategy -- EXC 2092 \textsc{CaSa} -- 390781972, and NSF grants CNS-2206865 and CNS-22 07008.
Any findings and opinions expressed in this material are those of the authors and do not necessarily reflect the views of the funding agencies.

\section*{Availability}
To allow full replication of our research as well as meta-research, we provide a replication package at~\url{https://doi.org/10.25835/xfc2h3pg}
. 

The replication package includes:
\begin{enumerate}[noitemsep]
    \item The full survey and interview recruitment materials (including Upwork post and invitation, as well as GitHub invite messages).
    \item The survey screening questions and interview pre-survey questionnaire.
    \item The survey and interview consent form.
    \item The survey questionnaire and interview guide.
    \item The survey and interview codebook.
    \item \revisionII{The background section on version control, source code platforms, and secret information.}
\end{enumerate}

\printbibliography

\appendix
\section{Analysis: Online Resources on Secret Handling}
\label{sec:guide-analysis}
As part of the exploration phase for this study, we researched online guides and documents in summer of 2021 that cover secret leakage prevention and remediation, \eg{} discussing best practices. While we mainly used the insights for the construction of survey questions and the interview guide, the guide analysis itself yielded some minor results that we therefore want to report here for completeness.

Interestingly, the online resources covered only a subset of approaches compared to the survey (cf.~Table~\ref{tab:findings}). 
Regarding leakage prevention, all but \emph{Code \& Secret Reviews} were mentioned. Contrary to the survey, \emph{Block Secrets} (\var{analysis.prevention.blocksecrets}~documents) was the most popular approach, followed by \emph{Monitoring} (\var{analysis.prevention.monitoring}). \emph{Externalizing} and \emph{encrypting} secrets both occurred in \var{analysis.prevention.externalizesecrets}~documents, and \emph{restricting access} in \var{analysis.prevention.restrictaccess}. The remaining approaches occurred only once.
Regarding leakage remediation, only four of the overall nine approaches from the survey could be found. As in the survey, \emph{Renew or Revoke Secret} is the most common approach (\var{analysis.remediation.reneworrevokesecret}~documents). This is followed by \emph{cleaning \gls{vcs} history} (\var{analysis.remediation.cleanupvcshistory}), \emph{analyzing leaks} in detail (\var{analysis.remediation.analyzeleak}), and \emph{access management} considerations (\var{analysis.remediation.accessmanagement}). 
That said, online resources are seemingly more incomplete for remediation of an incident than for prevention. However, the online guides covered the important approaches and lacked only minor ones.

Overall, we found only a few resources on handling code secrets. Many of them were incomplete (only covering one or a few approaches) or inconsistent with each other.
All this indicates a lack of helpful information sources in terms of secret handling for developers.

\section{Interview Questions}
\label{sec:interview-appendix}

In the following, we present the interview questions used to conduct the semi-structured interviews. The initial demographic and screening pre-survey including the consent form, and the complete interview guide  can be found in the replication package (\cf{}~\availability). We used a numbering format where \textit{S} stands for section and \textit{Q} for question.

{\small

\makeatletter
\renewcommand\paragraph{\@startsection{paragraph}{4}{0\parindent}{3pt plus 2pt minus 1pt}{0pt}
  {\noindent\normalfont\bfseries\maybe@addperiod}*%
}
\makeatother

\hypertarget{code-secrets}{%
\subsection*{S1 Code Secrets}\label{code-secrets}}

\paragraph{S1Q1}
\textbf{Secret Types:} Please tell us about the kind of secret information you come into contact with when writing and maintaining your source code.\\
\textbf{Definition Code Secret:} A code secret is any secret information that your program needs to access without user-input. For example, this could be an API key or a private key.
\begin{itemize}[noitemsep]
\renewcommand{\labelitemi}{\scriptsize$\square$}
    \item
      \textbf{S1Q1.1 Use Cases:} Please provide some typical examples of where you used code secrets before?
    \item
      \textbf{S1Q1.2 Sensibility:} What do you think about the “sensibility” of these secrets? How confidential/critical are these secrets?
    \item
        \textbf{S1Q1.3 Access:} How do you decide on components or users that can access these code secrets?  

    \item
        \textbf{S1Q1.4 Participants:} Who (components and people) typically had access to these secrets?  

\end{itemize}

\subsection*{S2 Code Secret Leakage \& Recommendation}
You are here because you experienced code secret leakage in the past.\\
\textbf{Definition: Code Secret Leakage} refers to leaking code secrets, for example through source code repositories, build scripts or CI and similar approaches to source code sharing.\\
\textbf{Examples:} Also, a push of an API-Key to a publicly available source code platform without consequences is considered as a code secret leak. Or if the leak is inside a company platform (e.g., pushing to an internal repo where only other members  of the company have access).

To better understand the prevalence of code secret leakage, we would like to know how often you have experienced it.

\paragraph{S2Q1}
  \textbf{Prevalence:} How often did you experience secret leakage?

Please elaborate on your most impactful or latest code secret leakage and the experiences you had with it.

\paragraph{S2Q2}
  \textbf{Becoming aware:} How did you recognize that a secret leakage happened?

\paragraph{S2Q3}
  \textbf{Experience}
      \begin{itemize}[noitemsep]
        \renewcommand{\labelitemi}{\scriptsize$\square$}
        
            \item
              \textbf{S2Q3.1 Reason:} How did the code secret leak happen?
            \item
            \textbf{S2Q3.2 Consequences:} Can you describe the immediate consequences?
                \begin{itemize}[noitemsep]
                \renewcommand{\labelitemi}{\scriptsize$\square$}
                
                    \item
                      \textbf{S2Q3.2.1 For the Company:} What consequences were there for the company? Cyberattacks? Monetary damage?
                    \item
                        \textbf{S2Q3.2.2 Authorities (If they got attacked):} Did you attempt to contact authorities/ prosecute the attackers?
                    \item
                        \textbf{S2Q3.2.3 For external Stakeholders:} Data Leakage/Monetary Damage/ Other inconvenience for the client or other parties?
                
                \end{itemize}
                
            \item
                \textbf{S2Q3.3 Changes:} Were there any new measures/approaches introduced to prevent secret leakage in the future?

        \end{itemize}

    \paragraph{S2Q4}
      \textbf{ Remediation:} How did you react to the incident? How did you remediate the code secret leak?

        \begin{itemize}[noitemsep]
        \renewcommand{\labelitemi}{\scriptsize$\square$}
        
            \item
              \textbf{S2Q4.1 Experiences:} What were your experiences when applying the approach(es)? What went well?
            \item
            \textbf{S2Q4.2 Involved Roles:} Who was actually involved in remediating the code secret leak?

            \item
            \textbf{S2Q4.3 Challenges:} Did you encounter any challenges? Why?

            \item
                \textbf{S2Q4.4 Resources:} Please tell us about the information sources you used to remediate the code secret leak. In example, online blogs, official documentation, other developers, or the it-security team / incident response team.
            \item
                \textbf{S2Q4.5 Satisfaction:} If a code secret leak happen again, would you like to apply the same process for future code secret leakage remediation, or is there something you would like to change or improve?

        \end{itemize}

\subsection*{S3 Prevention Approaches}
Let's talk about your experiences regarding code secret leakage prevention approaches.

\paragraph{S3Q1}
\textbf{Used approaches:} If you used prevention approaches before, please provide an example of approaches or tools you used to prevent code secret leakage. \textit{(If missing, provide examples: Externalize Secrets, Block Secrets, Monitoring, Restrict Access)}

\begin{itemize}[noitemsep]
\renewcommand{\labelitemi}{\scriptsize$\square$}

    \item
       \textbf{S3Q1.1 Understanding:} Please describe how you used the approach/tool in your project and why?

    \item
      \textbf{S3Q1.2 Reason:} What was the rationale, or reason, why this approach was introduced in your processes?
    \item
      \textbf{S3Q1.3 Experiences:} Please tell us about the experiences with  these approaches and tools. Would you consider it easy to use, effective, or secure?
    \item
        \textbf{S3Q1.4 Challenges:} Please tell us about any challenges or problems you faced.
    \item
        \textbf{S3Q1.5 Satisfaction:} Does the approach fulfill your needs, or do you desire changes to improve the approach?

\end{itemize}

\paragraph{S3Q2}
\textbf{Used approach and failed with:} Have you had problems trying to apply an approach or failed with an approach? Please tell us about them.

\begin{itemize}[noitemsep]
\renewcommand{\labelitemi}{\scriptsize$\square$}

    \item
      \textbf{S3Q2.1 Reason:} What was the reason you decided for the approach you failed with?
    \item
        \textbf{S3Q2.2 Causes of Failure:} What caused the approach to fail (in your specific accident)?
    \item
        \textbf{S3Q2.3 Improvements:} What changes do you suggest could improve the concept?
\end{itemize}

\paragraph{S3Q3}
\textbf{Known, but unused approaches:} Can you name any further approaches that you did not use so far and why?

\begin{itemize}[noitemsep]
\renewcommand{\labelitemi}{\scriptsize$\square$}
    \item
      \textbf{S3Q3.1 Reason}: Why did you decide not to use those? 
        \begin{itemize}[noitemsep]
        \renewcommand{\labelitemi}{\scriptsize$\square$}
            \item
              \textbf{S3Q3.1.1 Future Use:} Do you want to try them for future projects?
             \item
              \textbf{S3Q3.1.2 Experiences:} Do you think the approach would be easy-to-use?

            \item
              \textbf{S3Q3.1.3 Challenges:} Do you think you might encounter any challenges or problems along the way? What are these challenges? 

        \end{itemize}
    \item
    \textbf{S3Q3.2 Satisfaction:} Do you think the approach would fulfill your needs, or are you requesting changes to improve the approach?
\end{itemize}

\paragraph{S3Q3 [IF in S3Q1 stated]}
\textbf{Secret Scanner:} A popular approach to detect secret leaks, are so-called secret scanners, that scan code, repositories, or commits for any contained secrets. What are your experiences if you have any?

\begin{itemize}[noitemsep]
\renewcommand{\labelitemi}{\scriptsize$\square$}

    \item
      \textbf{S3Q3.1 Utilization:} When is this scanner executed in your project? (CI/CD, pre-commit, regular, etc.?)
      \item
      \textbf{S3Q3.2 False Positives:} Do you think the scan results are reliable? Why?
      \item
      \textbf{S3Q3.3 Challenges:} Did you encounter any challenges or problems along the way? What are these challenges? Have you solved them?

\end{itemize}

\paragraph{S3Q4}
\textbf{Not used: [If not used any approaches/tools]:} Have you ever considered using an approach to prevent code secret leakage? Can you elaborate on your decision?

\paragraph{S3Q5}
\textbf{Needs:} Looking back at all your experiences with code secret leakage and its prevention, what do you currently miss or think might be helpful to successfully prevent code secret leakage? \textit{(Then if nothing comes to their mind, we could specifically ask for resources, tools, approaches.)}

}

\end{document}